\documentclass[a4paper]{article}

 \usepackage[sort&compress,numbers,comma,merge]{natbib}		

 \usepackage[margin=2cm]{geometry}	

 \usepackage{hyperref}
 \hypersetup{
  colorlinks   = true, 	
  urlcolor     = blue, 	
  linkcolor    = blue, 	
  citecolor   = red, 		
    linktoc	= page		
}

\usepackage{amsmath}		
\usepackage{amssymb}	
\usepackage{slashed}
\usepackage{bbold}		
\usepackage{relsize }		

\usepackage[dvipsnames]{xcolor}	
\usepackage[normalem]{ulem}		

\usepackage{graphicx}		
\usepackage[font=small,labelfont=bf]{caption}
\usepackage{subcaption}	

\usepackage{footmisc}				
\usepackage{tabu, multirow}			

\usepackage[compat=1.1.0]{tikz-feynman}		
\usetikzlibrary{positioning,arrows.meta}

\tikzfeynmanset{ with arrow/.style = { 			
   decoration={
     markings,
     mark=at position 0.5
          with {   \node[
            transform shape,
            xshift=-0.5mm,
            fill,
            dart tail angle=100,
            inner sep=.8pt,
            draw=none,
            dart		 ] { };     }
     },
   postaction=decorate}, 
   with reversed arrow/.style = { 					
     decoration={
       markings,
       mark=at position 0.5
          with {   \node[
            transform shape,
            xshift=-0.5mm,
            rotate=180,
            fill,
            dart tail angle=100,
            inner sep=.8pt,
            draw=none,
            dart		 ] { };     }
     },
   postaction=decorate},  
  fermion4/.style={ 	/tikz/postaction={ 	/tikz/decoration={ 			
        markings,
        mark=at position 0.4 with {
          \node[
            transform shape,
            xshift=-0.5mm,
            fill,
            dart tail angle=100,
            inner sep=.8pt,
            draw=none,
            dart
          ] { };
        },	},
      /tikz/decorate=true,
    },	},
  anti charged boson/.style={ 		 /tikzfeynman/boson,		/tikz/postaction={ 	/tikz/decoration={ 		
        markings,
        mark=at position 0.5 with {
          \node[
            transform shape,
            xshift=-.2 mm,
            rotate=180,
            fill,
            dart tail angle=100,
            inner sep=.8pt,
            draw=none,
            dart
          ] { };
        },	},
      /tikz/decorate=true,
    },	},
  charged boson4/.style={ 		 /tikzfeynman/boson,		/tikz/postaction={ 	/tikz/decoration={ 		
        markings,
        mark=at position 0.4 with {
          \node[
            transform shape,
            xshift=-.7 mm,
            fill,
            dart tail angle=100,
            inner sep=.8pt,
            draw=none,
            dart
          ] { };
        },	},
      /tikz/decorate=true,
    },	},
  anti charged boson4/.style={ 		 /tikzfeynman/boson,		/tikz/postaction={ 	/tikz/decoration={ 		
        markings,
        mark=at position 0.4 with {
          \node[
            transform shape,
            xshift=-.5 mm,
            rotate=180,            
            fill,
            dart tail angle=100,
            inner sep=.8pt,
            draw=none,
            dart
          ] { };
        },	},
      /tikz/decorate=true,
    },	},
}

\DeclareMathOperator{\tr}{tr}
\DeclareMathOperator{\acot}{Arccot}
\DeclareMathOperator{\atg}{Arctg}

\title{Direct Detection of Electroweak Dark Matter}
\author{Ramtin Amintaheri
\footnote{Ramtin.Amintaheri@gmail.com}
}	
\date{\normalsize \emph{School of Physics, Physics Road, The University of Sydney, NSW 2006 Camperdown, Australia.}}

\begin{document}

\maketitle

\begin{abstract}

TeV-scale dark matter is well motivated by notions of naturalness as the new physics threshold is expected to emerge in the TeV regime. 
We extend the Standard Model by adding an arbitrary SU(2) dark matter multiplet in non-chiral representation. 
The pseudo-real representations can be viable DM candidates providing that one includes a higher dimensional mass-splitting operator, which avoids the tree-level inelastic scattering through Z-boson exchange. 
These effective operators give rise to sizable contributions from Higgs mediated dark matter interactions with quarks and gluons. 
A linear combination of the effective couplings named $\lambda$ is identified as the critical parameter in determining the magnitude of the cross-section. 
When $\lambda$ is smaller than the critical value, the theory behaves similar to the known renormalisable model, and the scattering rate stays below the current experimental reach. 
Nevertheless, above the criticality, the contribution from the higher dimensional operators significantly changes the phenomenology. The scattering amplitude of pseudo-real models will be coherently enhanced, so that it would be possible for next generation  large-exposure experiments to fully probe these multiplets. 
We studied the parameter space of the theory, taking into account both indirect astrophysical and direct search constraints. It is inferred that multi-TeV mass scale remains a viable region, quite promising for forthcoming dark matter experiments.

\end{abstract}


\tableofcontents


\section{Introduction}

Astronomical measurements from sub-galactic to large cosmological scales require the existence of huge amount of obscure non-luminous matter in the universe which is not contained in the Standard Model (SM) of particles \cite{DM1,DM2}. \emph{Dark matter} (DM) constitutes great majority of the total mass density of the cosmos \cite{Planck}, and takes a key role in the characteristics of large and small astrophysical structures \cite{LSS}. 

Among various hypotheses concerning the nature of DM and its interactions, it is compelling to couple the dark sector to the SM in a such minimal way that no new gauge field is introduced. In this method, the positive features of the Standard model is preserved and no new forces are added to the physics. Since weak bosons are the only fundamental force carriers that can mediate interactions with dark matter, this construction can be realised simply through introducing a non-trivial electroweak multiplet.

We generalise the Glashow-Weinberg-Salam (GWS) Theory by an extra fermionic $n$-tuplet of SU(2)\texttimes U(1)\textsubscript{$y$} symmetry group in the non-chiral representation, whose neutral component is referred to as \emph{electroweak dark matter} (EWDM). 
The important feature of the fermionic version is that it limits the renormalisable interactions of the dark sector with the SM only through the electroweak gauge bosons. 
This property allows the limited number of free parameters of the model to be determined robustly; so that the theory can provide accurate phenomenological predictions for a wide range of DM experiments. 
It also avoids emergence of dangerous decay operator that make the system unstable.

Extensions of the SM involving an additional weak multiplet have been studied in a range of contexts such as supersymmetry \cite{Susy_DM}, little Higgs model \cite{Little_Higgs}, inert Higgs models \cite{Inert_2et, Inert_3et}, neutrino mass generation \cite{SeesawIII}, Kaluza-Klein theory \cite{KK1},  etc. These extensions mostly contain lower dimensional representations of SU(2) doublet and triplet. However, larger multiplets have been proposed in recent works such as inert Higgs doublet-septuplet \cite{Inert_2et_7et}, exotic Higgs quintuplet \cite{Higgs_5et}, quintets in neutrino mass mechanism \cite{neutrino_5et_1, neutrino_5et_2}, SO(10) and E\textsubscript{6} scalar and fermionic multiplets in grand unified theories stabilised by a remnant discrete symmetry \cite{GUT_DM,SO10_DM,E6_DM}; and finally fermionic quintuplet and scalar septuplet in minimal dark matter model. The latter remains stable due to an accidental symmetry in the SM gauge groups and Lorentz representations leaving no renormalisable decay mode for DM \cite{MDM, MDM_2009}.

\emph{Direct detection} (DD) as one of the primary means in the search for dark matter, looks for nuclear recoil as the signal of an exotic particle collision with the detector. DM interactions with visible matter are so weak that there is a small probability to detect dark matter scattering off nucleus in the target volume. Nevertheless, if such a rare event is observed, DD experiment would reveal important properties of dark matter including its mass and coupling strength with the SM.

The past literature on direct detection mainly focused on the real representations of the electroweak dark matter, because the phenomenology was straight forward in comparison with the complex models. In addition, the effect of the mass-splitting and higher-dimensional couplings on the DD cross-section has been overlooked in the past. 

In this work, an Effective Field Theory (EFT) approach is employed to describe the non-renormalisable interactions of dark sector with the SM at low energies. We include the lowest-dimensional effective operators that are allowed by the symmetries of the electroweak theory. These operators encapsulate the effects of heavy particles, non-perturbative contributions and ultraviolet completions at higher energies.
The coefficients of these operators can be constrained by the available data from today's relic density, indirect searches, direct detection and collider experiments.

These effective terms in the Lagrangian break the initial U(1)\textsubscript{D} symmetry of the dark sector down to Z\textsubscript{2}. Introducing new off-diagonal components to the mass matrix, they split the pseudo-Dirac dark matter into two Majorana fermions. This mechanism eliminates the dangerous tree-level DM coupling to the nucleon and therefore recovers the pseudo-real representations of EWDM theory, which otherwise would have been ruled out by current constraints. 

In the previous work \cite{Amintaheri:2021xbh}, the thermal masses of all possible EWDM models were computed using freeze-out mechanism. We also studied the gamma-ray probes of the theory in a variety of astronomical sources including the Milky Way's black hole, inner Galaxy and dwarf satellites for continuum and line spectra.

In the current paper, we include the non-renormalisable interactions in the full theory to study the impact of the higher dimensional operators on the effective scattering of EWDM off nucleus. 
The coupling of SU(2) multiplet to Higgs boson induced by five-dimensional operators, gives rise to new scattering diagrams. We evaluate the Wilson coefficients corresponding to these processes, and analyse the behaviour of the spin independent (SI) cross-section with respect to the changes in the higher dimensional coupling constants. Finally, the phenomenological results are confronted with the latest experimental data as well as projected sensitivities of the future DD experiments.

This article is organised as follows:

In the next section, we provide a generalised framework to assess how DM particles can couple to the SM electroweak gauge bosons, and will introduce the electroweak theory of dark matter. The chiral and pseudo-real representations of such DM theory are reviewed, and mass splitting between components of EWDM multiplet is analysed. It will be explained that, as a general rule, any pseudo-real model can avoid bounds from direct detection experiments, by including a dimension five operator that splits the neutral Dirac state into two Majorana components.

After a brief review of the low-energy effective interactions of electroweak dark matter with nuclei in section \ref{sec:Scattering}, we evaluate matrix elements of effective operators at parton level, and discuss the elastic spin-independent cross-section for the scattering process. In section \ref{sec:Wilson}, using Feynman diagram matching, we provide a detailed calculation of the Wilson coefficients for both quark and gluon interactions, and explain the scale dependence of these coefficients. 
Section \ref{sec:constraints} is devoted to numerical computation of the spin-independent cross-section for EWDM scattering off nuclei. The results will be compared with current experimental bounds and projected sensitivities. Next, we combine the constraints from direct detection with astronomical indirect probes to explore the parameter spaces for real and complex models. 
Finally, we conclude the study in section \ref{sec:conclusion}.

Details of the interaction Lagrangian of the dark sector and relevant Feynman rules are presented in the Appendix \ref{app:Feynman}. 
We provide in outline, the loop integrals required to compute the Wilson coefficients of the effective theory in the Appendix \ref{app:Loop}.


\section{EWDM Theory}

In this section, we study the extension of the SM by adding an arbitrary fermionic $ n$-tuplet charged under SU(2)\texttimes U(1)\textsubscript{$y$}. 
A fourth chiral generation is severely disfavoured by electroweak precision \cite{PDG,4th_gen_constraints} and Higgs production experiments \cite{gluon_fusion,Combined_EW_H}, so it is conceivable that dark matter belongs to a non-chiral representation of the electroweak sector. We will briefly review the pseudo-real representations of this theory, investigate the mass split between different components, and explain the phenomenological consequences.


\subsection{Non-chiral Dark Matter}

\emph{Non-chiral} or \emph{vector} fermion has the property that its right and left chirality components transform in the same way under gauge symmetry \cite{Vector1, Vector2}. As a result a gauge-invariant mass term $\bar{\chi}\chi$ is allowed. Their masses are unbounded as the mass is not obtained through EWSB mechanism \cite{Vector3}. Their coupling to the electroweak bosons $Z$ and $W^\pm$ are purely vector $\bar{\chi}\gamma^\mu\chi$ hence having the left and right currents on an equal footing.

Vector fermions are not subject to bounds resulting from the electroweak precision data. A degenerate non-chiral multiplet gives no contribution to the value of $S$, $T$ and $U$ oblique parameters \cite{STU_multiplet}. 

The multiplets are labelled by dimension of the representation and the hypercharge $(n,y)$. 
Considering the normalisation convention $q= y +t^{(3)}$ for Gell-Mann–Nishijima relation, the hyper-charge is the offset of the electric charge from the range of the weak isospin values. 
So the electric charge of the $i^\mathrm{th}$ element in the multiplet reads:
\begin{equation}
	q_i= \frac{1}{2} (n+1)-i+y
\end{equation}

Since dark matter is electrically neutral, then its weak isospin must have the same value as the hyper charge $-t^{(3)}_0 =y$. 
Therefore a generic electroweak gauge n-tuplet takes the form: 
\begin{equation}
	\label{eq:nplet}
	X=\left( \chi^{t+y}, \ldots \chi^q,\dots \chi^0, \ldots \chi^{-t+y} \right)^T
\end{equation}

where $\chi^q$ denotes the element with electric charge $q$. The $\frac{1}{2}(n+1)+y$ neutral component $\chi^0$ is the actual dark matter candidate.%
\footnote{It can be seen that the element n+1-i, in the multiplet with hypercharge -y, has the opposite charge of -(n+1)/2+i-y. So, the opposite hypercharge multiplet can be considered as the reversed order multiplet with elements having opposite electric charges. As an example compare the dark matter quadruplet 
$\left( \chi^{++}, \chi^+, \chi^0, \chi^- \right)^T$
with hypercharge 1/2 with the one 
$\left( \chi^+, \chi^0, \chi^-, \chi^{--} \right)^T$ 
with $y=-1/2$. This becomes a useful property when the conjugate multiplet is defined in \eqref{eq:Xc} as the later is proportional to the conjugate of the former.}

This means that for representations of odd dimension, $y$ takes integer values; while even-dimensional multiplets have half-integer hypercharge.
In addition, the total number of n-tuplets containing a DM candidate equals the dimension of the representation $n=2t+1$ where $t$ is the highest isospin weight.%
\footnote{For example a DM quadruplet with highest weight $t=3/2$ could refer to four possible multiplets 
$\left( \chi^{+++}, \chi^{++}, \chi^+, \chi^0 \right)^T$, 
$\left( \chi^{++}, \chi^+, \chi^0, \chi^- \right)^T$, 
$\left( \chi^+, \chi^0, \chi^-, \chi^{--} \right)^T$, and 
$\left( \chi^0, \chi^-, \chi^{--}, \chi^{---} \right)^T$ 
with hypercharges $y=$ 3/2, 1/2, -1/2 and -3/2 respectively.}

The Standard Model is believed to be a self-consistent description of the physics up to the \emph{Plank scale} $M_{pl}$. As a result the gauge couplings need to remain perturbative up to that cut-off scale. Adding extra SU(2) multiplets will accelerate running of these marginal couplings to the non-perturbative regime which might lead into appearance of \emph{Landau pole} (LP) before the Plank scale. Landau pole is thought to be associated with some new physics mechanisms that violate accidental symmetries of the SM. So we demand LP to be above $M_{pl}$ which will lead into an upper bound on dimensionality of the fermionic EWDM multiplet to $n \le 5$. This result holds true for renormalisation group equations solved up to two-loop level \cite{MDM, RGE}.

Due to different theoretical properties and phenomenological consequences, non-chiral dark matter is usually classified into real and complex representations. We intend to keep focus of this discussion on the pseudo-real models.%
\footnote{Extra information about the real modules can be found in the Appendix B of the reference \cite{Amintaheri:2021xbh}.}


\subsection{Pseudo-real Representation}

For DM in pseudo-real representation of SU(2)$\times$U(1)\textsubscript{y}, the hypercharge is non-vanishing $y \neq 0$. In this case, all components of the multiplet including the neutral DM are Dirac fermions.

The SM is extended by adding the dark sector Lagrangian:
\begin{align}
\label{eq:L_C}
	\mathcal{L}_\mathrm{D} 	&= \overline{X} \left( i \slashed{D} - m \right) X	
				-\frac{\lambda_c}{\Lambda} ( H^\dagger \mathcal{T}^a H) (\overline{X}T^a X)	\\
				& -\frac{\lambda_0}{\Lambda} ( H^\dagger \mathcal{T}^a H^\mathfrak{C}) (\overline{X^\mathfrak{C}}T^a X )
					- \frac{\lambda_0^*}{\Lambda} (H^{\mathfrak{C}\dagger} \mathcal{T}^a H) (\overline{X} T^a X^\mathfrak{C})		
							\nonumber
\end{align}

where $\Lambda$ is the mass scale, $\lambda_0$ and $\lambda_c$ are coefficients, and $T^a$ and $\mathcal{T}^a=\sigma^a /2$ are SU(2) generators in the n-dimensional and fundamental representations respectively.
The covariant derivative reads 
$ D_\mu =	\partial_\mu 	+ iy g_y B_\mu 	+ i g_w	W_\mu ^a T^a$.

The conjugate representation is defined as $X^\mathfrak{C} \equiv C X^c$ where $X^c$ denotes the multiplet with charge conjugated fields. $C$ is an anti-symmetric off-diagonal matrix with alternating $\pm1$ entries, normalised so that it equals $-i\sigma^2$ in 2 dimensions i.e. 
$H^\mathfrak{C} = -i\sigma^2 H^*$. 
More precisely:
\begin{equation}
	C_{i,j} \equiv \delta_{i, n+1-j}	\ (-1)^{\frac{n+1}{2} -i -y}
\end{equation}

So for the generic multiplet \ref{eq:nplet} the conjugate multiplet reads:
\begin{equation}
	\label{eq:Xc}
	X^\mathfrak{C}= \left( (-1)^{-t+y} (\psi^{-t+y})^c, \ldots 	(\psi^0)^c, \ldots 	(-1)^q (\psi^q)^c,\dots 	 (-1)^{t+y} (\psi^{t+y})^c \right)^T
\end{equation}

Note that the scalar product of the adjoint vectors in the the second line, will preserve electric charge only for hyper-charge $y=\frac{1}{2}$, (c.f. equation \ref{eq:neutral}). 
This also means that such an operator only exists  for even-dimensional multiplets. 
It can also be seen that other terms in the Lagrangian respect a U(1)\textsubscript{D} symmetry of the form 
$X \to e^{i\theta} X$ where $\theta$ is an arbitrary real parameter.
The mentioned effective operator enables coupling of two conjugated DM multiplets $X^*$ with Higgs boson, and therefore explicitly breaks this symmetry down to Z\textsubscript{2} under which dark matter particles are odd $X \to -X$.%
\footnote{ Such U(1)\textsubscript{D} or Z\textsubscript{2} invariances can potentially originate from some unknown fundamental local gauge symmetry masqueraded as discrete global symmetries to an observer probing at low energies \cite{Discrete_Sym}.
These discrete symmetries provide interesting phenomenological applications, for example in supersymmetric extensions of the SM with $Z_n$ generalised matter parities and $R$-parity \cite{Zn_SUSY1, Zn_SUSY2, R_SUSY} or grand unified theories with an extra U(1) symmetry which guarantees the stability \cite{Z_GUT, U1_GUT}.}


\subsection{Mass Splitting}

\begin{figure} [t] 				
	\begin{subfigure}{.5\linewidth}
		\centering
		\includegraphics[width=\linewidth]{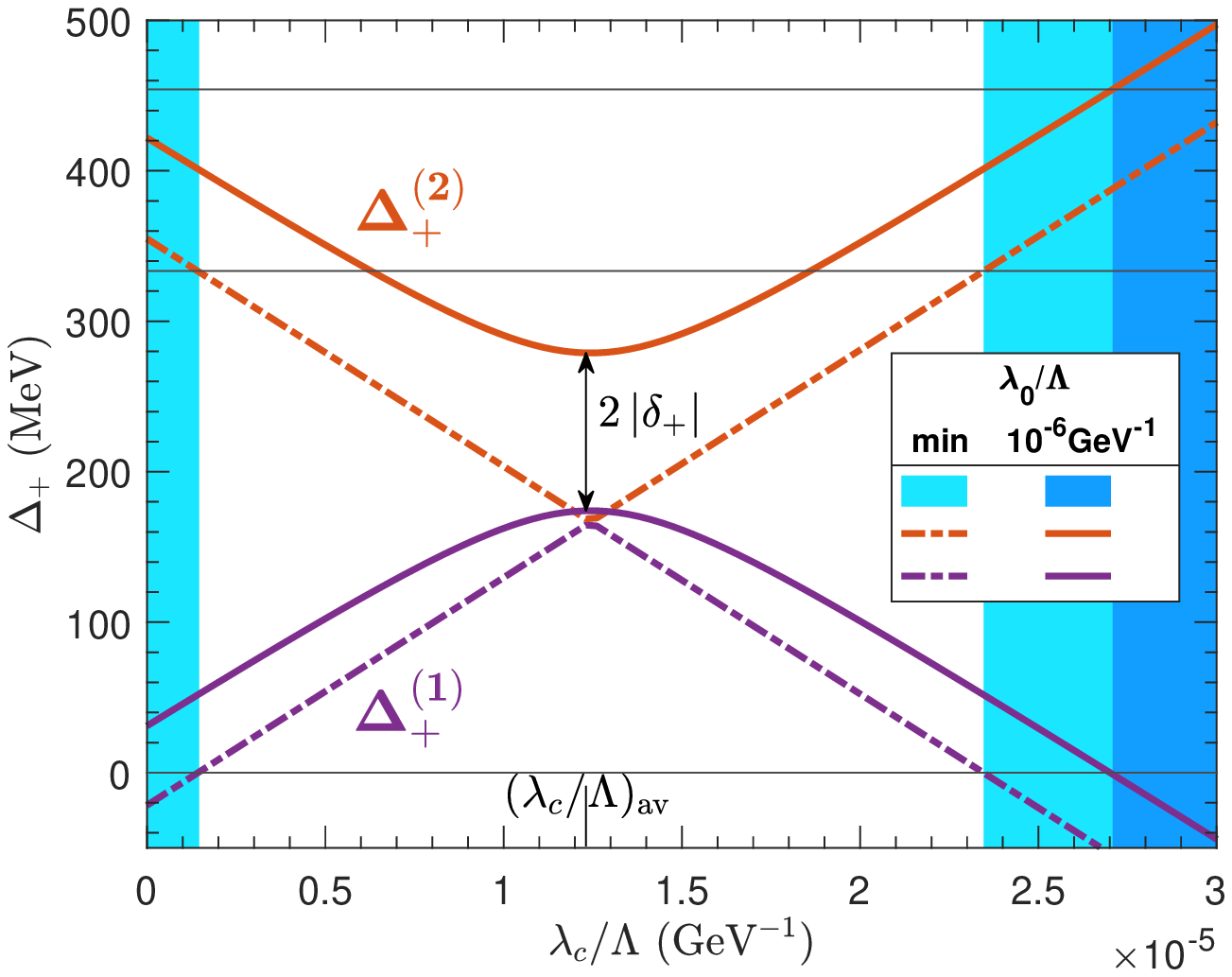}
		\subcaption{}
		\label{fig:Dm_lc}
	\end{subfigure}%
	\begin{subfigure}{.5\linewidth}
		\centering
		\includegraphics[width=\linewidth]{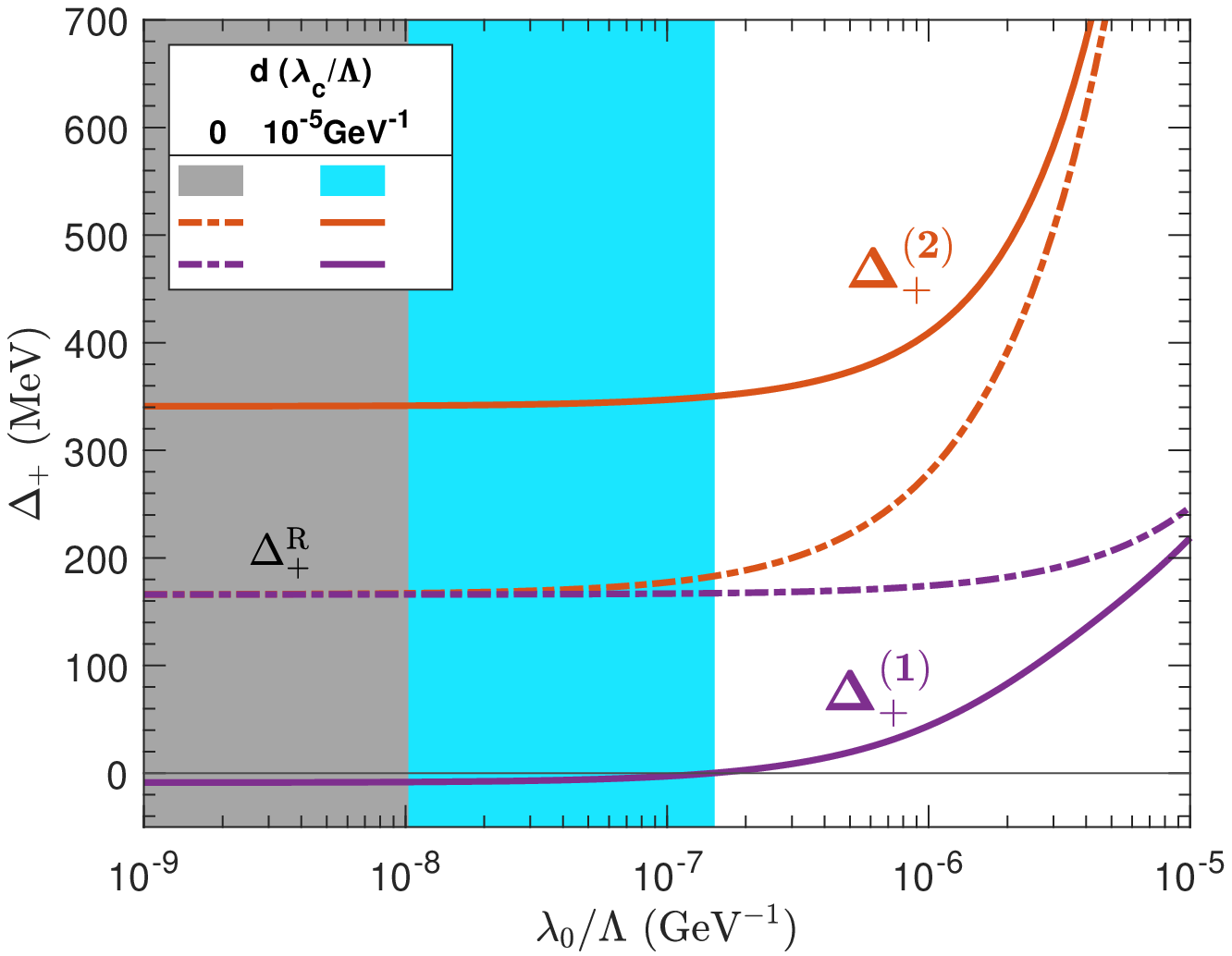}
		\subcaption{}
		\label{fig:Dm_l0}
	\end{subfigure}
	\caption{Left \ref{fig:Dm_lc}: Mass splitting between the dark matter candidate $\chi^0$ and the singly charged states $\chi^+_1$ (in blue) and $\chi^+_2$ (in red) as a function of the non-renormalisable coupling $\lambda_c/\Lambda$ for the complex quadruplet. The figure shows mass-splitting for two values of the other non-renormalisable coefficient $\lambda_0/\Lambda$. The curves corresponding to 
$\lambda_0/\Lambda = 10^{-6} \,\mathrm{GeV}^{-1}$
are in solid lines, while those at the minimum 
$(\lambda_0 /\Lambda)_\mathrm{min} = 10^{-8} \, \mathrm{GeV^{-1}}$ 
\eqref{eq:l0_min} are plotted in dashed-dotted style. The light (dark) blue shaded regions are excluded at minimum ($10^{-6} \,\mathrm{GeV}^{-1}$) value of $\lambda_0/\Lambda$ as the theory cannot be considered as a dark mater model 
$\Delta_+^{(1)} <0$ according to \ref{eq:lc_min} \eqref{eq:lc}.\\
Right \ref{fig:Dm_l0}: The right panel illustrates the mass splitting of these fields $\Delta_+^{(1)}$ (blue) and $\Delta_+^{(2)}$ (red) with respect to the effective coupling $\lambda_0/\Lambda$ for the same quadruplet. The solid lines show the result in case the other coupling has 
$\mathrm{d}\lambda_c/\Lambda = 1.1\times10^{-5} \,\mathrm{GeV}^{-1}$ 
 deviation from the mean value, and the dashed-dotted curves correspond to the average value of 
$(\lambda_c / \Lambda)_\mathrm{av}	=	1.2\times10^{-5} \mathrm{GeV}^{-1}$. 
In the light blue coloured region, the $\chi^+_1$ field becomes the lightest particle 
$\Delta_+^{(1)} <0$ as shown in \eqref{eq:l0}, 
when $\lambda_c$ deviates from the mean, beyond the range specified in \eqref{eq:lc_min}. The grey shaded area is ruled out by DD experiments due to inelastic scattering of DM off target nucleus via exchange of $Z$ boson at tree-level \eqref{eq:l0_min}. \\
Note that in both figures, the blue coloured regions still remain a valid beyond the Standard Model (BSM) proposal. }
	\label{fig:Dm}
\end{figure}

The effective operator proportional to $\lambda_c$ is responsible for mass splitting between the neutral and charged components at tree-level. After electroweak symmetry breaking,  $H\to \left(0, (h+\nu)/\sqrt{2}) \right)^T$, the mass splitting can be cast as:
\begin{equation}
	\Delta_q^{(\mathrm{t})}	\equiv	m_q^{(\mathrm{t})} -m_0^{(\mathrm{t})}		=	- \frac{\lambda_c}{4\Lambda} \nu^2 q
\end{equation}

where $m_q^{(\mathrm{t})}$ indicates the tree-level mass of the component $\psi^q$, and $m_0^{(\mathrm{t})}=m + \frac{\lambda_c}{4\Lambda} \nu^2 y$.

In addition, self energy corrections via coupling to the SM gauge bosons, at one loop order, induce radiative mass splitting between charged and neutral particles which takes the form ~\cite{MDM_2009}:
\begin{equation}
	\Delta_q^{(\ell)}	\equiv	\quad	m_q^{(\ell)} -m_0^{(\mathrm{t})}		=	\quad	q\, ( q +\frac{2}{c_w} y )\, \Delta
\end{equation}

here $m_q^{(\ell)}$ is the loop-induced mass of $\psi^q$ field, and radiative mass splitting $\Delta$ is given by $\Delta = \alpha_w m_w \sin^2 (\theta_w/2) \approx$ 166 MeV \cite{dm_R}.

Interactions of DM with gauge fields are encoded in the kinetic term of the Lagrangian \ref{eq:L_C}. The covariant derivative after EWSB takes the form 
$D_\mu = \partial_\mu 	+i eQA_\mu 	+i \frac{g_w}{c_w} \left( T^{(3)} -s_w^2 Q \right) Z_\mu 	+ i \frac{g_w}{\sqrt{2}}  \left( W^-_\mu T^- +W^+_\mu T^+ \right)$.

Due to DM vector coupling to the neutral weak gauge field at tree-level, it can scatter coherently off the nuclei by $Z$-boson exchange \cite{Witten_DM}. The resultant cross-section is so large that the pseudo-real electroweak dark matter would  be excluded, based on current experimental data \cite{XENON_2018}.

However, complex dark matter scenario can be resurrected if DM-Z boson interaction at tree level could be avoided, by adding a new mechanism. This can be done through introducing an effective operator which violates the would-be U(1)\textsubscript{D} symmetry and allows for splitting of the Dirac neutral state, as will be explained. 

After EWSB, the non-renormalisable operator proportional to $\lambda_0$ reduces to:
\begin{equation}
\label{eq:neutral}
	-\frac{\lambda_0}{\Lambda} ( H^\dagger \mathcal{T}^a H^\mathfrak{C}) (\overline{X^\mathfrak{C}}T^a X )		+ \mathrm{cc}	
	\quad = \quad
	-\frac{\lambda_0}{\Lambda}	\frac{(h+\nu)^2}{4}		\sum_{q=-t+\frac{1}{2}}^{t-\frac{1}{2}}	(-1)^q		\sqrt{ n^2 -4q^2} \ 
		(\overline{\psi}{}^{-q})^c  \,\psi^q 		+ \mathrm{cc}
\end{equation}

Vacuum expectation value (VEV) of Higgs field makes an additional contribution to the mass matrix of the neutral components, and their mass term will change to:
\begin{equation}
	\begin{pmatrix}		(\overline{\psi}{}^0)^c		&\overline{\psi}{}^0 \ 	\end{pmatrix}
	\begin{pmatrix}		\delta_0	&\frac{m_0^{(\mathrm{t})}}{2} 		\\	\frac{m_0^{(\mathrm{t})}}{2} 	&\delta_0^*	\end{pmatrix}
	\begin{pmatrix}		\psi^0	\\ 	(\psi^0)^c		\end{pmatrix}	=	
	\frac{1}{2}	\begin{pmatrix}		\overline{\widetilde{\chi}^0} 	& 	\overline{\chi}^0	\end{pmatrix}
				\begin{pmatrix}		\widetilde{m}_0  	&0	 \\	0	&m_0	\end{pmatrix}
				\begin{pmatrix}		\widetilde{\chi}^0	\\	\chi^0	\end{pmatrix}.
\end{equation}

So, the the scalar product of the adjoint vectors \ref{eq:neutral} splits the masses of the neutral components through the 
$\delta_0 = n \, \nu^2 \lambda_0  / 8\Lambda$ 
term. As will be discussed $|\delta_0| \ll  m$, and DM mass eigenstates up to zeroth order in $\mathcal{O}(\Im \, \delta_0/m)$ can be written as:
\begin{equation}
	\widetilde{\chi}^0 =	\frac{1}{\sqrt{2}}		\left(	  \psi^0   + (\psi^0)^c	 \right)\,,			\qquad
	\chi^0 = 				\frac{1}{\sqrt{2}i}		\left( \psi^0   - (\psi^0)^c 		\right)\,.
\end{equation}

with masses $\widetilde{m}_0 = m_0^{(\mathrm{t})} + 2\Re\,\delta_0$, and $m_0 = m_0^{(\mathrm{t})} - 2\Re\,\delta_0$ respectively. The mass eigenstates $\widetilde{\chi}^0$ and $\chi^0$ are Majorana fermions into which pseudo-Dirac state $\psi^0$ is split. Without loss of generality, we take the imaginary field $\chi^0$ to be the lightest DM candidate.

Equation \ref{eq:neutral} shows that $\lambda_0$ term also introduces an off-diagonal contribution proportional to \\
$\delta_q \equiv (-1)^q \, \nu^2 \sqrt{ n^2 -4q^2} (\lambda_0/8\Lambda)$ 
causing mixing of the charged states $\psi^q$ and $(\psi^{-q})^c$.
\footnote{It should be emphasised that for a Dirac particle $\psi^q$ and $(\psi^{-q})^c$ are not the same.}
Then using the relations $\overline{\psi}{}^{-q} \psi^{-q} = (\overline{\psi}{}^{-q})^c (\psi^{-q})^c$ and $(\overline{\psi}{}^q)^c \psi^{-q} = (\overline{\psi}{}^{-q})^c \psi^q$, the mass term for the charged particles with $q>0$ can be cast as:
\begin{equation}
	\begin{pmatrix}		\overline{\psi}{}^q		&(\overline{\psi}{}^{-q})^c \ 	\end{pmatrix}
	\begin{pmatrix}		\mathfrak{m}_q + 	d_q		&2\delta^*_q	\\
					2\delta_q		&\mathfrak{m}_q - d_q	 	\end{pmatrix}
	\begin{pmatrix}		\psi^q	\\ 	(\psi^{-q})^c		\end{pmatrix}	=	
		\begin{pmatrix}		\overline{\chi}^q_2 	& 	\overline{\chi}^q_1	\end{pmatrix}
		\begin{pmatrix}		m^{(2)}_q  	&0	 \\	0	&m^{(1)}_q	\end{pmatrix}
		\begin{pmatrix}		\chi^q_2	\\	\chi^q_1	\end{pmatrix}.
\end{equation}

After diagonalisation of the mass matrix, the following new eigenstates are obtained:
\begin{equation}
	\chi^q_1 = -e^{i\widehat{\lambda}_0/2} s_q  \ \psi^q		+ e^{-i\widehat{\lambda}_0/2}	c_q  \ (\psi^{-q})^c \,,			\qquad
	\chi^q_2 = e^{i\widehat{\lambda}_0/2}	c_q	\ \psi^q	+e^{-i\widehat{\lambda}_0/2} s_q	\ (\psi^{-q})^c \,.
\end{equation}

where the hat in the exponent should be understood as argument $\widehat{\lambda}_0  \equiv  \arg \lambda_0$.

With corresponding masses:
\begin{equation}
	m_q^{(1),(2)}= \mathfrak{m}_q	\mp \sqrt{ 4 |\delta_q|^2	+d^2_q}	
				\ \approx \ 	 \mathfrak{m}_q  \mp   |d_q|	\,,
\end{equation}

here $m_q$ is the mass of the charged field $\chi^q$, and without loss of assumption, we take $m_q^{(1)}<m_q^{(2)}$. 
We also define $\mathfrak{m}_q \equiv m^{(\mathrm{t})}_0 +  \Delta_q^\mathbb{R}$, and 
$d_q \equiv \left[ 2y c_w^{-1} \Delta -(\lambda_c /4 \Lambda) \nu^2 \right] q$. 
The radiative mass splitting takes the form 
$\Delta_q^\mathbb{R}	\equiv	m_q -m_0		=	q^2 \Delta$ \cite{MDM}. 
The splitting between the neutral and $\chi^\pm$ components is given by $\Delta = \alpha_w m_w \sin^2 (\theta_w/2) \approx$ 166 MeV \cite{dm_R}.

The mixing angle $\phi_q$ is defined such that $\sin \phi_q = |\delta_q| / \sqrt{  |\delta_q|^2	+d^2_q /4 }$. 
For ease of notation we denote $c_q \equiv \cos(\phi_q /2)$ and $s_q \equiv \sin(\phi_q /2)$.

The dark matter candidate $\chi^0$ needs to be the lightest member of the multiplet $m_q > m_0$. 
This condition does not always hold true for the lighter charged field $\chi^q_1$. Demanding $m^{(1)}_q$ to be less than DM mass i.e. 
$4 |\delta_q|^2	+d^2_q		<	 \left( \Delta_q^\mathbb{R}		+2\Re \delta_0 \right)^2$,	
further imposes constraints on the value of the non-renormalisable coupling constants $\lambda_c$ and $\lambda_0$. It turns out that if $\lambda_c$ stays within the following range, then there will be no limit on the value of the neutral coupling $\lambda_0$:
\begin{equation}
	\label{eq:lc_min}
	4 \left(c_w^{-1} -1 \right)	 \frac{\Delta}{\nu^2}	\approx	1.5 \times 10^{-6} \, \mathrm{GeV}^{-1}
		\quad	<	 \frac{\lambda_c}{\Lambda}	<	\quad
	4 \left(c_w^{-1} +1 \right) \frac{\Delta}{\nu^2}	\approx	2.3 \times 10^{-5} \, \mathrm{GeV}^{-1}	\,,
\end{equation}

with an average of 
$(\lambda_c / \Lambda)_\mathrm{av}	=	4 \Delta / c_w \nu^2		\approx 1.2\times10^{-5} \mathrm{GeV}^{-1}$.

This requirement further sets the limit for the scale of new physics that induces the charged-neutral splitting to $\Lambda/\lambda_c \sim 10^5$ GeV.

Figure \ref{fig:Dm_lc} illustrates the mass splitting of the charged states $\chi^+_1$ and $\chi^+_2$ as a function of the non-renormalisable coupling $\lambda_c$, for pseudo-real quartet. 
Note that the range of acceptable values of the charged coupling 
$ \mathrm{d}(\lambda_c/\Lambda) 	\equiv		\left| \lambda_c -\lambda_c^\mathrm{av} \right| / \Lambda $:
\begin{equation}
	\label{eq:lc}
	\mathrm{d}(\lambda_c/\Lambda) 		\le		\ \frac{4}{\nu^2}		\sqrt{ ( 2\Re \delta_0	+\Delta_+^\mathbb{R})^2		-4 |\delta_+|^2}
\end{equation}
 
increases as the other coupling $\lambda_0$ gets stronger. 
It can be seen that the mass difference between the states 
$\Delta_+^{(2)} - \Delta_+^{(1)}$ 
increases as the coupling deviates from the mean value. It has a minimum of $4|\delta_+|^2$ at the average $\lambda_c^\mathrm{av}$, and the particles with the same charge $q$ become degenerate 
$ m_q^{(1)} = m_q^{(2)} = \Delta_q^\mathbb{R}$ 
if $\lambda_0 \approx 0$.

If $\lambda_c$ exceeds the range \eqref{eq:lc_min}, then there will be a lower bound on the value of the neutral coupling:
\begin{equation}
	\label{eq:l0}
	\frac{\lambda_0}{\Lambda}	>	\frac{1}{\nu^2}		\left[ -n \Delta	+ \sqrt{ 	(n^2 -4) \Delta^2	+4 d_+^2 } \right] \,.
\end{equation}

The left panel \ref{fig:Dm_l0} of the figure shows changes in the mass splitting of theses states with respect to the coupling $\lambda_0$. It can be observed that difference between the masses of particles increases with coupling strength going up. The minimum allowed value of $\lambda_0$ grows as the other coupling $\lambda_c$ stays away from the average.

The highest weight particle $\chi^{n/2}$ is unique and has a mass of:
\begin{equation}
	m_\frac{n}{2}	=	\quad	m_0^{(\mathrm{t})} 	+\Delta_\frac{n}{2}^{(\mathrm{t})}	+\Delta_\frac{n}{2}^{(\ell)}
					=	\quad	m_0^{(\mathrm{t})} 	+\Delta_{\frac{n}{2}}^\mathbb{R}		+d_{\frac{n}{2}}
\end{equation}

The requirement for $\chi^{n/2}$ to be lighter than DM candidate $\chi^0$ demands:
\begin{equation}
	\label{eq:l_hw}
	\frac{\lambda_c}{\Lambda}	-2 \Re \frac{\lambda_0}{\Lambda}	<		\frac{8}{n \nu^2}	\Delta_\frac{n}{2}^{(\ell)}
\end{equation}

This imposes an upper bound on the strength of $\lambda_c$. 
If the charged coupling is less than 
$\lambda_c/\Lambda		<		4 \left(c_w^{-1} +1 \right) 	\Delta/\nu^2	\approx	2.3 \times 10^{-5} \, \mathrm{GeV}^{-1}	$, 
there is no restriction on the value of the other coefficient $\lambda_0$. 
In the event that $\lambda_c$ exceeds this limit, the neutral coupling $\lambda_0$ will be bound from below.

In all the complex scenarios except the doublet case, the bounds on the non-renormalisable couplings from the lighter charged state \eqref{eq:lc}, \eqref{eq:l0} prove to be stronger than that of the highest weight \eqref{eq:l_hw}. So in practice, the conditions above is only applicable to C2 model.

\begin{figure} [t] 				
	\begin{subfigure}{.5\linewidth}
		\centering
		\includegraphics[width=\linewidth]{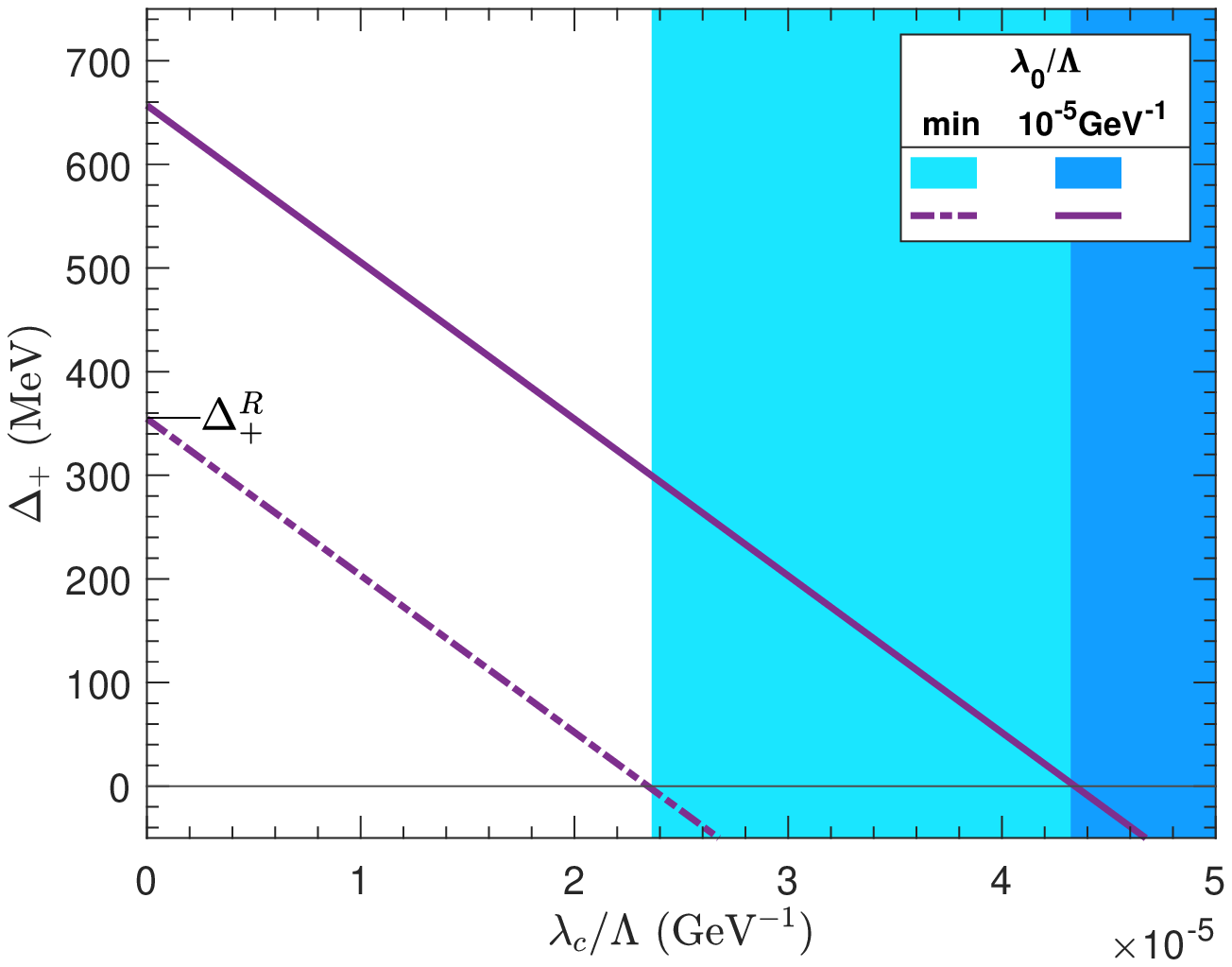}
		\subcaption{}
		\label{fig:Dm_C2_lc}
	\end{subfigure}%
	\begin{subfigure}{.5\linewidth}
		\centering
		\includegraphics[width=\linewidth]{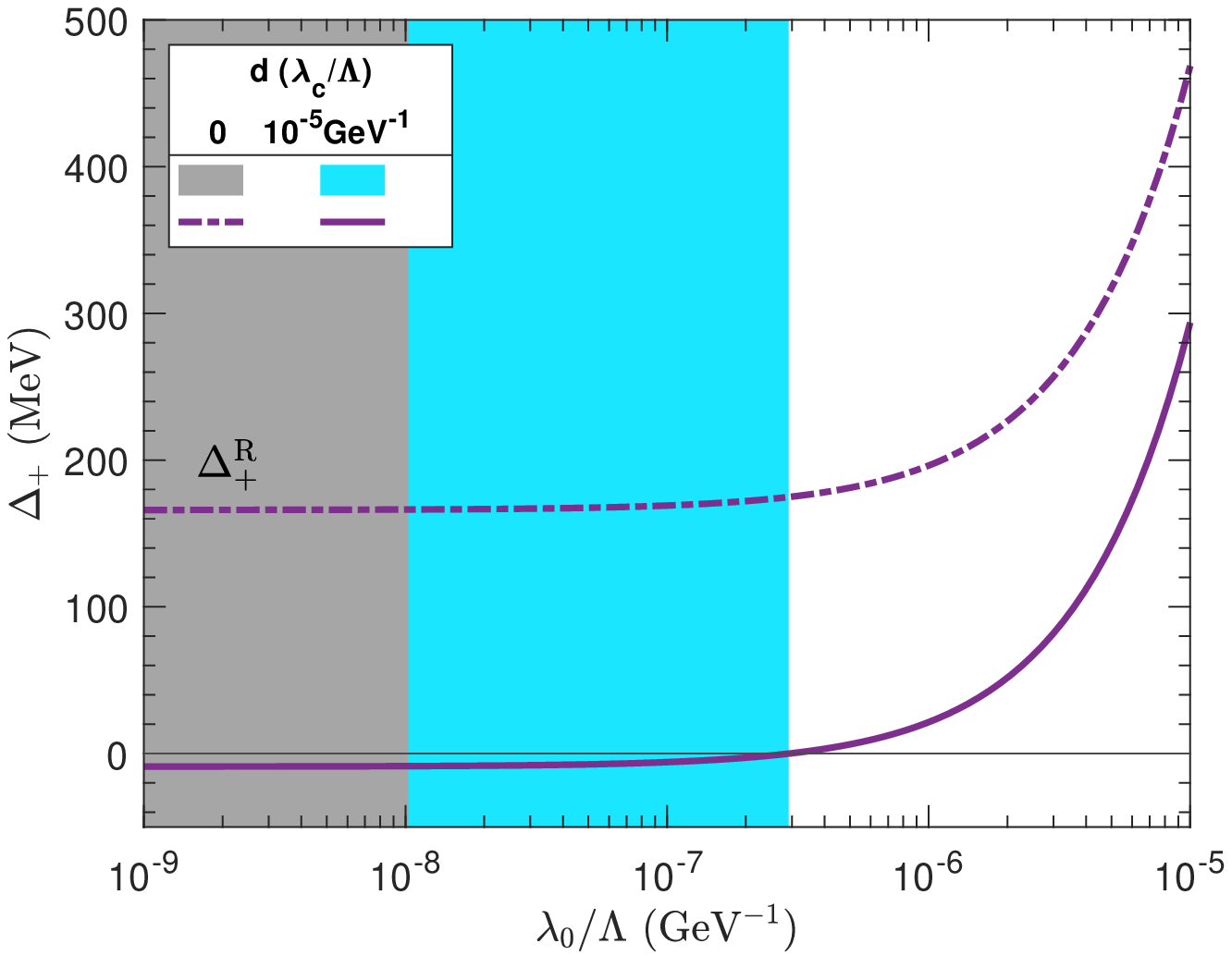}
		\subcaption{}
		\label{fig:Dm_C2_l0}
	\end{subfigure}
	\caption{Left \ref{fig:Dm_C2_lc}: Mass splitting of the highest weight state $\chi^+$ with respect to $\lambda_c/\Lambda$ coupling in the pseudo-real doublet model. The neutral coefficient is fixed to its minimum value of 	
	$(\lambda_0/\Lambda)_\mathrm{min} = 10^{-8} \mathrm{GeV}^{-1}$ and also 
	$\lambda_0/\Lambda = 10^{-5} \mathrm{GeV}^{-1}$ 
	for solid and dashed-dotted lines respectively. The light (dark) blue region is excluded at minimum ($10^{-5} \mathrm{GeV}^{-1}$) due to the requirement \eqref{eq:l_hw} for DM $\chi^0$ to be the lightest particle.\\
	Right \ref{fig:Dm_C2_l0}: Mass difference of the $\chi^+$ field as a function of $\lambda_0/\Lambda$ coupling. The dashed-dotted line correspond to the average strength of 
	$(\lambda_c / \Lambda)_\mathrm{av}	=	1.2\times10^{-5} \mathrm{GeV}^{-1}$, and solid line corresponds to the deviation 
	$\mathrm{d}\lambda_c/\Lambda = 1.2\times10^{-5} \,\mathrm{GeV}^{-1}$.
	The grey patch is ruled out as a result of the direct detection constraint \eqref{eq:l0_min}. The light blue area is excluded since $\chi^+$ becomes heavier than DM candidate, according to \eqref{eq:l_hw}.}
	\label{fig:Dm_C2}
\end{figure}

Figure \ref{fig:Dm_C2_lc} plots the mass splitting of the $\chi^+$ state as a function of the charged coupling in the pseudo-real doublet. It can be seen that the maximum allowed value of $\lambda_c$ increases with strength of the other coupling $\lambda_0$.

The right panel of this figure \ref{fig:Dm_C2_l0} illustrates the changes in the mass splitting of the highest weight field with respect to the neutral coupling $\lambda_0$. The lower bound on $\lambda_0$ goes up as $\lambda_c$ gets further from its mean value.


\section{EWDM--Nucleon Scattering}
\label{sec:Scattering}

The scattering event rate in direct detection experiment is very sensitive to the actual form of coupling between dark matter particle and parton constituents of the hadron. At the beginning of this section, we provide a brief review on inelastic scattering process, and the restrictions imposed on the coupling strength. Then, we consider the full set of relativistic operators that contribute to the effective interactions of EWDM with quarks as well as gluons at the leading order. Finally, the matrix elements of these operators are evaluated at different scales, and the spin-independent scattering cross-section will be discussed.


\subsection{Inelastic Scattering}

Interactions of the neutral dark particles are obtained from expansion of the gauge-fermion kinetic term in \ref{eq:L_C}:
\begin{align}
	\mathcal{L}^0_\mathrm{int} & =		
	\frac{i}{2} \, g_\mathrm{z} \ \overline{\widetilde{\chi}^0} \gamma^\mu \chi^0 	Z_\mu 		\\	\nonumber
	& + \frac{g_\mathrm{w}}{4} \left[ 	
	\left(   n \ e^{-\frac{i}{2} \widehat{\lambda}_0} c_+   - \sqrt{ n^2 -4} \ e^{\frac{i}{2} \widehat{\lambda}_0} s_+ \right)	\overline{\widetilde{\chi}^0} \gamma^\mu \chi_2^+
	-i 	\left(   n \ e^{-\frac{i}{2} \widehat{\lambda}_0} c_+   + \sqrt{ n^2 -4} \ e^{\frac{i}{2} \widehat{\lambda}_0} s_+ \right)	\overline{\chi^0} \gamma^\mu \chi_2^+  
					\right.	\\	\nonumber
	& \left. - \left(   n \ e^{-\frac{i}{2} \widehat{\lambda}_0} s_+   + \sqrt{ n^2 -4} \ e^{\frac{i}{2} \widehat{\lambda}_0} c_+ \right)	\overline{\widetilde{\chi}^0} \gamma^\mu \chi_1^+
	+i 	\left(   n \ e^{-\frac{i}{2} \widehat{\lambda}_0} s_+   - \sqrt{ n^2 -4} \ e^{\frac{i}{2} \widehat{\lambda}_0} c_+ \right)	\overline{\chi^0} \gamma^\mu \chi_1^+  	
	\right]	W^-_\mu 		+ \mathrm{cc}
\end{align}

where $g_\mathrm{z} \equiv g_\mathrm{w}/c_\mathrm{w}$. 
The full Lagrangian of the dark sector is lengthy and is presented in Appendix \ref{app:Feynman}, together with the Feynman rules for couplings to EW gauges and scalar Higgs.

It can be seen that dark matter particle $\chi^0$ has no interaction with Z-boson, so elastic scattering off the nuclei is forbidden at tree level. Elastic scattering still can occur via non-renormalisable DM-Higgs interactions \eqref{eq:L_H} or through different loop-induced processes. The relevant cross-section is either suppressed by the mass scale of the theory or loop-suppressed, and therefore expected to be small. 
In the remainder of this paper, we perform an explicit evaluation of the Feynman diagrams, and exact calculation of the cross-section to qualitatively and quantitatively predict the direct search results for these processes.

However, there exists a coupling between $\chi^0$ and $\widetilde{\chi}^0$ via Z-boson exchange which leaves a possibility for DM-nucleon inelastic scattering $\chi^0 N\to \widetilde{\chi}^0 N$ that would be excluded by many orders of magnitude by direct search experiments.%
\footnote{If the coupling constant $\lambda_0$ is not purely real, then there exists a tree-level inelastic scattering process that is mediated by scalar Higgs field (c.f. coupling \eqref{eq:C_h3}). The amplitude for this reaction is, nevertheless, suppressed by 
$(m_\mathfrak{q}/\Lambda) (\Im \lambda_0 /\alpha_\mathrm{w}) $
compared with Z vector induced diagram.}

Inelastic scattering of dark matter off nucleus is an \emph{endothermic} process in which DM kinetic energy loss 
$ \Delta K_\chi = ( m_0 v_\chi^2 - \widetilde{m}_0 {v'_\chi}^2 ) /2$ 
is converted into the mass difference of the outgoing odd particle $\Delta_{\widetilde{0}}$ and recoil energy of the target nucleus $E_R$:
\begin{equation}
	\label{eq:iDM}
	\Delta K_\chi	=	\Delta_{\widetilde{0}}	+ E_R
\end{equation}

One needs to Impose momentum conservation 
$ 2m_\mathcal{N} E_R 	=	(m_0 v_\chi)^2	+(\widetilde{m}_0 v'_\chi)^2	 	-2 m_0 \widetilde{m}_0 v_\chi v'_\chi \cos \theta$, 
in \ref{eq:iDM}, to eliminate $E_R$ which is a difficult to measure quantity. Solving for the unknown parameter, outgoing DM velocity $v'_\chi$ we obtain:
\begin{equation}
	\widetilde{m}_0 ( \widetilde{m}_0 + m_\mathcal{N} ) \,{v'_\chi}^2		-2 \,m_0 \widetilde{m}_0 v_\chi \cos \theta \,v'_\chi
				+2 m_\mathcal{N} \Delta_{\widetilde{0}}		+ m_0 (m_0 -m_\mathcal{N} )  v_\chi^2		=	0
\end{equation}

where $\theta$ is the scattering angle.
For an endothermic reaction, there exist a \emph{threshold energy} $K_\chi^{(\mathrm{th})}$ below which no solution is available for the equation above \cite{dm_inelastic, iDM}, and thus scattering is not possible:
\begin{equation}
	K_\chi^{(\mathrm{th})}		=		\left( 1 +	\frac{ m_\chi }{ \Delta_{\widetilde{0}} +m_\mathcal{N} }	\right)	\Delta_{\widetilde{0}}
\end{equation}

So keeping the kinetic energy of the incoming particle less than the threshold 
$m_0 v_\chi^2 /2	<	K_\chi^{(\mathrm{th})}$,
dark matter $\chi_0$ cannot up-scatter to the excited state $\widetilde{\chi}_0$. In the high mass regime, we finally arrive at the condition:
\begin{equation}
	\Delta_{\widetilde{0}}	>	\frac{1}{2} m_{\mathcal{N} \chi}	v_\chi^2
\end{equation}

It means if the mass splitting $\Delta_{\widetilde{0}}$ is sufficiently large, then the inelastic nucleonic scattering will be kinematically forbidden. Therefore, by setting 
$ \Delta_{\widetilde{0}}		 > 	\mathcal{O}(100)$ keV 
the tree-level coupling to nuclei is completely suppressed.

In addition, this restricts the value of the neutral pseudo-Dirac splitting coefficient to 
\begin{equation}
	\label{eq:l0_min}
	\Re \, \lambda_0 /\Lambda > 10^{-8} \, \mathrm{GeV^{-1}} \,. 
\end{equation}

In other words, the scale of the new physics responsible for breaking U(1)\textsubscript{D} symmetry is set to $\Lambda/\lambda_0 \sim 10^8$ GeV.

So, in the following discussion, we focus our study on elastic direct detection processes.


\subsection{Effective Interactions}

The effective Lagrangian describing EWDM -- nucleon scattering at parton level is composed of two parts: 
$\mathcal{L}_\mathrm{R}$ which is constructed only from the renormalisable interactions of the UV theory, while $\mathcal{L}_\mathrm{NR}$ allows for non-renormalisable couplings of DM to the SM Higgs \eqref{eq:L_H}:
\begin{equation}
	\mathcal{L}_\mathrm{eff}	=	\mathcal{L}_\mathrm{R}			+\mathcal{L}_\mathrm{NR}	\,.
\end{equation}

The low-energy Lagrangian of the electroweak dark matter composed of renormalisable couplings, has already been studied in the literature. We provide a review in the appendix \ref{app:L_Ren} to calibrate our results with previous publication.

The effective Lagrangian which includes non-renormalisable interactions, containing DM bilinear operators of up to dimension--4, in leading order, is given by:%
\footnote{We only use those operators which are not velocity suppressed in leading contribution. Also, recall that vector and tensor bilinears vanish for a Majorana DM.}
\begin{equation}
	\label{eq:L_NR}
	\mathcal{L}_\mathrm{NR}	=	
		\sum_{\mathfrak{q}=d}^{b}	m_\mathfrak{q}		\left( C_\mathfrak{q}^\mathrm{h3}	+	C_\mathfrak{q}^\mathrm{h4} \right)		\overline{\chi^0} \chi^0	\,\overline{\mathfrak{q}} \mathfrak{q}
		+	\frac{\alpha_\mathrm{s}}{\pi}		\left( C_\mathfrak{g}^\mathrm{h3}	+	C_\mathfrak{g}^\mathrm{h4} \right) 		\overline{\chi^0} \chi^0		G_{\mu\nu}^a G^{\mu\nu a}		\,,
\end{equation}

where $\mathfrak{q}$ and $G_{\mu\nu}^a$ denote quark field and gluon field strength tensor. 
$C_\mathfrak{q}^\mathrm{h3}$ $\left( C_\mathfrak{q}^\mathrm{h4} \right)$ and $C_\mathfrak{g}^\mathrm{h3}$ $\left( C_\mathfrak{g}^\mathrm{h4} \right)$ 
are respectively \emph{Wilson coefficients} for EWDM--quark and EWDM--gluon scatterings that include non-renormalisable DM-Higgs cubic \eqref{eq:C_h3} (quartic \eqref{eq:C_h4}) coupling. 

As will be explained, the dominant loop momenta in the Feynman diagrams of the full theory are generally around the weak scale. So, we set the UV scale at $\mu_\mathrm{UV} \approx m_\mathrm{z}$, and therefore consider an effective theory with $n_f = 5$ active quarks lighter than this characteristic mass $m_\mathfrak{q} < \mu_\mathrm{UV}$, that are $d$, $u$, $s$, $c$, and $b$.

Since the quark scalar operator breaks the chiral symmetry of QCD, it is suppressed by quark mass $m_\mathfrak{q}$.
Electroweak dark matter scatters off gluons at loop level, therefore the effective interactions are suppressed by strong structure constant $\alpha_\mathrm{s} \equiv g_s^2 / 4\pi$.
As will be discussed later, in order to make the quark and gluon Wilson coefficients $C_\mathfrak{q}$ and $C_\mathfrak{g}$ comparable, we multiply them by factors of $m_\mathfrak{q}$ and $\alpha_\mathrm{s} / \pi$ respectively.%
\footnote{Note that any such factors as $m_\mathfrak{q}$ and $\alpha_\mathrm{s} / \pi$ will be compensated when matching the effective and full theories.}

All the operators in the Lagrangian above, are scalar type, and thus only generate Spin-Independent (SI) interactions.


\subsection{Matrix Elements}

\begin{table}[t]
	\centering
	{\setlength{\tabulinesep}{3pt}
	\begin{tabu}{|[1pt]c|[1pt]c|c|[1pt]}
		\tabucline[1pt]{-}
		$f_{\mathrm{T}\mathfrak{q}}$		&	Proton				&		Neutron			\\
		\tabucline[1pt]{-}
		$f_{\mathrm{T}d}$					&		0.0234(23)			&		0.0298(23)			\\
		\hline
		$f_{\mathrm{T}u}$					&		0.0149(17)			&		0.0117(15)			\\
		\hline
		$f_{\mathrm{T}s}$					&		0.0440(88)			&		0.0440(88)			\\
		\hline
		$f_{\mathrm{T}c}$					&		0.085(22)			&		0.085(22)			\\
		\tabucline[1pt]{-}							
	\end{tabu}}
	\caption{Mass fraction parameter for different quark flavours \cite{fTc}. The digits in parentheses show the statistical uncertainty.}
	\label{tab:fTq}
\end{table}

Nucleon indeed obtains the bulk of its mass $M_N$ through spontaneous chiral symmetry breaking, even in the limit of vanishing quark masses. However, a small fraction is attributed to the quark $\sigma$-\emph{terms} which cause explicit breaking of the chiral symmetry. The contribution of valance and sea quarks to the nucleon mass is parametrised by the quark \emph{mass fraction}:
\begin{equation}
	f_{\mathrm{T}\mathfrak{q}}		=	\frac{m_\mathfrak{q}}{m_N}		\langle \bar{\mathfrak{q}} \mathfrak{q} \rangle	\,,
\end{equation}

where the matrix element 
$\langle \bar{\mathfrak{q}} \mathfrak{q} \rangle	\equiv	 \langle N | \bar{\mathfrak{q}} \mathfrak{q} | N \rangle$, 
evaluates the scalar operator on the nucleon state $| N \rangle$.

For light quarks, this quantity can be determined experimentally from $\sigma$ terms of the nucleon-pion scattering (for up and down) and kaon-nucleon scattering (for strange). 
The pion-nucleon sigma term 
$\sigma_{\pi N} = (m_u + m_d) \langle \bar{u} u +\bar{d} d \rangle /2$, 
can be read off $\pi-N$ scattering amplitude at \emph{Cheng–Dashen point} \cite{Cheng_Dashen_pt} using dispersive analysis \cite{sigma_KH,sigma_GW}.

Alternatively, we can use chiral perturbation theory ($\chi$PT) where $\sigma_{\pi N}$ depends on a set of low energy constants which can be determined by fitting to the experimental $\pi-N$ scattering data \cite{BXPT}. 
Different extensions of this theory to the baryonic sector have been studied for this purpose, including heavy baryon $\chi$PT \cite{HBXPT1,HBXPT2}, infrared B$\chi$PT \cite{IR_HBXPT1,IR_HBXPT2,IR_HBXPT3}, and covariant B$\chi$PT with extended-on-mass-shell scheme \cite{EOMS_HBXPT}.

Determination of strangeness content of the nucleon is theoretically more involved \cite{Bernard_2008}. One can input the value of $\sigma_{\pi N}$ to B$\chi$PT, and use the relationship between SU(3) flavour violation parameter and strangeness fraction to obtain $f_{Ts}$ \cite{Meissner_97,Gasser_82}.

For heavy quarks, neither a theoretical framework nor any phenomenological experiment exist, and hence a need for lattice QCD non-perturbative simulations. Lattice calculations are performed using two different techniques \cite{Sigma_AnthonyThomas}.
 
In the indirect method, the quark matrix element is obtained from variation of the nucleon mass with respect to the quark mass \cite{Indirect_1,Indirect_2}
$\langle \bar{\mathfrak{q}} \mathfrak{q} \rangle	=	\partial M_N / \partial m_\mathfrak{q}$ 
through the Feynman-Hellmann theorem \cite{Feynman,Hellmann}.

In an alternative way, one can obtain the scalar matrix element from the ratio of the three-point to two-point functions of the nucleon. The 3-point function arises from two types of diagrams. The connected part contains the propagator of the valence up and down quarks. On the other hand, in the disconnected diagram, the sea quarks form a vacuum blob. The $\sigma_{\pi N}$ receives contribution from both parts, but for the sigma term of other quarks, only the disconnected piece is present \cite{Engelhardt_2011,QCDSF,XQCD}.

In this work, we use the quark mass fractions listed in table \ref{tab:fTq} which are computed using lattice QCD simulation by \cite{fTc}. 
It can be observed that heavier flavours have larger fraction of the nucleon mass.

Finally, worth noting is that scalar quark operator is independent of the scale to all orders 
$(\partial / \partial \mu ) \, m_\mathfrak{q}	\overline{\mathfrak{q}} \mathfrak{q} =0$.

QCD symmetric and gauge-invariant \emph{energy-momentum tensor} (EMT) can be derived from the Noether current associated with space-time translation invariance \cite{EMT1}:
\begin{equation}
	\Theta^{\mu\nu}		=		-G^{a\mu\lambda} \, G^{a \nu}_\lambda 		+\frac{1}{4} \eta^{\mu\nu} G^{a\lambda \rho} G^a_{\lambda \rho}
								+ \sum_\mathfrak{q}	\frac{i}{2} \overline{\mathfrak{q}}   \left( \gamma^\mu D^\nu 	+\gamma^\nu D^\mu  \right)	\mathfrak{q}
\end{equation}

This tensor gives rise to the same physical momentum and generators of the Lorentz symmetry as the canonical one does \cite{EMT2}. 
It can be shown that $\Theta_{\mu\nu}$ is identified with derivative of the dilatation current which corresponds to the scaling transformation \cite{dilatation}.

Like any rank-two tensor, energy-momentum tensor can be decomposed into traceless and trace parts, in $d$ dimensions as \cite{Ji94,Ji95}:
\begin{subequations}
\begin{align}
	\Theta^{\mu\nu}	&	=	\mathcal{O}^{\mu\nu}		+\widehat{\Theta}^{\mu\nu}		\\
						&\equiv		\left( \Theta^{\mu\nu}		-\frac{\eta^{\mu\nu}}{d}	\Theta^\lambda_\lambda \right)		
																+\frac{\eta^{\mu\nu}}{d}	\Theta^\lambda_\lambda
\end{align}
\end{subequations}

where the symmetric traceless parts \cite{Wilson_70}:
\begin{subequations}
\begin{align}
	\mathcal{O}^\mathfrak{q}_{\mu\nu}		&=		\frac{i}{2}	\bar{\mathfrak{q}} 	
				\left(	\gamma_\mu D_\nu 		+	\gamma_\nu D_\mu 		-\frac{1}{2} \eta_{\mu\nu} \slashed{D}		\right)	\mathfrak{q}		\\
	\mathcal{O}^\mathfrak{g}_{\mu\nu}		&=			G_{\nu\rho}^a G_\mu^{a\rho} 		-\frac{1}{4} \eta_{\mu\nu} G_{\rho\sigma}^a G^{a\rho\sigma}
\end{align}
\end{subequations}

are known as \emph{spin-2 twist-2} operators for quark and gluon respectively. Here, \emph{twist} is defined as the difference between mass-dimension and spin. The QCD covariant derivative reads 
$D_\mu = \partial_\mu 	+i g_\mathrm{s} A_\mu$, 
where $A_\mu \equiv A_\mu^a \gamma^a /2$
with $\gamma^a$ meant to be Gell-Mann matrices here.%
\footnote{In general one can define \emph{spin-$n$ twist-2} operators for quark and gluon by symmetrised traceless relations \cite{Twist2}:\\
$\mathcal{O}^\mathfrak{q}_{{\mu_1} \dots {\mu_n}}		\equiv		
		\bar{\mathfrak{q}} 	\left(	\gamma_{\{ \mu_1} i D_{\mu_2} \dots i D_{\mu_n \}}	\right)	\mathfrak{q}	\ - $ trace\,,  \qquad 
$\mathcal{O}^\mathfrak{g}_{{\mu_1} \dots {\mu_n}}		\equiv		
		- G_{ \nu \{ \mu_1}^a		\left(	i D_{\mu_2} \dots i D_{\mu_{n-1}}		\right)	G_{\mu_n  \}}^{a\nu} 		 \ - $ trace.\\
where $A_{\{\mu} B_{\nu\}}	\equiv	\left( A_\mu B_\nu	+	A_\nu B_\mu \right) /2$ for arbitrary operators $A$ and $B$. Note that spin-$n$ twist-2 gluon operator equals a total derivative for odd $n$.
}

Using equation of motion, the classical trace of EMT simplifies to 
$\Theta^\mu_\mu = 	\sum_\mathfrak{q} 	m_\mathfrak{q}	\overline{\mathfrak{q}} \mathfrak{q}$, 
which vanishes in the limit of zero quark mass. This, in fact, indicates that dilatation current is conserved and QCD is therefore \emph{scale invariant} at classical level. 

However, scale symmetry is broken due to running of the coupling constant which is an intrinsic quantum effect. Working in $d = 4 - \epsilon$ dimensions, the quantised $\Theta^\mu_\mu$ differs from the classical version by a divergent term \cite{FF_renorm} that should be renormalised 
$ - (\epsilon /4) G_{\mu\nu} G^{\mu\nu}	=	(\beta / 4 \alpha_\mathrm{s} ) ( G_{\mu\nu} G^{\mu\nu} )_\mathrm{R} 	
-\gamma_m \sum_\mathfrak{q} 	m_\mathfrak{q}	\overline{\mathfrak{q}} \mathfrak{q}$. 
This is known as \emph{trace anomaly} and is composed of contributions proportional to beta-function $\beta$, and mass anomalous dimension $\gamma_m$ \cite{EMT_Nielsen}.

Therefore, the renormalised trace of the full energy momentum can be expressed as \cite{SVZ_78}:
\begin{equation}
	\label{eq:trace}
	\Theta^\mu_\mu  =		 \frac{ \beta }{ 4 \alpha_\mathrm{s} } \, G_{\mu\nu} G^{\mu\nu}		
							+ \left( 1 -\gamma_m \right)	 \sum_\mathfrak{q} 	m_\mathfrak{q}	\overline{\mathfrak{q}} \mathfrak{q}	\,,
\end{equation}

where the \emph{beta function} and \emph{quark mass anomalous dimension} to leading order are given by:
\begin{align}
	\beta (\alpha_\mathrm{s})		&\equiv	\frac{\partial \alpha_\mathrm{s}}{\partial \ln \mu}
				=	\frac{\alpha_\mathrm{s}^2}{2\pi}	\left(	- \frac{11}{3} N_\mathrm{c}		+ \frac{2}{3} n_f 	\right)		\,,		\\	\nonumber
	\gamma_m 	(\alpha_\mathrm{s})		&\equiv		\frac{\partial \ln m_\mathfrak{q}}{\partial \ln \mu}
				=	- \frac{3 \,\alpha_\mathrm{s}}{2 \pi}	\frac{N_\mathrm{c}^2 -1}{2 N_\mathrm{c}}		\,.	
\end{align}

with $N_\mathrm{c} = 3$ being the number of colours.

Either by applying the viral theorem in a stationary state \cite{Manoukian,Kashiwa}, or using Poincare invariance in four-momentum eigenstates \cite{Donoghue_Nucleon}, one can show that matrix element of the trace of energy-momentum tensor generates the nucleon mass $\langle \Theta^\mu_\mu \rangle	=	m_N $ \cite{EMT_QCD}.

Taking the expectation value of the operator expression \eqref{eq:trace} within the nucleon state, for $n_f=3$ flavours, and to the leading order $\mathcal{O}(\alpha_\mathrm{s}^0)$, we get:
\begin{equation}
	- \frac{9}{8} \langle	 \frac{\alpha_\mathrm{s}}{\pi}	G_{\mu\nu}^a G^{a\mu\nu}		\rangle	
	+ \sum_\mathfrak{q} 	\langle	m_\mathfrak{q}	\overline{\mathfrak{q}} \mathfrak{q}		\rangle		=		m_N
\end{equation}

It can be seen that the terms on the left hand side has order of $\mathcal{O}(m_N)$, hence factors of $m_\mathfrak{q}$ and $\alpha_\mathrm{s} / \pi$ for quark and gluon scalar operators in the effective Lagrangian \eqref{eq:L_NR}.

Moreover, as mentioned before, the quark scalar operator $m_q \bar{q}q$ is renormalisation group invariant. The nucleon mass $m_N$ is a physical quantity and hence scale independent. As a consequence, only using the current choice of factors, the gluon operator 
$ ( \alpha_\mathrm{s} / \pi )	\, G_{\mu\nu}^a G^{a\mu\nu} $ can be scale-invariant to the leading order of $ \mathcal{O} ( \alpha_\mathrm{s}^0 )$.

Therefore, the Gluon matrix element can be expressed as:
\begin{equation}
	\langle	 \frac{\alpha_\mathrm{s}}{\pi}	\, G_{\mu\nu}^a G^{a\mu\nu}		\rangle		=	- \frac{8}{9}	m_N 	f_{\mathrm{T}\mathfrak{g}}	\,,
\end{equation}

where $f_{\mathrm{T}\mathfrak{g}}	\equiv	1 -\sum_{\mathfrak{q}=u,d,s} 	f_{\mathrm{T}\mathfrak{q}}$.

By differencing the trace anomaly expression \eqref{eq:trace}, the matrix element of the heavy quark $Q$ can be shown to induce scalar gluon interactions of the form \cite{SVZ_80}:
\begin{equation}
	\label{eq:Heavy_Q}
	\langle m_Q \, \overline{Q} Q \rangle	=	- \frac{1}{12}	\langle	 \frac{\alpha_\mathrm{s}}{\pi}	\, G_{\mu\nu}^a G^{a\mu\nu}		\rangle	\,,
\end{equation}

which is independent of the heavy flavour mass. This is equivalent to closing the heavy quark external loops in the scattering diagrams, and replacing them by one-loop coupling to gluons.

In case of charm quark, we can see that the theoretical prediction of \eqref{eq:Heavy_Q} is close to the numerical value shown in table \ref{tab:fTq}, which is computed in  by lattice QCD.%
\footnote{Because charm flavour mass is close to QCD scale, the effect of higher dimensional operators should be taken into account in lattice simulations. 
Theses operators are suppressed by powers of $m_c$, and can provide a correction up to a few percent \cite{HD_Opr_Gloun}.}


\subsection{Scattering Cross section}

The non-renormalisable interactions only contribute to the scalar amplitude at the nucleon level:
\begin{equation}
	\mathcal{L}_\mathrm{NR}	=	f_N^\mathrm{NR}	\, \overline{\chi^0} \chi^0	\,\overline{N} N
\end{equation}

The effective non-relativistic amplitude for dark matter--nucleon scattering is derived from evaluation of the effective Lagrangian between initial and final nucleonic states:
\begin{equation}
	f_N^\mathrm{NR}	=		\langle \mathcal{L}_\mathrm{NR} \rangle 	=		
						m_N \left[	\sum_{\mathfrak{q}	=d,u,s}	\left( C_\mathfrak{q}^\mathrm{h3}	+	C_\mathfrak{q}^\mathrm{h4} \right)	f_{\mathrm{T}\mathfrak{q}}
						- \frac{8}{9}	\left( C_\mathfrak{g}^\mathrm{h3}	+	C_\mathfrak{g}^\mathrm{h4} \right)	f_{\mathrm{T}\mathfrak{g}}		\right]
\end{equation}

As discussed, the scalar matrix elements are evaluated at hadronic scale where only the three light quarks $d$, $u$ and $b$ are active.

Pseudo-real EWDM scattering off nuclei proceeds through two kinds of interactions. $f_N^\mathrm{R}$ amplitude only contains renormalisable couplings of DM and the SM, whereas $f_N^\mathrm{NR}$ includes higher dimensional coupling to Higgs boson:
\begin{equation}
	\label{eq:fN}
	f_N =	 f_N^\mathrm{R}	+ f_N^\mathrm{NR}
\end{equation}

The elastic cross-section for spin-independent interactions of EWDM and nucleon $N$ can be written as \cite{Susic_DM}:
\begin{equation}
	\sigma_N 	=	\frac{4}{\pi}	m_{\chi N}^2	\left| f_N \right|^2	\,,
\end{equation}

where $m_{\chi N} = M_\chi m_N / (M_\chi + m_N)$ is dark matter -- nucleon reduced mass.%
\footnote{In the rest of this report, we denote DM candidate mass by $M_\chi \equiv m_0$.}


\section{Wilson Coefficients}

\label{sec:Wilson}

In what follows, the Wilson coefficients of EWDM - nucleon scattering will be computed in leading order of non-renormalisable couplings $\lambda_0$ and $\lambda_c$. 
Since the relic has a very slow speed, only a small fraction of the incident DM momentum is transferred to the target nuclei. Therefore we also assume a zero momentum transfer.

In order to fix the value of Wilson coefficients, we use Feynman diagram matching. It involves computing the scattering amplitude corresponding to each diagram in the full theory, and then comparing them with the same amplitude in the effective Lagrangian at UV scale. We find the coefficients for both effective operators of quark and gluon by integrating out the mediators in the full EWDM-quark and  EWDM-gluon scattering processes.

Our understanding of the nuclear physics matrix element of the scalar operator is restricted to the hadronic scale. However, the Wilson coefficients of the quark and gluon scalar interactions are scale independent to leading order in the strong coupling constant, so there is no need to evolve them down to $\mu_\mathrm{had}$ using the renormalisation group equations (RGE's). Notwithstanding, as we cross the heavy flavour masses when integrating them out, the threshold corrections should be included in the gluon Wilson coefficient. The quark and gluon coefficients are matched by making a comparison between the two effective theories at ultraviolet and nucleon scales at $\mathcal{O} (\alpha_\mathrm{s}^0)$ \cite{Lecture_Hisano}:
\begin{subequations}
\begin{align}
	C^h_\mathfrak{q} 	(\mu_\mathrm{had})		&=		C^h_\mathfrak{q} 	(\mu_\mathrm{uv}) \,,		\qquad		\mathfrak{q} = d, u, s 			\\
	C^h_\mathfrak{g} 	(\mu_\mathrm{had})		&=		C^h_\mathfrak{g} 	(\mu_\mathrm{uv})		
													-\frac{1}{12}	\sum_{\mathfrak{q} = c, b, t}		C^h_\mathfrak{q} 	(\mu_\mathrm{uv}) \,.
\end{align}
\end{subequations}

Electroweak DM mixes with the SM Higgs through dimension-six operators introduced in \eqref{eq:L_C}. Since these interactions include both cubic and quartic DM-Higgs couplings, we need to consider the leading order diagrams which contain these two types of coupling.


\subsection{EWDM -- Quark Scattering}

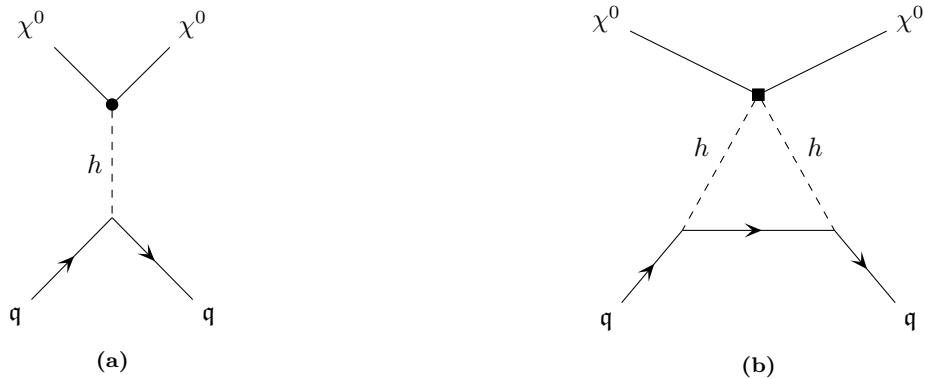
\begin{figure} [t] 				
	\begin{subfigure}{.5\linewidth}
		\centering

		\begin{tikzpicture}
		\begin{feynman}
			\vertex [dot] (a) {};
			\vertex [above left=of a] (i1) {$\chi^0$};
			\vertex [above right=of a] (f1) {$\chi^0$};
			\vertex [below =of a] (b);
			\vertex [below left=of b] (i2) {$\mathfrak{q}$};
			\vertex [below right=of b] (f2) {$\mathfrak{q}$};
		\diagram* {
			(i1) -- [plain] (a)  -- [plain] (f1),
			(a) -- [scalar, edge label'=$h$] (b),
			(i2) -- [fermion] (b) -- [fermion] (f2),
		};				
		\end{feynman}
		\end{tikzpicture}

		\subcaption{}
		\label{fig:Xq_h3}
	\end{subfigure}%
	\begin{subfigure}{.5\linewidth}
		\centering
		
		\begin{tikzpicture}
		\begin{feynman}
			\vertex [square dot] (a) at (0,0) {};
			\vertex (i1) at (-2,1) {$\chi^0$};
			\vertex (f1) at (2,1)  {$\chi^0$};
			\vertex (b) at (-1,-2+.2);
			\vertex (c) at (1,-2+.2);			
			\vertex (i2) at (-2,-3) {$\mathfrak{q}$};
			\vertex (f2) at (2,-3) {$\mathfrak{q}$};
		\diagram* {
			(i1) -- [plain] (a)  -- [plain] (f1),
			(a) -- [scalar, edge label'=$h$] (b),
			(a) -- [scalar, edge label=$h$] (c),
			(i2) -- [fermion] (b) -- [fermion] (c)-- [fermion] (f2),
		};				
		\end{feynman}
		\end{tikzpicture}

		\subcaption{}
		\label{fig:Xq_h4}
	\end{subfigure}
	\caption{Feynman diagrams which generate the effective coupling of electroweak dark matter with quarks at leading order. The EWDM-Higgs cubic and quartic vertices are represented by dot ($\bullet$) and square ($\blacksquare$) respectively.}
	\label{fig:Xq}
\end{figure}

For cubic interactions \eqref{eq:C_h3}, dark matter couples to quarks at tree level through exchange of Higgs boson (Fig. \ref{fig:Xq_h3}). After integrating the scalar mediator out, the effective coefficient is derived as:
\begin{equation}
	C_\mathfrak{q}^\mathrm{h3}	=	\frac{\lambda}{4 \, m_h^2}
\end{equation}

where we have defined the linear combination of the non-renormalisable constants as:
\begin{equation}
	\label{eq:lambda}
	\lambda	\equiv	y \,	\frac{\lambda_c}{\Lambda}	- n \, \Re \frac{\lambda_0}{\Lambda} \,.
\end{equation}

The effective interaction of EWDM with quarks induced by quartic Higgs coupling is generated at one loop level in leading order as depicted in figure \ref{fig:Xq_h4}.
This diagram gives rise to the effective coefficient:
\begin{align}
	C^\mathrm{h4}_\mathfrak{q}			&=		i \left( \frac{m_\mathfrak{q}}{\nu} \right)^2	\ C ( \overline{\chi^0}, \chi^0, h, h )		
				\int 		\frac{1}{ [ \left( \ell +k \right)^2	  -m_h^2 ]^2	\left( \slashed{\ell}	-m_\mathfrak{q} \right)	}		\frac{\mathrm{d}^4 \ell}{(2\pi)^4}		\nonumber	\\
				&=		\frac{i}{32}	 g_\mathrm{w}^2	\,\lambda \,		\frac{ m_\mathfrak{q}^3}{m_\mathrm{w}^2 }
						\left( B_0^{(2,1)} + B_1^{(2,1)} \right) 	\left( m_\mathfrak{q}^2	|	m_h, m_\mathfrak{q} \right)
\end{align}

where $k$ denotes momentum of quark, and the vertex $C ( \overline{\chi^0}, \chi^0, h, h )$ is defined in \eqref{eq:C_h4}. 
The two point functions $B_0^{(2,1)}$ and $B_1^{(2,1)}$ are evaluated in \eqref{eq:B12} and \eqref{eq:B1_1n}.

\begin{figure} [t] 				
	\centering
	\includegraphics[width=.6\linewidth]{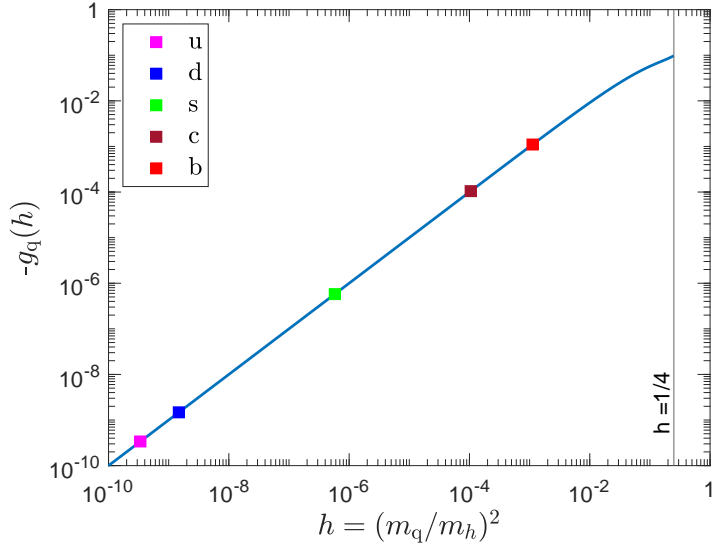}
	\caption{The absolute value of quark mass function $g_\mathfrak{q}$ in dependence of the parameter $h \equiv (m_\mathfrak{q} / m_h)^2$. 
	For $h > 1/4$, this function is not defined. The data points correspond to the values at quark mass scales, namely up (light blue), down (dark blue), strange (light green), charm (dark green) and bottom (red).}
	\label{fig:g_q}
\end{figure}

After performing the loop integration, we arrive at the Wilson coefficient:
\begin{equation}
	C^\mathrm{h4}_\mathfrak{q}	=	\frac{\alpha_\mathrm{w}}{2^8 \pi}	\frac{\lambda}{m_\mathrm{w}^2}	\ g_\mathfrak{q} (h)	\,,
\end{equation}

where $\alpha_\mathrm{w} \equiv g_\mathrm{w}^2 /4\pi$ is the  weak structure constant, $h \equiv (m_\mathfrak{q} / m_h)^2$, 
and the quark mass function is defined as:
\begin{equation}
	g_\mathfrak{q}	(x)	=		2	+ \left( 1-x \right)	\frac{\ln x}{x}		+  \frac{1-3x}{x} 	K(x)	\,,
\end{equation}

where $K$-function is defined in \eqref{eq:K_}. 
Figure \ref{fig:g_q} illustrates the changes in the absolute value of the mass function with respect to the scaling variable $h$. It also shows the data-points for the masses of different flavours. 
It can be noticed that in the limit of vanishing quark mass $h \to 0$ which corresponds to the active light flavours, $g_\mathfrak{q}$ approaches zero. Therefore $C^\mathrm{h4}_\mathfrak{q}$ cannot make a significant contribution to the total EWDM-nucleon scattering amplitude.

\subsection{EWDM -- Gluon Scattering}

Since by definition, electroweak DM is colourless under SU(3)\textsubscript{c}, it cannot scatter off gluon fields at tree-level. The interactions of dark matter with gluon is therefore loop-induced.

Although gluon loop-level interactions generate a factor of $\alpha_\mathrm{s}$, as discussed, due to order counting, it will be absorbed into the definition of the gluon operator $(\alpha_\mathrm{s}/\pi) \, G_{\mu\nu} G^{\mu\nu}$. Consequently, these loop diagrams are not only suppressed, but can even dominate over the DM - quark reactions.

In general, the loop momentum is characterised by the masses of the virtual particles running in the loop as well as the external momenta. Accordingly, we classify the DM-gluon scattering diagrams into two types. If the momentum scale is dominated by mass of heavy particles like EWDM, gauge vectors and Higgs, then the process is referred to as \emph{short distance}. On the other hand, \emph{long distance} contributions arise from loop integrals whose momenta are governed by the quark masses \cite{Neutralino_nucleon}.

Short distance diagrams should be evaluated explicitly using perturbative quantum chromodynamics machinery. This is also true of the long distance integrals involving the top quark. However, when light quarks i.e. up, down and strange run in the long-distance loops, both mass and momentum are below the QCD scale, so the process is characterised by the confinement dynamics. This contribution is already included in the quark scalar matrix element $\langle N| \bar{\mathfrak{q}} \mathfrak{q} |N\rangle$ when computing the quark mass fraction $f_{T\mathfrak{q}}$. Charm and bottom flavours are also close to $\Lambda_\mathrm{QCD}$, so the strong coupling is still large and non-perturbative effects are significant at their mass scale.	Therefore, the contributions from softer quarks $d$, $u$, $s$, $c$ and $b$ should not be incorporated in the long-distance gluon Wilson coefficient $C_\mathfrak{g}$, otherwise we would count them twice in the calculations \cite{Gluon}.
\begin{equation}
	C_\mathfrak{g}	=	\sum_{\mathfrak{q}=\mathrm{All}} C_\mathfrak{g}^{\mathrm{SD}}   (m_\mathfrak{q})		
						+ \ C_\mathfrak{g}^{\mathrm{LD}}   (m_t)
\end{equation}


Note that $C_\mathfrak{g}$ is of leading order of strong structure constant $\mathcal{O}(\alpha_\mathrm{s}^0)$ in power counting, despite loop-suppression of the DM-gluon scattering \cite{DD_Wino}. As discussed, this is due to the factor of $\alpha_\mathrm{s}/\pi$ for the gluon scalar operator in the effective Lagrangian.

Computation of the effective interactions of gluonic operators is a tedious task that requires constructing the tensor structure of the gluon field strength. 
When the field is \emph{weak} which means the external momentum is much smaller than the characteristic scale of the process, then the gluon field strength can be treated as background field. 
In this case, it is more convenient to choose the \emph{Fock–Schwinger gauge} \cite{FS_Gauge_Fock,FS_Gauge_Schwinger}:
\begin{equation}
	\label{eq:gauge}
	x^\mu A^a_\mu =0 \,.
\end{equation}

In this gauge, the Wilson coefficient for gluon interactions can be easily extracted, since the coloured propagators are already defined in terms of the background field strength tensors. 
The origin is singled out in relation \eqref{eq:gauge}, and the gauge condition is not translational invariant. Accordingly, one should be careful when computing gauge dependent quantities like propagators, because for example, forward $S(x,0)$ and backward $S(0,x)$ propagation have different forms, as will be shown explicitly. However, the translation symmetry will be restored in gauge-independent physical quantities like correlation functions \cite{21_SVZ}.

The most important property is that the gauge field can be directly replaced by the field strength tensor \cite{Shifman_chapter}:
\begin{equation}
		A^a_\mu 	=	\frac{i}{2}	(2\pi)^4	\, G^a_{\mu\nu}		\, \partial_{k_\nu} \delta^{(4)} (k)
\end{equation}

where the higher order derivative terms are irrelevant and therefore neglected. 
As a result, the gluon field strength bilinear $G^a_{\alpha\mu} G^a_{\beta\nu}$ will appear in the amplitude of the effective scalar interactions with EWDM. 
The gluon scalar operator can be easily projected out of the field strength bilinear, using the identity \\
$G^a_{\alpha\mu} G^a_{\beta\nu}	=	G^a_{\rho \sigma} G^{a \rho \sigma}	 (\eta_{\alpha \beta} \eta-{\mu\nu}	-\eta_{\alpha \nu} \eta_{\beta \mu} ) /12		+\ldots \,,$ 
where other terms in ellipsis are not relevant and thus omitted.

\begin{figure} [t] 				
	\begin{subfigure}{.3\linewidth}
		\centering

		\begin{tikzpicture}
		\begin{feynman}
			\vertex [dot] (a) at (0,0) {};
			\vertex (i1) at (-2,1) {$\chi^0$};
			\vertex (f1) at (2,1)  {$\chi^0$};
			\vertex (b) at (0,-1.5+.1);
			\vertex (c) at (-1,-3+.2);
			\vertex (d) at (1,-3+.2);			
			\vertex (i2) at (-2,-4) {$\mathfrak{g}$};
			\vertex (f2) at (2,-4) {$\mathfrak{g}$};
		\diagram* {
			(i1) -- [plain] (a)  -- [plain] (f1),
			(a) -- [scalar, edge label'=$h$] (b),
			(b) -- [fermion] (d) -- [fermion, edge label'=$\mathfrak{q}$] (c)-- [fermion] (b),	
			(i2) -- [gluon] (c),
			(f2) -- [gluon] (d),
		};		
		\end{feynman}
		\end{tikzpicture}

		\subcaption{}
		\label{fig:Xg_h3}
	\end{subfigure}%
	\begin{subfigure}{.8\linewidth}
		\centering
		
		\begin{subfigure}{.4\linewidth}
		\centering
		
		\begin{tikzpicture}
		\begin{feynman}
			\vertex [square dot] (a) at (0,0) {};
			\vertex (i1) at (-2,1) {$\chi^0$};
			\vertex (f1) at (2,1)  {$\chi^0$};
			\vertex (b) at (-1,-1.5+.1);
			\vertex (c) at (1,-1.5+.1);			
			\vertex (d) at (-1,-3+.2);
			\vertex (e) at (1,-3+.2);			
			\vertex (i2) at (-2,-4) {$\mathfrak{g}$};
			\vertex (f2) at (2,-4) {$\mathfrak{g}$};
		\diagram* {
			(i1) -- [plain] (a)  -- [plain] (f1),
			(a) -- [scalar, edge label'=$h$] (b),
			(a) -- [scalar, edge label=$h$] (c),
			(b) -- [fermion] (c) -- [fermion] (e)-- [fermion, edge label'=$\mathfrak{q}$] (d)-- [fermion] (b),	
			(i2) -- [gluon] (d),
			(f2) -- [gluon] (e),
		};					
		\end{feynman}
		\end{tikzpicture}

		\end{subfigure}%
		\begin{subfigure}{.4\linewidth}
		\centering
		
		\begin{tikzpicture}
		\begin{feynman}
			\vertex [square dot] (a) at (0,0) {};
			\vertex (i1) at (-2,1) {$\chi^0$};
			\vertex (f1) at (2,1)  {$\chi^0$};
			\vertex (b) at (-1,-1.5+.1);
			\vertex (c) at (1,-1.5+.1);			
			\vertex (d) at (-1,-3+.2);
			\vertex (e) at (1,-3+.2);			
			\vertex (i2) at (-2,-4) {$\mathfrak{g}$};
			\vertex (f2) at (2,-4) {$\mathfrak{g}$};
		\diagram* {
			(i1) -- [plain] (a)  -- [plain] (f1),
			(a) -- [scalar, edge label'=$h$] (b),
			(a) -- [scalar, edge label=$h$] (c),
			(b) -- [fermion4] (e) -- [fermion] (c)--  [fermion4] (d)--[fermion, edge label=$\mathfrak{q}$] (b),	
			(i2) -- [gluon] (d),
			(f2) -- [gluon] (e),
		};		
		\end{feynman}
		\end{tikzpicture}
		
		\end{subfigure}%

		\subcaption{}
		\label{fig:Xg_h4}
	\end{subfigure}
	\caption{Feynman diagrams for EWDM - gluon effective scattering induced by non-renormalisable Higgs interactions. The EWDM-Higgs cubic and quartic vertices are represented by dot ($\bullet$) and square ($\blacksquare$) respectively.}
	\label{fig:Xg}
\end{figure}
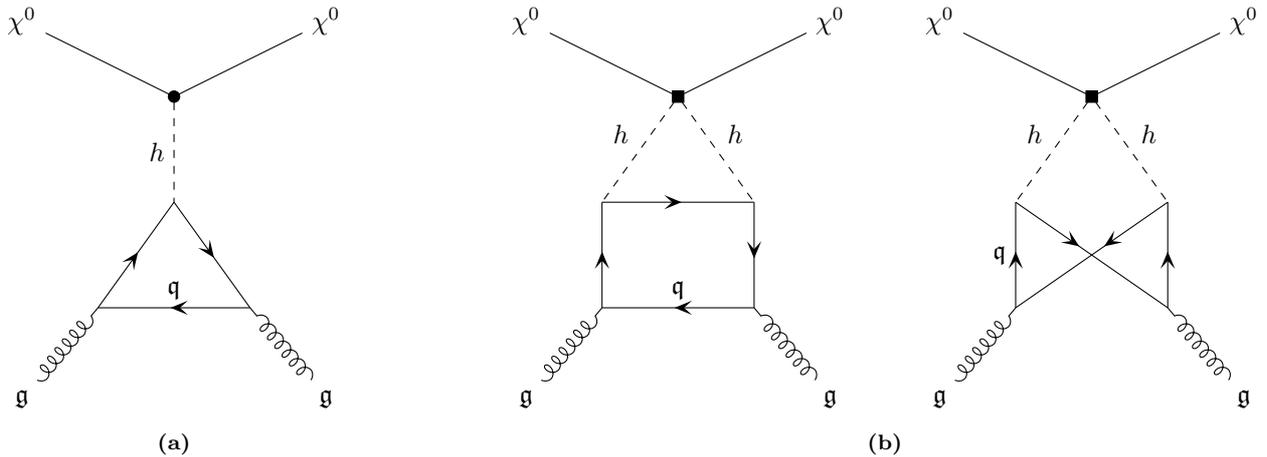

The DM-gluon interactions involving the cubic coupling are generated by the triangle-loop diagram of figure \ref{fig:Xg_h3}. 
To compute the scattering amplitude in this gauge, we need the EWDM two-point function in the gluon background. This requires computing the contribution of Higgs tadpole $\Gamma_\mathfrak{q}$ in the gluon external field.

Clearly the triangle loop in this diagram where virtual quarks are circulating will only give rise to a long-distance integral.

Higgs tadpole has the form:
\begin{align}
	\Gamma_\mathfrak{q} 	&= 		- \tr \int S_\mathfrak{q}^{(2)} (\ell)		\frac{\mathrm{d}^4 \ell}{(2\pi)^4}		\\
							&=		\frac{-i}{4\pi^2}		\, G_{\rho\sigma}^a G^{a\rho\sigma}	m_\mathfrak{q}		
									\left(	A_0^{(3)}(m_\mathfrak{q} )	- m^2_\mathfrak{q}	\, A_0^{(4)}(m_\mathfrak{q} )		\right)		\nonumber
\end{align}

The scalar one-point functions $A_0^{(3)}$ and $A_0^{(4)}$ are defined in \eqref{eq:A0}.
The second order correction to the propagator of a coloured fermion when two external gluons are inserted reads \cite{6c_SVZ}:
\begin{equation}
	S_\mathfrak{q}^{(2)} (\ell)	=		S_\mathfrak{q}^{(0)} (\ell)	\int 	\left(	-i g_\mathrm{s} \slashed{A}(k_1)	\right)		S_\mathfrak{q}^{(0)} (\ell -k_1)
											\int 	\left(	-i g_\mathrm{s} \slashed{A}(k_2)	\right)		 S_\mathfrak{q}^{(0)} (\ell -k_1 -k_2)
											\frac{\mathrm{d}^4 k_1}{(2\pi)^4}	\frac{\mathrm{d}^4 k_2}{(2\pi)^4}
\end{equation}

$S_\mathfrak{q}^{(0)} (\ell)	=	i/(\slashed{\ell} -m_\mathfrak{q})$ 
is the usual fermionic Feynman propagator in the vacuum.
Since the gluon field contains derivative of Dirac delta function, the integration over the background field momenta can be carried out trivially. In practice, it reduces to differentiation of the propagators located after the vertex.

Using the relation 
$ C_\mathfrak{g}^{h3}	=	 i \, C ( \overline{\chi^0}, \chi^0, h ) \, m_t \Gamma_t / ( \nu \, m_h^2 )$, 
with the vertex factor $C ( \overline{\chi^0}, \chi^0, h )$ defined in \eqref{eq:C_h4}, the amplitude of Feynman diagram \ref{fig:Xg_h3} gives rise to the following coefficient:
\begin{equation}
	C_\mathfrak{g}^{h3}	=	\frac{-\lambda}{48 \, m_h^2}
\end{equation}

It can be checked that DM - gluon coupling arising from the hard triangle loop is related to the top quark contribution through
$ C_\mathfrak{g}^{h3} = - C_t^{h3} /12$.
This behaviour can be explained by \emph{heavy quark expansion} of the trace anomaly of the energy-momentum tensor \cite{Witten_heavyQ}. In short distances of order $1/m_t$, one can expand the virtual top state in powers of $m_t^{-2}$. To the first order in $\alpha_\mathrm{s}$, top scalar operator converts to the gluon operator as 
$m_t \bar{t}t 	\to 		( - \alpha_\mathrm{s} /12\pi) \, G_{\mu\nu}^a G^{\mu\nu a}$ \cite{SVZ_1975}.

Now, we move on to the EWDM-gluon interactions that include the quartic non-renormalisable Higgs coupling. As shown in figure \ref{fig:Xg}, the diagrams are generated at two-loop level.

At first step, one needs to evaluate the quantum corrections to the Higgs self energy $\Pi_\mathfrak{q}$ induced by virtual quarks, in the gluon external field. 

When each quark propagator emits one gluon which is the case for diagram \ref{fig:Xg_h4} (right), the self energy can be written as:
\begin{align}
\label{eq:Pi_1}
	\Pi_\mathfrak{q}^{(1)} (q^2)	&=	\left( \frac{g_\mathrm{w}}{2 \,m_\mathrm{w}} \right)^2	m_\mathfrak{q}^2
			\ \tr	\int 		S_\mathfrak{q}^{(1)} (\ell +q)	\, \widetilde{S}_\mathfrak{q}^{(1)} (\ell )	\frac{\mathrm{d}^4 \ell}{(2\pi)^4}	\\	\nonumber
	& =		- \left( \frac{g_\mathrm{w} g_\mathrm{s}}{2 \,m_\mathrm{w}} \right)^2		G_{\mu\nu}^a G^{a\mu\nu} 	m_\mathfrak{q}^2	
			\left( 3 m_\mathfrak{q}^2 	B_0^{(2,2)}		+q^2 B_1^{(2,2)}	+B_0^{(2,1)} 	\right)	 ( q^2 | m_\mathfrak{q}, m_\mathfrak{q} ) 		
\end{align}

The loop integrals $B_0^{(1,2)}$ and $B_0^{(2,2)}$ are explicitly defined in \eqref{eq:B12_m} and \eqref{eq:B22}.

The first order correction to the fermionic propagator which corresponds to one gluon field in the background, is given as:
\begin{equation}
		S_\mathfrak{q}^{(1)} (\ell)	=		S_\mathfrak{q}^{(0)} (\ell)	
								\int 	\left(	-i g_\mathrm{s} \slashed{A}(k_1)	\right)		S_\mathfrak{q}^{(0)} (\ell -k_1)		\frac{\mathrm{d}^4 k_1}{(2\pi)^4}	\,.
\end{equation}

Due to violation of the translation invariance in the gauge condition \eqref{eq:gauge}, propagation of an antiparticle in the opposite direction has a different form \cite{21_SVZ}:
\begin{equation}
		\widetilde{S}_\mathfrak{q}^{(1)} (\ell)	=		\int  	S_\mathfrak{q}^{(0)} (\ell +k_2)		\left(	-i g_\mathrm{s} \slashed{A}(k_2)	\right)		\frac{\mathrm{d}^4 k_2}{(2\pi)^4}
											\  S_\mathfrak{q}^{(0)} (\ell)	\,.						
\end{equation}

It can be seen that the quark masses and the external momentum $q$ contribute on equal footing to the loop momenta in \eqref{eq:Pi_1}. Since the dominant value of the external momentum is at Higgs mass scale, the quark box receives short-distance contributions, and therefore all quark flavours should be taken into account.

Executing the integrals, we finally find:
\begin{equation}
	\Pi_\mathfrak{q}^{(1)} (y)	=	-\frac{i}{4}	\frac{\alpha_\mathrm{w} \alpha_\mathrm{s}}{m_\mathrm{w}^2}		G^a_{\mu\nu} G^{a\mu\nu}
				\frac{y}{1-4y}	\left[	8y 	\left( 1-3y \right)	K(y)			+\left( 1-6y \right)	\right]
\end{equation}

where $y \equiv (m_\mathfrak{q} / q)^2$, and $K$-function is defined in \eqref{eq:K_}. When the momentum is smaller than twice quark mass $y>1/4$, one can use the identity \eqref{eq:K_} to avoid root of negative numbers and logarithm with complex arguments.

\begin{figure} [t] 				
	\begin{subfigure}{.5\linewidth}
		\centering
		\includegraphics[width=\linewidth]{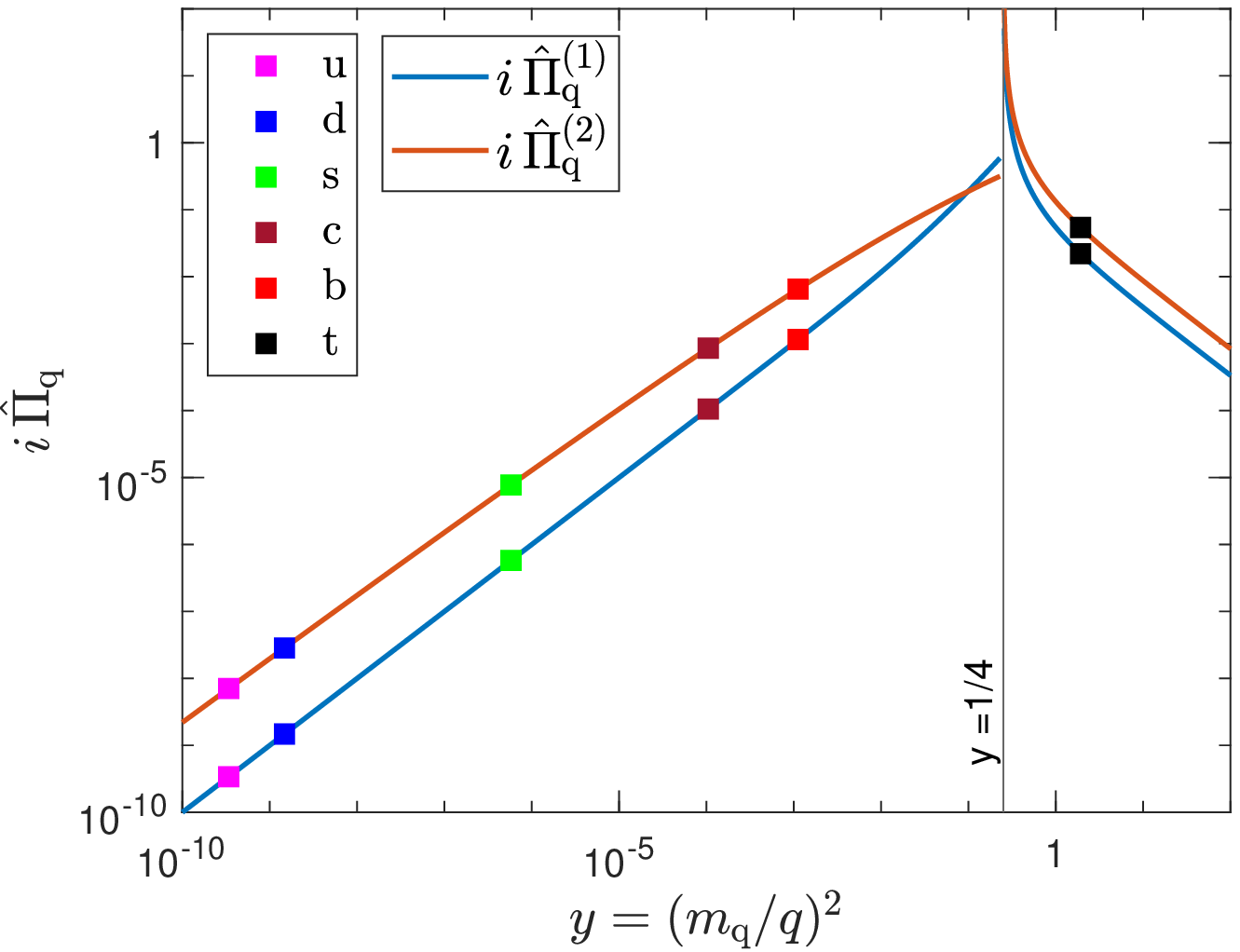}
		\subcaption{}
		\label{fig:2pt_2g}
	\end{subfigure}%
	\begin{subfigure}{.5\linewidth}
		\centering
		\includegraphics[width=\linewidth]{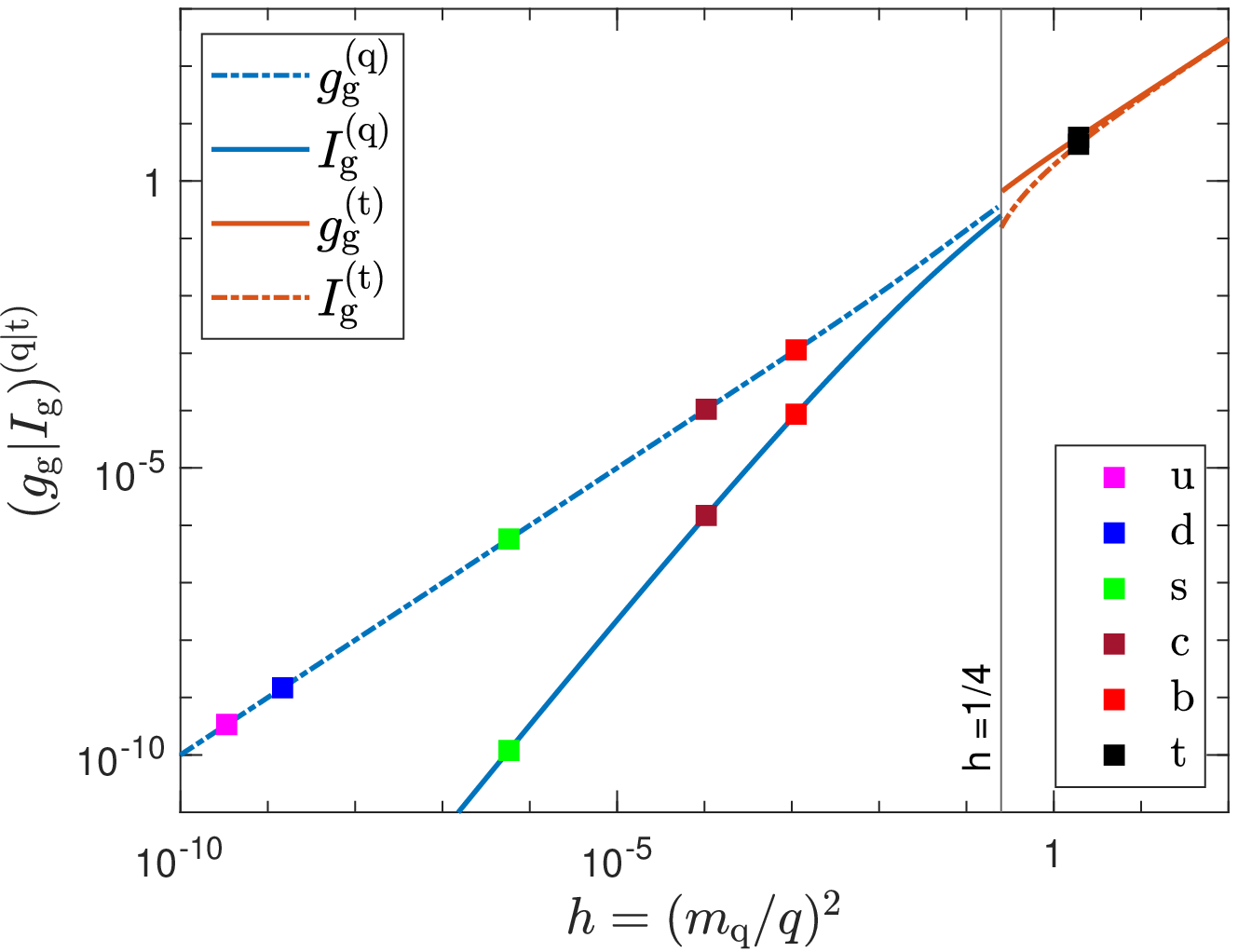}
		\subcaption{}
		\label{fig:mfunc_2g}
	\end{subfigure}
	\caption{(left \ref{fig:2pt_2g}) The normalised two-point function of Higgs boson in two gluon background fields with respect to $y = (m_\mathfrak{q}/q)^2$ with $q$ being the external momentum. 
	The blue curve corresponds to $i \widehat{\Pi}^{(1)}_\mathfrak{q}$ correlation function, where each internal quark propagator emits one gluon (c.f. figure \ref{fig:Xg_h4}-right). While the red curve represents $i \widehat{\Pi}^{(2)}_\mathfrak{q}$ function, in which two gluons are attached to the same internal quark propagator (c.f. figure \ref{fig:Xg_h4}-left).	
	The data-points present the value of the two-point functions at the mass of up (light blue), down (dark blue), strange (light green), charm (dark green), bottom (red) and top (black) quarks.\\
	(right \ref{fig:mfunc_2g}) Changes in the gluon mass functions in dependence of $h =(m_\mathfrak{q}/m_h)^2$. 
	The blue curves indicate the analytical and numerical gluon mass functions $g^{(\mathfrak{q})}_\mathfrak{g}$ and $I^{(\mathfrak{q})}_\mathfrak{g}$ where softer i.e. up, down, strange, charm and bottom flavours run in the loops. Whereas when top quark is the only virtual particle, the relevant mass functions  $g^{(t)}_\mathfrak{g}$ and $I^{(t)}_\mathfrak{g}$ are plotted in red. 
	Negative-valued functions are shown in dash-dotted style.}
	\label{fig:2g}
\end{figure}

Figure \ref{fig:2pt_2g} depicts the behaviour of the normalised Higgs self-energy 
$\Pi_\mathfrak{q}		\equiv 	( \alpha_\mathrm{w} \alpha_\mathrm{s}	/4 m_\mathrm{w}^2 )		G^a_{\mu\nu} G^{a\mu\nu}		\,\widehat{\Pi}_\mathfrak{q} $, 
with respect to $y$. The data-points corresponds to the values for different quark masses at the dominant momentum $q \approx m_h$. 
The two-point function vanishes when the momentum is either much larger $y\to0$, or much smaller $y\to\infty$ than the quark mass. It is finite at $y=1/4$ if approaching from below, but explodes when taking the limit from right.

In case two gluon fields are attached with the same propagator as in diagram \ref{fig:Xg_h4} (left), the self energy is properly obtained as:
\begin{align}
	\label{eq:Pi_2}
	\Pi_\mathfrak{q}^{(2)} (q^2)	&=	\left( \frac{g_\mathrm{w}}{2 \,m_\mathrm{w}} \right)^2	m_\mathfrak{q}^2
			\ \tr	\int 		S_\mathfrak{q}^{(0)} (\ell +q)	\, \widetilde{S}_\mathfrak{q}^{(2)} (\ell )	\frac{\mathrm{d}^4 \ell}{(2\pi)^4}	\\	\nonumber
	& =		\left( \frac{g_\mathrm{w} g_\mathrm{s}}{2 \,m_\mathrm{w}} \right)^2		G_{\mu\nu}^a G^{a\mu\nu} 	m_\mathfrak{q}^2	
			\left( 	B_0^{(1,2)}		- m_\mathfrak{q}^2 	B_0^{(1,3)}		+q^2 B_1^{(1,3)}	-2 m_\mathfrak{q}^4 B_0^{(1,4)}	-m_\mathfrak{q}^2 q^2 B_1^{(1,4)}	\right) 	 
					( q^2 | m_\mathfrak{q}, m_\mathfrak{q} )							
\end{align}

The loop integrals $B_0^{(1,3)}$ and $B_0^{(1,4)}$ are evaluated in \eqref{eq:B13} and \eqref{eq:B14}.

The fermionic propagator for the antiparticle emitting two gluons is given by:
\begin{equation}
	\widetilde{S}_\mathfrak{q}^{(2)} (\ell)	=			\iint 		S_\mathfrak{q}^{(0)} (\ell +k_1 +k_2)		\left(	-i g_\mathrm{s} \slashed{A}(k_2)	\right)		
						\frac{\mathrm{d}^4 k_2}{(2\pi)^4}		 	\ S_\mathfrak{q}^{(0)} (\ell +k_1)			\left(	-i g_\mathrm{s} \slashed{A}(k_1)	\right)		 
						\frac{\mathrm{d}^4 k_1}{(2\pi)^4}			\ S_\mathfrak{q}^{(0)} (\ell)
\end{equation}

The dominant contribution to the box integral \eqref{eq:Pi_2} is provided by the quark mass. As discussed before, when computing such long-distance diagrams, due to the implicit infrared cut-off at $\Lambda_\mathrm{QCD}$ in the loop momenta, only top quark should be considered. 

Carrying out the integrals, we arrive at the following expression:
\begin{equation}
	\Pi_\mathfrak{q}^{(2)} (y)	=	-\frac{i}{8}	\frac{\alpha_\mathrm{w} \alpha_\mathrm{s}}{m_\mathrm{w}^2}		G^a_{\mu\nu} G^{a\mu\nu}
				\frac{y}{1-4y}	\left[	2 	\left( 1-2y \right)	K(y) 		-1	\right]
\end{equation}

For $y<1/4$, using the identity \eqref{eq:K_}, the result can be expressed in terms of the inverse cotangent.

As illustrated in figure \ref{fig:2pt_2g}, the Higgs two-point function in two gluon background field $\Pi_\mathfrak{q}^{(2)}$ has the same asymptotic behaviour as that of $\Pi_\mathfrak{q}^{(1)}$. It has only a one-sided limit at $y=1/4$ when approaching from left.

It is also in clear view that we only need to take into account the input from the three heaviest quarks $c$, $b$ and $t$, in the calculations, and can safely ignore other lighter flavours.

The total Higgs self energy therefore can be written as:
\begin{equation}
	\Pi^h =	\sum_{\mathfrak{q}=\mathrm{All}} 	\Pi^{(1)}_\mathfrak{q}	+  \Pi^{(2)}_t 
			=	\sum_{\mathfrak{q}=c,b} 	\Pi^{(1)}_\mathfrak{q}	+	\Pi_t
\end{equation}

where $\Pi_t 	\equiv	\Pi^{(1)}_t 		+ \Pi^{(2)}_t \,.$


At this stage, we need to evaluate the second loop using the Higgs correlator $\Pi^h$ in order to find the Wilson coefficient of the gluon operator.
\begin{equation}
	\frac{\alpha_\mathrm{s}}{\pi}	\ G_{\mu\nu}^a G^{\mu\nu a}	\ C^\mathrm{h4}_\mathfrak{g}	=		i \ C ( \overline{\chi^0}, \chi^0, h, h )		
		\int 		\left( \sum_{\mathfrak{q}=c,b} 	\Pi^{(1)}_\mathfrak{q}	+	\Pi_t \right)		\frac{1}{(q^2 -m_h^2)^2}		\frac{\mathrm{d}^4 q}{(2\pi)^4}
\end{equation}

Using the Higgs self energies we can finally compute the effective coefficient as:
\begin{equation}
	\label{eq:Cg_h4}
	C^\mathrm{h4}_\mathfrak{g}	=		\frac{\alpha_\mathrm{w}}{2^9 \pi}	\frac{\lambda}{m_\mathrm{w}^2}
						\left[	\left(	I_\mathfrak{g}^{(t)} 	+g_\mathfrak{g}^{(t)}	\right)	(h_t)			
						+\sum_{\mathfrak{q}=c,b}	\left(	I_\mathfrak{g}^{(\mathfrak{q})} 	+g_\mathfrak{g}^{(\mathfrak{q})}	\right)	(h_\mathfrak{q})	\right]
\end{equation}

The first two terms are contributions of top quark, and the last two terms are the input from charm and bottom. 

The numeric integral $I_\mathfrak{g}^{(t)}$ is defined as:
\begin{equation}
		I_\mathfrak{g}^{(t)} (x)		=		2x 	\int_0^\infty		\frac{ l^2 -6xl -24x^2 }{ \sqrt{l} 	\left( l +4x \right)^\frac{3}{2}		\left( l+1\right)^2}
												\ \ln \frac{ \sqrt{l +4x} +\sqrt{l} }{ 2 \sqrt{x} }			\ \mathrm{d} l 		\,,
\end{equation}

where $l \equiv (q^{(\mathrm{E})} / M_h)^2$, and Euclidean momentum is related to the actual loop momentum through a Wick rotation $q_0 = i q^{(\mathrm{E})}_0$. 
The typical momentum for this integral is around the Higgs mass $q \approx M_h$.

$ g_\mathfrak{g}^{(t)} $ has the analytical form of:
\begin{equation}
		g_\mathfrak{g}^{(t)} (x)		=		x \left[	\frac{-1}{4x-1}		- \frac{8x}{\left( 4x-1 \right)^2} \left( 24 x^2 -12x +1 \right)	\ln 4x		
											\ + 12x 	\ln (4x+1)		\right]		\,.
\end{equation}

In addition, $I_\mathfrak{g}^{(\mathfrak{q})}$ is expressed by the following integral:
\begin{equation}
			I_\mathfrak{g}^{(\mathfrak{q})} (x)		=		8 x^2 	\int_0^\infty		\frac{ l + 3x }{ \sqrt{l} 	\left( l +4x \right)^\frac{3}{2}		\left( l+1\right)^2}
												\ \ln \frac{ \sqrt{l +4x} +\sqrt{l} }{ 2 \sqrt{x} }			\ \mathrm{d} l 		\,.
\end{equation}

This loop integral is also dominated by momenta at Higgs mass scale $q \approx M_h$.

The mass function $g_\mathfrak{g}^{(\mathfrak{q})}$ is given by:
\begin{equation}
		g_\mathfrak{g}^{(\mathfrak{q})} (x)		=		x \left[	\frac{-1}{1-4x}		+ \frac{2x}{\left( 1-4x \right)^2} \left( 1 -24x +48 x^2 \right)	\ln 4x		
											\ + 6x 	\ln (1 +4x)		\right]	\,.
\end{equation}

Figure \ref{fig:mfunc_2g} illustrates changes of the mass functions $I_\mathfrak{g}^{(\mathfrak{q})}$, $g_\mathfrak{g}^{(\mathfrak{q})}$, $I_\mathfrak{g}^{(t)}$ and $g_\mathfrak{g}^{(t)}$, involved in gluonic interactions through quartic coupling. The lighter flavour functions $I_\mathfrak{g}^{(\mathfrak{q})}$ and $g_\mathfrak{g}^{(\mathfrak{q})}$ vanish in the massless quark limit, and the top flavour ones $I_\mathfrak{g}^{(t)}$ and $g_\mathfrak{g}^{(t)}$  approach infinity in the heavy quark region. 

We can also safely ignore the non-analytical light quark term $I_\mathfrak{g}^{(\mathfrak{q})}$ in $C^\mathrm{h4}_\mathfrak{g}$.

It is noticed that the computational $I_\mathfrak{g}$ analytical $g_\mathfrak{g}$ terms in the effective coupling \eqref{eq:Cg_h4} has the same order of magnitude, but opposite signs. This leads to an accidental cancellation which deceases the contribution of $C^\mathrm{h4}_\mathfrak{g}$ coefficient to the effective amplitude by an order of magnitude.


\section{Direct Detection Constraints}
\label{sec:constraints}

\begin{figure} [t] 				
	\begin{subfigure}{.5\linewidth}
		\centering
		\includegraphics[width=\linewidth]{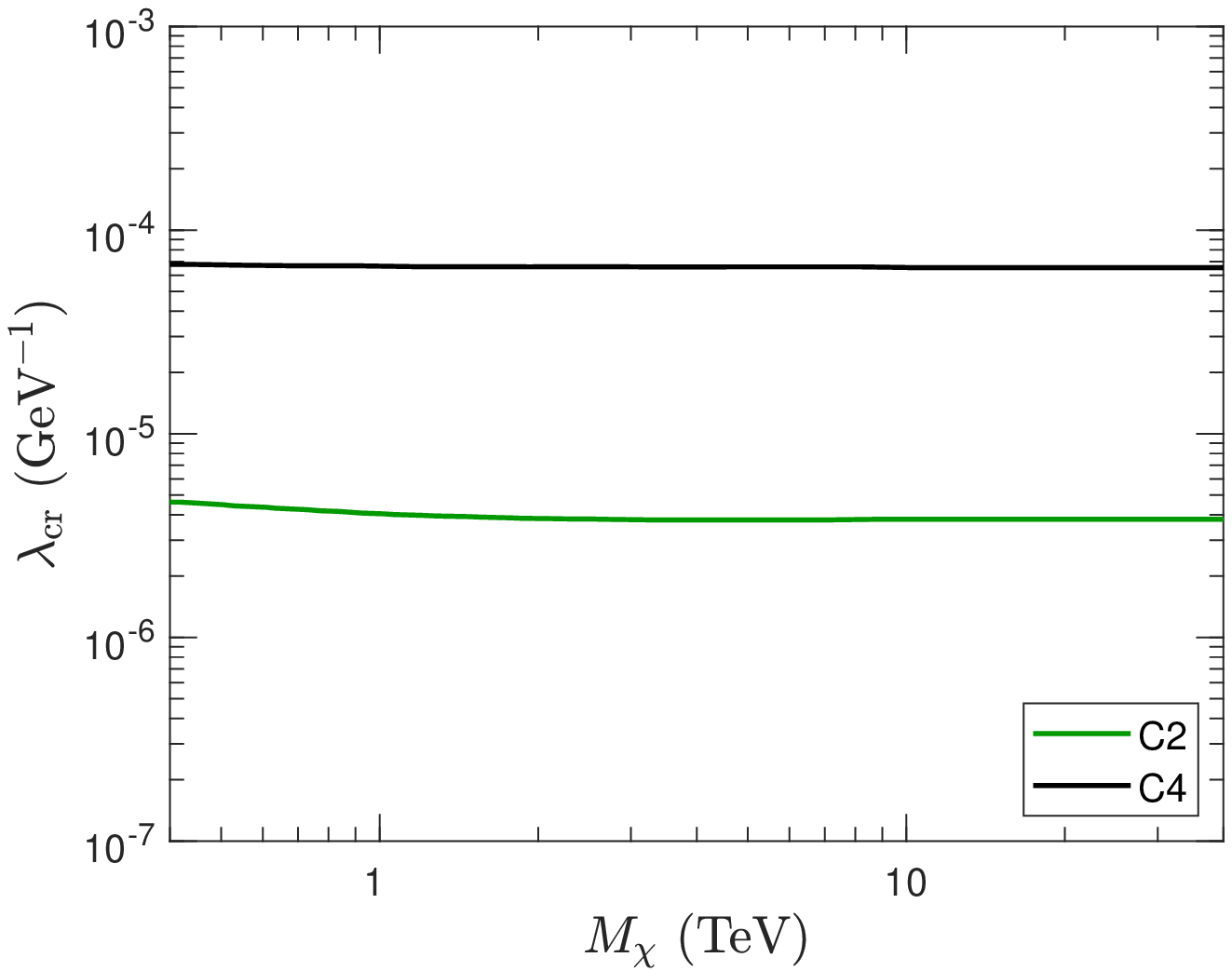}
		\subcaption{}
		\label{fig:l_cr}
	\end{subfigure}%
	\begin{subfigure}{.5\linewidth}
		\centering
		\includegraphics[width=\linewidth]{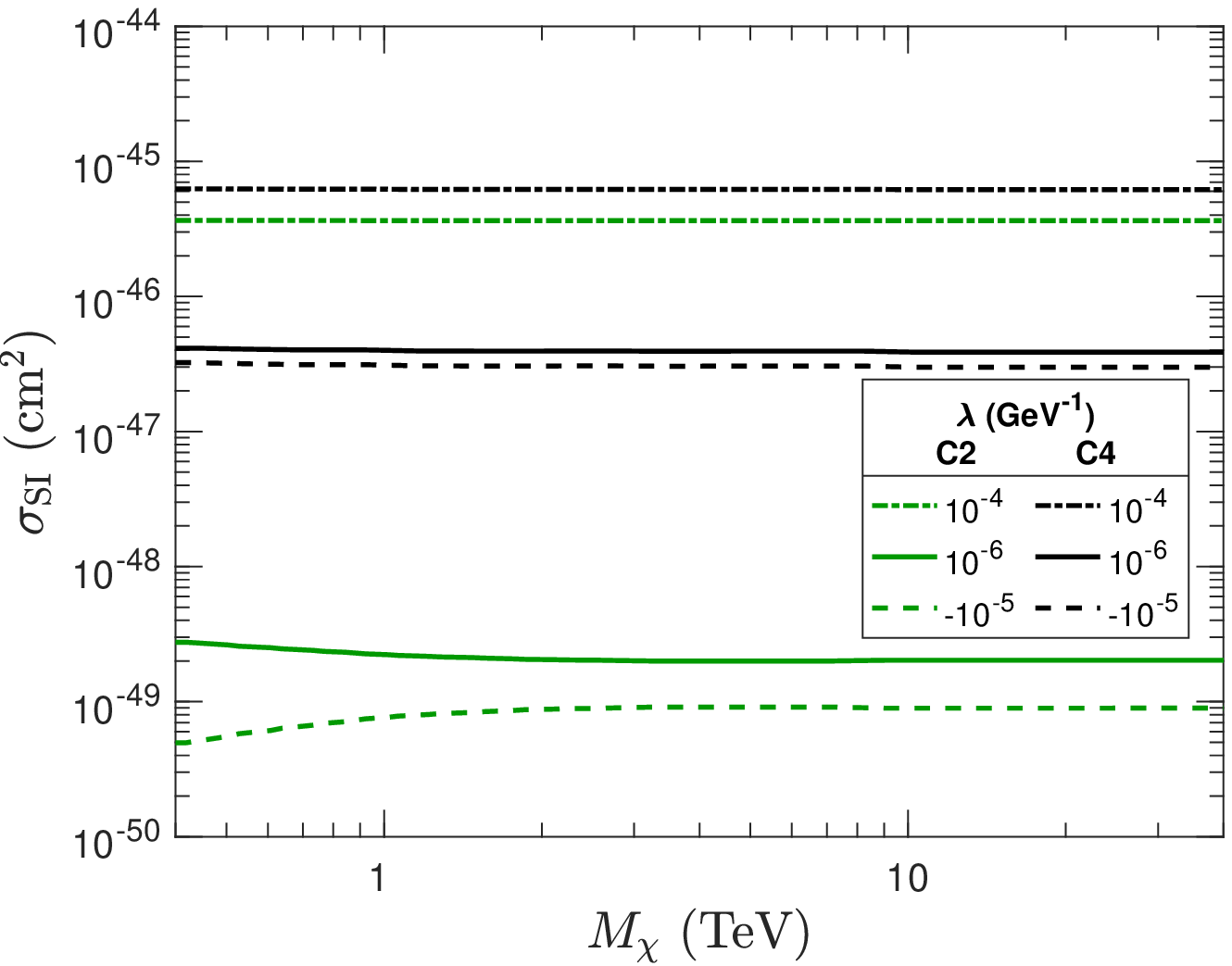}
		\subcaption{}
		\label{fig:Xn_Complex}
	\end{subfigure}
	\caption{(Left panel \ref{fig:l_cr}): Changes in the critical value of the coupling constant $\lambda_\mathrm{cr}$ with respect to EWDM mass for SU(2) pseudo-real representations namely doublet (green) and quartet (black).\\
	(Right panel \ref{fig:Xn_Complex}): Spin independent scattering cross-section off nuclei for doublet (green) and quartet (black) complex representations as a function of dark matter mass. The pairs of curves are plotted at three indicative values of the coupling constant that are $\lambda =-$10\textsuperscript{-5} GeV\textsuperscript{-1} (dashed), 10\textsuperscript{-6} GeV\textsuperscript{-1} (solid), and 10\textsuperscript{-4} GeV\textsuperscript{-1} (dash-dotted).}
	\label{fig:lcr_Xn} 
\end{figure}

Having derived all the required theoretical ingredients, in this section we numerically compute the EWDM-nucleon SI cross section, and compare our results with the latest direct search data and future sensitivities. 

In the previous section, it was found that the effective amplitude for all the main channels have positive sign, therefore all the diagrams containing the non-renormalisable couplings contribute constructively to the total cross-section through \eqref{eq:fN}.

At leading order, all the effective couplings induced by higher dimensional operators have a factor of $\lambda$. We can therefore define the \emph{critical coupling constant} $\lambda_\mathrm{cr}$ where the non-renormalisable amplitude equates the renormalisable one $f_N^\mathrm{NR} = f_N^\mathrm{R}$, so that:
\begin{equation}
	\lambda_\mathrm{cr}	\equiv	\frac{f_N^\mathrm{R}}{f_N^\mathrm{NR} /\lambda}
\end{equation}

Figure \ref{fig:l_cr} presents the critical coupling for the two pseudo-real EWDM representations. It can be observed that $\lambda_\mathrm{cr}$ is almost independent of DM mass, particularly in TeV mass region. In addition, the value of the critical parameter in quartet model is about an order of magnitude above that of doublet dark matter. That is due to the factor of $[n^2 - (4y^2 +1)]/8$ in the charged weak induced amplitude \eqref{eq:Cq_R} and lack of the light charged EWDM mediated diagrams in $\mathbb{C}2$ model.

In fact the behaviour of the pseudo-real models of the electroweak dark matter crucially depends on the strength of the coupling $\lambda$. In order to further study this, in figure \ref{fig:Xn_Complex}, we compare the performance of the complex models with three indicatory values of the coupling. 

Below the critical coupling, the direct search observables of the model behave similar to those of the renormalised theory, as shown for $\lambda = 10^{-6} \,\mathrm{GeV}^{-1}$ case. Since the coupling is less than the minimum possible critical value $\lambda < \lambda_\mathrm{cr}^{\mathbb{C}2}$, there are distinct spectra for different representations.

Obviously this can happen if both $\lambda_0$ and $\lambda_\mathrm{c}$ are small. In addition, if the relationship between the two couplings is in direct proportion that is $\lambda_\mathrm{c} / \lambda_0 = 2n$, then effect of the higher dimensional operators cancel out each other. In such cases, the effective theory will produce the same signal for direct detection experiments as the renormalisable low-energy Lagrangian \eqref{eq:L_Ren}.

In the event that $2n \,\Re \lambda_0 > \lambda_\mathrm{c}$, the coupling constant will take negative values $\lambda<0$. It causes a measurable destructive interference between the renormalisable and non-renormalisable amplitudes, which leads to the suppression of the total SI cross-section. This is illustrated in the figure, for $\lambda = -10^{-5} \,\mathrm{GeV}^{-1}$ curve. 
At the critical point of $\lambda = - \lambda_\mathrm{cr}$, the scattering amplitude will totally vanish in the leading order.

On the contrary, above $\lambda_\mathrm{cr}$, the scattering amplitude is governed by the non-renormalisable terms. 
Since the effective coupling induced by the higher dimensional operators is independent of the representation, the cross-section curves of all pseudo-real models converge. This behaviour can be verified for the indicative value of $\lambda = 10^{-4} \,\mathrm{GeV}^{-1}$. 
As the coupling to higher dimensional operators strengthens, the non-renormalisable couplings enhance the scalar effective amplitude, and thus the total scattering cross-section of pseudo-real modules increases significantly.

\begin{figure} [t] 				
	\centering
	\includegraphics[width=.6\linewidth]{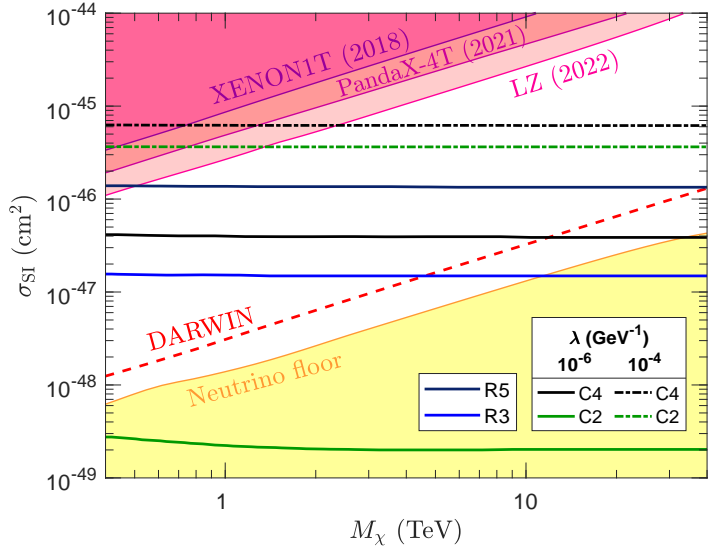}
	\caption{Current and projected direct detection constraints on four fermionic multiplets of electroweak dark matter. 
	The spin-independent scattering cross-sections are plotted in blue for the real triplet, navy for the real quintet, green for the complex doublet, and black which corresponds to the complex quartet. The predicted curves for pseudo-real models are presented at two indicatory values of the coupling constant $\lambda=10^{-6} \,\mathrm{GeV}^{-1}$ (solid) and $\lambda=10^{-4} \,\mathrm{GeV}^{-1}$ (dash-dotted).
	The overlaid lines from top to bottom represent the experimental upper bounds from XENON1T (2018) \cite{XENON1T_2018}, PandaX-4T (2021) \cite{PandaX_4T_2021} and LUX-ZEPLIN (LZ) (2022) \cite{LZ_2022}. The red dash line indicates the projected sensitivity of DARWIN future experiment \cite{DARWIN}.
	Where the predicted cross-section curves enter the shaded areas, the corresponding mass values are excluded. 
	The upper boundary of the yellow shaded area which is labelled Neutrino floor corresponds to discovery limit for dark matter \cite{Neutrino_BG_2014}. If scattering cross-section drops to this region, it would be unlikely to detect EWDM particle due to neutrino background.}
	\label{fig:Xn_Reps}
\end{figure}

The SI cross-sections for various representations of EWDM have been compared with current experimental data from XENON1T (2018) \cite{XENON1T_2018}, PandaX-4T (2021) \cite{PandaX_4T_2021} and LUX-ZEPLIN (LZ) (2022) \cite{LZ_2022}, and future projection for DARWIN detector \cite{DARWIN} in figure \ref{fig:Xn_Reps}, for two values of the coupling $\lambda$.

In general, the SI cross-section has small dependence on EWDM mass, especially above TeV scale. That is due to the fact that the non-renormalisable effective couplings are totally independent of dark matter mass, and the renormalisable contribution becomes mass independent in the heavy dark matter limit (c.f. section \ref{app:L_Ren_Xn}).

The solid lines illustrate the cross-section curves of the real theories as well as the complex models for $\lambda = 10^{-6} \,\mathrm{GeV}^{-1}$ which is an indicative value for the coupling strength being below the criticality.

In this region, the predicted SI cross-section is far below the present direct detection bound. 
The future DARWIN experiment might fully probe the multiplets of dimension $n>3$ in TeV mass scale, although the higher mass range will remain unconstrained.

At very small cross-sections the potential DM signal would be saturated by the background atmospheric, solar, and diffuse supernova neutrinos colliding with target nuclei \cite{Neutrino_BG_2007}. 
The discovery limit is defined as the cross-section where there is 90\% probability that experiment can detect the true DM with a minimum significance of 3-$\sigma$ \cite{Discovery_limit}. 
We refer to this lower limit as \emph{neutrino floor} which is shown in figure \ref{fig:Xn_Reps} as the border of the yellow shaded area \cite{Neutrino_BG_2014}. 
Within this parameter region, it would be difficult to discover dark matter events.

It can be seen that, below the clerical value $\lambda_\mathrm{cr}$, scattering cross-section enhances with the dimension of representation. The reason is that $W$-boson mediated renormalisable interactions are proportional to $[n^2 - (4y^2 +1)]/8$ (c.f. \eqref{eq:Cq_R},\eqref{eq:Cg_R}). This factor increases when going either from a complex representation $\mathbb{C}n$ to a higher dimensional real model $\mathbb{R}(n+1)$ or from a real $n$-tuplet $\mathbb{R}n$ to a larger complex multiplet $\mathbb{C}(n+1)$. The $Z$-boson induced interactions rise in proportion to $y^2/4$; however, the effective EWDM--$Z$ coupling is comparatively smaller than $W$ contribution.

In contrast, for $\lambda=10^{-4} \,\mathrm{GeV}^{-1}$, the coupling is above the maximum critical strength $\lambda > \lambda_\mathrm{cr}^{\mathbb{C}4}$. As discussed, the non-renormalisable operators, in this case, provide a constructive contribution to the scattering amplitude which could increase the total SI cross-section up to orders of magnitude. 
Therefore, for high values of non-renormalisable coupling, the pseudo-real EWDM can reach the detection limit, and produce signals that potentially could be observed by direct detection experiments.

The pseudo-real doublet has the smallest dimension of the SU(2) representations, and misses the light charged component. Consequently the scattering cross-section arising from renormalisable interactions falls below the neutrino background limit for this model. 
Nevertheless, the coupling to Higgs boson via non-renormalisable operators can increase the scattering rate to the current direct detection energy thresholds. As a result future experiments would be promising to detect the doublet EWDM scenario.

\subsection{Parameter Space}

\begin{figure} [!t] 				
	\begin{subfigure}{.5\linewidth}
		\centering
		\includegraphics[width=\linewidth]{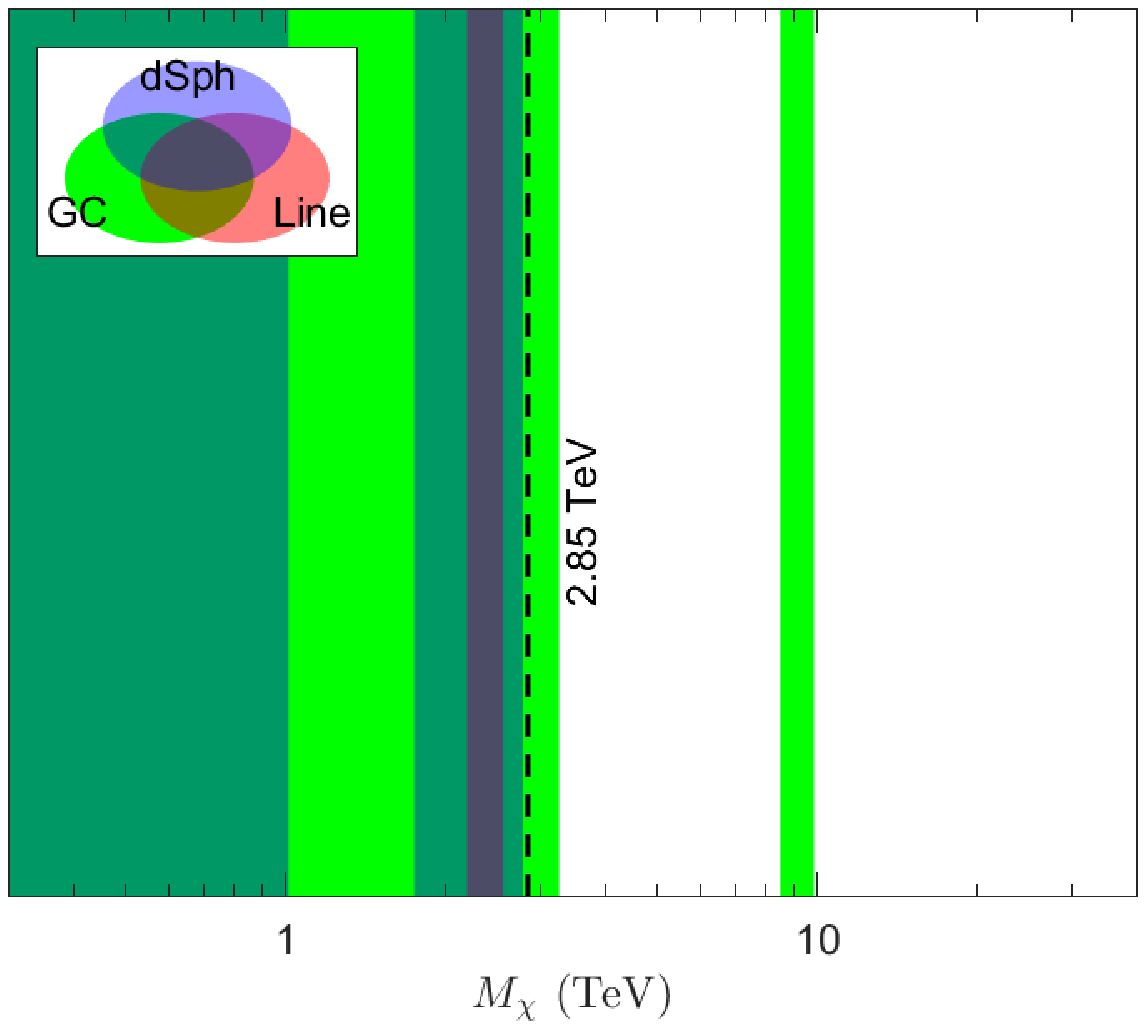}
		\subcaption{}
		\label{fig:Par_R3}
	\end{subfigure}%
	\begin{subfigure}{.5\linewidth}
		\centering
		\includegraphics[width=\linewidth]{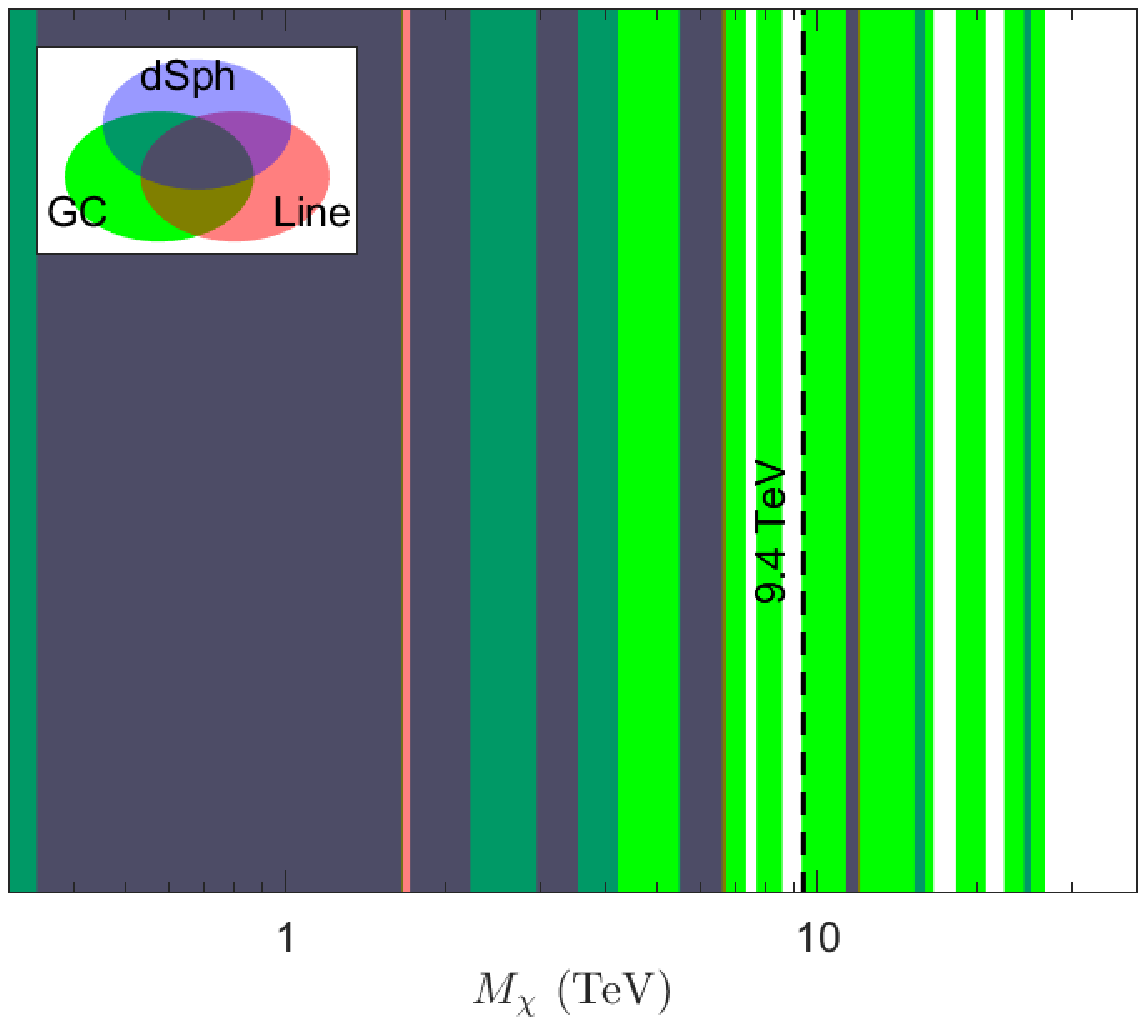}
		\subcaption{}
		\label{fig:Par_R5}
	\end{subfigure}
	\begin{subfigure}{.5\linewidth}
		\centering
		\includegraphics[width=\linewidth]{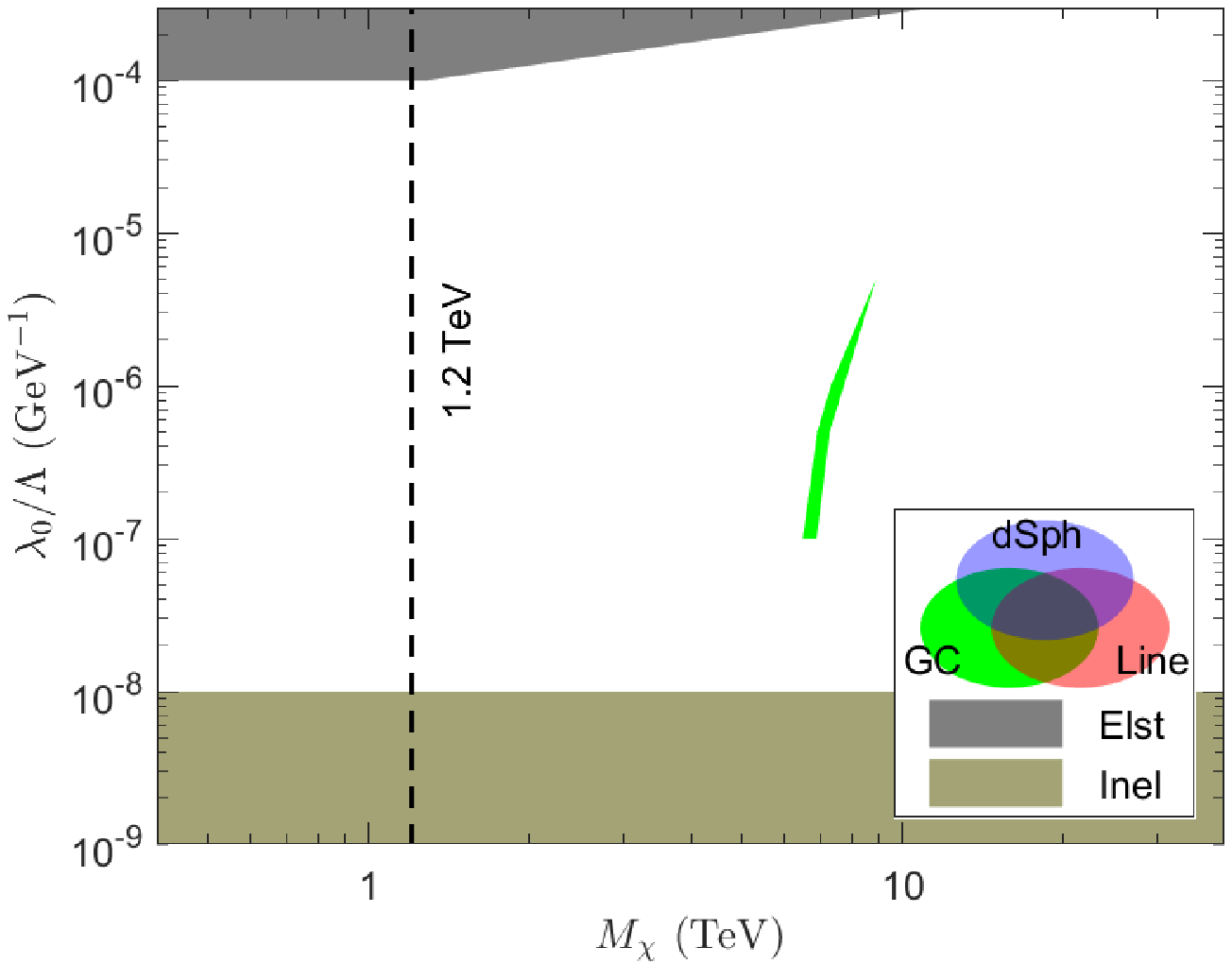}
		\subcaption{}
		\label{fig:Par_C2}
	\end{subfigure}%
	\begin{subfigure}{.5\linewidth}
		\centering
		\includegraphics[width=\linewidth]{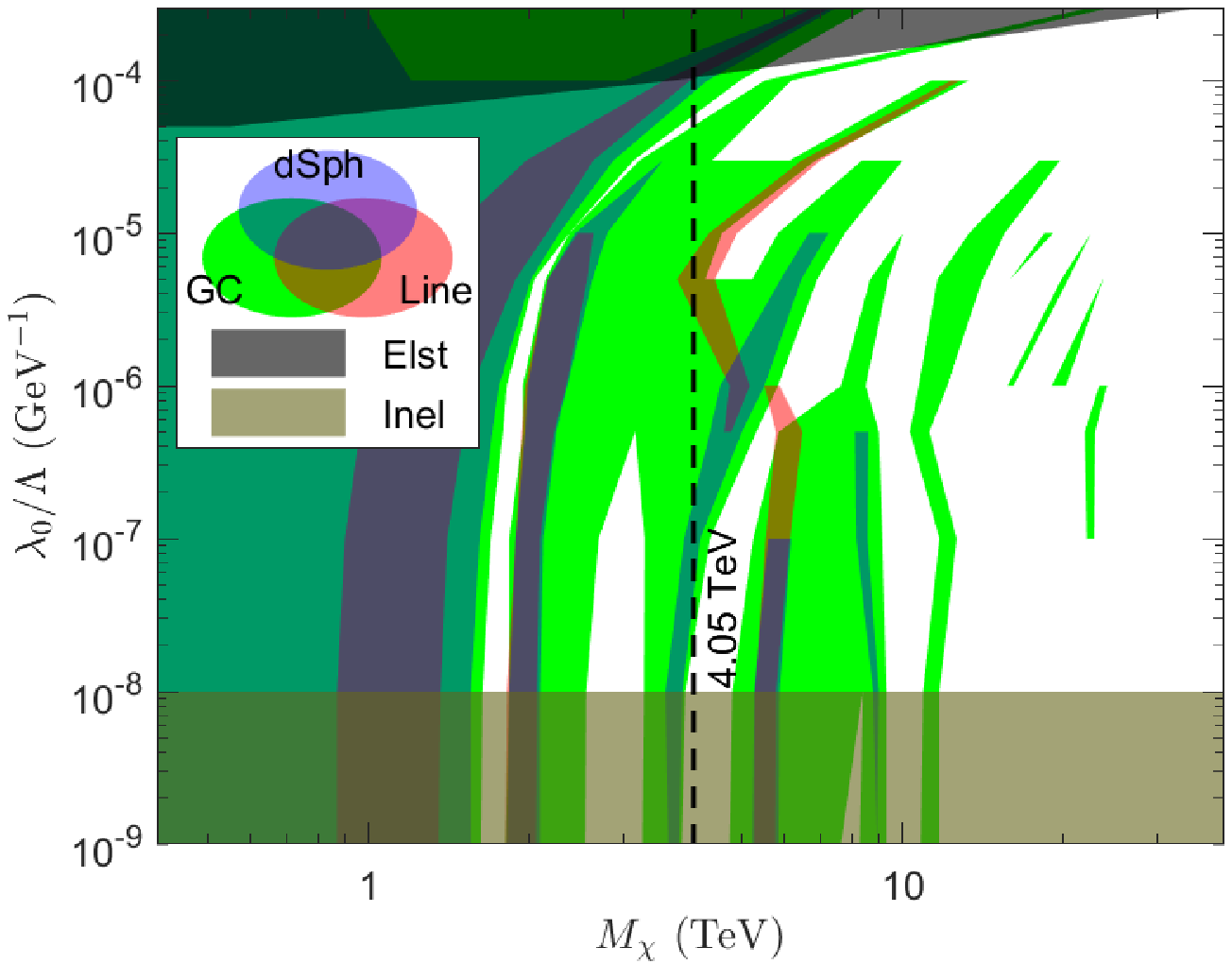}
		\subcaption{}
		\label{fig:Par_C4}
	\end{subfigure}
	\caption{(Upper panels) Summary charts for the real triplet (left \ref{fig:Par_R3}) and quintet (right \ref{fig:Par_R5}) showing the allowed values of DM mass as the only free parameter.
	The vertical bars are ruled out by experimental constraints from the inner Galaxy (green), dwarf satellites (blue), and gamma-ray line searches (red). The vertical dash-line indicates the thermal mass which is fixed by the freeze-out mechanism. The vertical axis does not report any physical variable. It is provided for the real cases to make the comparison with complex models possible.\\
	(Lower panels) The mass-coupling $m_\chi - \lambda_0$ two-dimensional parameter space of the pseudo-real doublet (left \ref{fig:Par_C2}) and quartet (right \ref{fig:Par_C4}). Here, the top horizontal grey region is excluded by elastic direct DM searches. In addition, the horizontal area at the bottom is disfavoured by inelastic EWDM-nucleon interactions through Z-boson exchange.}
	\label{fig:Par_ID_DD}
\end{figure}

In this section, we update the observational bounds on the free parameters of the electroweak theory of DM, to include the constraints from both direct detection and indirect searches.  

Figure \ref{fig:Par_ID_DD} illustrates the valid values of DM mass for the real models, as well as the neutral coupling - mass $\lambda_0 - m_\chi$ plane for the complex representations. 
The vertical shaded patches are excluded by the three gamma-ray observations that are inner galaxy, dwarf satellites and photon lines. In addition, the top horizontal areas in the pseudo-real representations are disfavoured by elastic direct detection results, while the horizontal regions at the bottom are excluded due to inelastic scattering mediated by Z-boson \eqref{eq:l0_min}.

Since nucleus recoiling constraints are weakened above 100 GeV scale, direct detection data favours TeV electroweak dark matter. 
Real models cannot be probed by present DD experiments, as their scattering amplitude does not receive any contribution from the non-renormalisable operators. 

As discussed, due to the effective interactions with Higgs boson induced by higher dimensional operators, the scattering amplitude can be significantly enhanced in the pseudo-real dark matter representations.
So, the direct search data will set an upper-limit for the value of non-renormalisable couplings. It can be seen that regions of parameter space above $\lambda_0 \approx 10^{-4} \ \mathrm{GeV}^{-1}$ are not accessible for complex models, with constraints on the pseudo-real doublet being slightly weaker.

Restrictions exerted by gamma-ray probes are noticeably representation dependant. For the real triplet ($\mathbb{R}3$), the mass interval including and below the thermal value is ruled out, although higher DM masses are still acceptable. It can be observed that the real quintuplet ($\mathbb{R}5$) is severely bounded by indirect detection, leaving only a few narrow mass ranges available. The complex quadruplet ($\mathbb{C}4$) allows for a wider favoured areas of the parameter space particularly at larger masses and stronger scalar coupling. The pseudo-real doublet ($\mathbb{C}2$) is not considerably constrained by gamma-ray searches of EWDM.

To summarise, the multi-TeV and higher mass regions  of the parameter space with an intermediate coupling strength are favoured by a combination of the direct and indirect experiments and could be explored further by the future probes.


\section{Conclusion}
\label{sec:conclusion}

The electroweak sector was extended by adding a fermion multiplet in non-chiral representation charged under the SU(2)$\times$U(1)\textsubscript{y} gauge group, so that successful features of the Standard Model were minimally impacted. The pseudo-real models are excluded by direct detection results due to the tree-level coupling to nuclei through the exchange of Z boson. 
We generalised the pseudo-real EWDM framework by introducing dimension five operators which couple the multiplet to Higgs boson. This mechanism revitalised the pseudo-real theory as the mentioned effective term splits the pseudo-Dirac dark matter into two Majorana states, therefore eliminating the tree-level Z-mediated interaction with nuclei.

While EWDM does not scatter off the nucleon at tree-level, it does through non-renormalisable couplings, in addition to loop diagrams. 
in this paper, we studied the direct detection of electroweak dark matter as a suitable method to probe effects of ultraviolet operators on the pseudo real models at TeV scale. 
We formulated the effective scalar theory of EWDM - nucleon scattering at parton level which is quite useful in evaluating the non-renormalisable couplings in a systematic way.

All the diagrams that make a contribution to the EWDM - nucleon scattering to the leading order of $\mathcal{O}(\Lambda^{-1})$ were taken into account. We evaluated the tree-level (one-loop) process that gives rise to the DM - quark collision through the non-renormalisable cubic (quartic) Higgs coupling, in addition to the one-loop (two-loop) processes generating interactions with gluons.

The effective amplitudes for the main scattering channels all have positive values which leads to a constructive contribution of the non-renormalisable interactions to the predicted detection rate.

In this framework, we studied the SI cross section of the electroweak DM arising from effective operators of the lowest dimension. 
The behaviour of the scattering cross-section across different pseudo-real models is determined by the (square of) parameter $\lambda$ which is a linear combination of the two non-renormalisable couplings \eqref{eq:lambda}.

There exist a critical value for this parameter at which the amplitudes for renormalisable and non-renormalisable effective couplings are the same. 
Below $\lambda_\mathrm{cr}$, different EWDM representations have distinct spectra which lies well below the present direct detection constraints. 
However, as $\lambda$ gets stronger than the criticality value, the spectral curves for complex modules tend to become degenerate. The DD cross-section keeps increasing up to orders of magnitude, and therefore will be bounded from above by the measurement data.

If the charged and neutral couplings are proportional to each other, then $\lambda$ approaches zero. At this minimum limit, the effective theory discussed in this paper, will behave similar to the renormalisable model from the viewpoint of experimentally observable results.

The pseudo-real doublet as the least constraint EWDM model lies far below the neutrino floor. It  is therefore difficult to detect this model in current direct detection searches. However, the non-renormalisable effects can raise the annihilation rate to the level that hopefully will be detectable for the next generation experiments.

Finally, we combined the astrophysical indirect search and direct detection bounds to find the allowed parameter space for all the electroweak dark matter representations. 
In general, pseudo-real theories support a wider range of viable values of the parameters. The higher mass intervals and intermediate coupling regime are overall favoured by a combination of constraints imposed by ID and DD data.

In conclusion, in case, the non-renormalisable coupling of DM to the SM through Higgs field is negligible, the resultant cross-section would stay below the current experimental constraints. 
However, if the new physics responsible for the mass splitting of the dark spices is closer to the electroweak scale, the effect of the higher dimensional operators will escalate the scattering amplitude. This would potentially open the window for detection of electroweak dark matter in the near future experiments.


\section{acknowledgement}

I would like to acknowledge Natsumi Nagata, Rouven Essig, and Thomas Hahn for their discussions and contributions to this project.

We used TikZ-Feynman 1.1.0 \cite{TikZ_Feynman} to generate Feynman diagrams , and FeynArts 3.11 \cite{FeynArts} for evaluating the diagrams.


\appendix

\section{Lagrangian and Feynman rules}
\label{app:Feynman}

The complete interaction Lagrangian describing the couplings of the pseudo-real multiplet with the SM particles can be decomposed as:
\begin{equation}
	\mathcal{L}_\mathrm{int}	=	\mathcal{L}_A 	+\mathcal{L}_Z 	+\mathcal{L}_W 	+\mathcal{L}_H
\end{equation}

Interactions of dark sector particles and gauge fields are derived from expansion of the kinetic term in the general Lagrangian \ref{eq:L_C}.
Electromagnetic interactions are given by the Lagrangian:
\begin{equation}
	\mathcal{L}_A	 =	-e \left[		\frac{n}{2}	\, \overline{\chi}^\frac{n}{2} \gamma^\mu \chi^\frac{n}{2} 
					+	\sum_{q = 1} ^{n/2-1}	q \left( 	\overline{\chi}^q_1 \gamma^\mu \chi^q_1	+\overline{\chi}^q_2 \gamma^\mu \chi^q_2	\right) 	\right]	A_\mu 	\,,
\end{equation}

Interactions mediated by the neutral weak boson can be written as:
\begin{align}
	\mathcal{L}_Z &=		- \frac{g_\mathrm{z}}{2} 	\left\{ 	\vphantom{\sum_{q = -n/2 +1} ^{n/2-1}}
					\left( nc_w^2 -1 \right) \overline{\chi}^\frac{n}{2} \gamma^\mu \chi^\frac{n}{2}	-i \overline{\widetilde{\chi}^0} \gamma^\mu \chi^0		\right.	\\	\nonumber
			& \left.	+ \sum_{q = 1} ^{n/2-1}	 \left[ 	\left( 2c_w^2 q + \cos \phi_q \right)	\overline{\chi}^q_1 \gamma^\mu \chi^q_1	
					+	\left( 2c_w^2 q - \cos \phi_q \right)	 	\overline{\chi}^q_2 \gamma^\mu \chi^q_2 
					+	\sin \phi_q	\left(	\overline{\chi}^q_1 \gamma^\mu \chi^q_2	+\overline{\chi}^q_2 \gamma^\mu \chi^q_1	\right)	\right]	\right\} Z_\mu \,,
\end{align}

The odd particles couple to the charged weak gauges through:
\begin{align}
	\mathcal{L}_W &=		- \frac{g_w}{2\sqrt{2}}	\left\{	
			2 \sqrt{n-1} \, e^{\frac{i}{2} \widehat{\lambda}_0}	\left[ c_{\frac{n}{2} -1}	 \,	\overline{\chi}^{\frac{n}{2} -1}_2 \gamma^\mu \chi^\frac{n}{2}	
									- s_{\frac{n}{2} -1} \,	\overline{\chi}^{\frac{n}{2} -1}_1 \gamma^\mu \chi^\frac{n}{2}		\right]	\right.	\\	\nonumber
		&+	\frac{1}{\sqrt{2}}		\left(   n \ e^{-\frac{i}{2} \widehat{\lambda}_0} c_+   
									- \sqrt{ n^2 -4} \ e^{\frac{i}{2} \widehat{\lambda}_0} s_+ \right)	\overline{\widetilde{\chi}^0} \gamma^\mu \chi_2^+
			- \frac{i}{\sqrt{2}} 	\left(   n \ e^{-\frac{i}{2} \widehat{\lambda}_0} c_+   
									+ \sqrt{ n^2 -4} \ e^{\frac{i}{2} \widehat{\lambda}_0} s_+ \right)	\overline{\chi^0} \gamma^\mu \chi_2^+  		\\	\nonumber
		&  - \frac{1}{\sqrt{2}}	\left(   n \ e^{-\frac{i}{2} \widehat{\lambda}_0} s_+   
									+ \sqrt{ n^2 -4} \ e^{\frac{i}{2} \widehat{\lambda}_0} c_+ \right)		\overline{\widetilde{\chi}^0} \gamma^\mu \chi_1^+
			+ \frac{i}{\sqrt{2}}	\left(   n \ e^{-\frac{i}{2} \widehat{\lambda}_0} s_+   
									- \sqrt{ n^2 -4} \ e^{\frac{i}{2} \widehat{\lambda}_0} c_+ \right)	\overline{\chi^0} \gamma^\mu \chi_1^+ 		\\	\nonumber
		& +	\sum_{q=2}^{\frac{n}{2} -1}	\left[		\left(	\sqrt{ n^2 -4(q-1)^2}	\, s_{q-1} s_q
									- \sqrt{ n^2 -4q^2} \, c_{q-1} c_q	\right)	\overline{\chi}^{q-1}_1 \gamma^\mu \chi^q_1	\right.	\\	\nonumber
		& \qquad	+ 	\left(	\sqrt{ n^2 -4(q-1)^2} \,		 c_{q-1} c_q	
									- \sqrt{ n^2 -4q^2} \,	 s_{q-1} s_q	\right)	\overline{\chi}^{q-1}_2 \gamma^\mu \chi^q_2		\\	\nonumber
		& \qquad	-		\left(	\sqrt{ n^2 -4(q-1)^2}	\, s_{q-1} c_q
									+ \sqrt{ n^2 -4q^2} \,	 c_{q-1} s_q	\right)	\overline{\chi}^{q-1}_1 \gamma^\mu \chi^q_2		\\	\nonumber
		& \qquad	-		\left(	\sqrt{ n^2 -4(q-1)^2}	\, c_{q-1} s_q
									+ \sqrt{ n^2 -4q^2} \,	 s_{q-1} c_q	\right)	\overline{\chi}^{q-1}_2 \gamma^\mu \chi^q_1		
										\left. \left.		\right]	\vphantom{\overline{\chi}^{\frac{n}{2} -1}_2}	\right\}		W^-_\mu 	+ \mathrm{cc} \,,
\end{align}

In special case of the complex doublet, the charged weak boson interactions read:
\begin{equation}
	\mathcal{L}_W^{\mathbb{C}2} =	-\frac{g_w}{2}	\left(	\overline{\widetilde{\chi}^0} \gamma^\mu \chi^+		
						-i \,	\overline{\chi^0} \gamma^\mu \chi^+	\right)	W^-_\mu 		+ \mathrm{cc}\,,
\end{equation}

Non-renormalisable interactions with Higgs boson can be cast as:
\begin{align}
\label{eq:L_H}
	\mathcal{L}_H &=	-\Delta^{(\mathrm{t})}_{\frac{1}{2} (n-1)}	\left(	\frac{2}{\nu} h 	+ \frac{1}{\nu^2} h^2	\right)	\overline{\chi}^\frac{n}{2}	\chi^\frac{n}{2}		\\		\nonumber
		& - \left(	\frac{2}{\nu} h 	+ \frac{1}{\nu^2} h^2	\right)		\left[	\left(	\frac{1}{2} \,\Delta^{(\mathrm{t})}_{-\frac{1}{2}} 	-\Re \delta_0	\right)	\overline{\chi^0} \chi^0	
			+ \left(	\frac{1}{2} \,\Delta^{(\mathrm{t})}_{-\frac{1}{2}} 	+\Re \delta_0	\right)	\overline{\widetilde{\chi}^0}		\widetilde{\chi}^0		
			- 	2 \, \Im \delta_0		\overline{\chi^0}	\widetilde{\chi}^0	\right]							\\	\nonumber
		& - \sum_{q=1}^{\frac{n}{2} -1}		\left(	\frac{2}{\nu} h 	+ \frac{1}{\nu^2} h^2	\right)		\left[	
			\left(	\Delta^{(\mathrm{t})}_{q-\frac{1}{2}} 	- (-1)^q \, 2 |\delta_q| \sin \phi_q	\right)		\overline{\chi}^q_1  \chi^q_1	
			+ \left(	\Delta^{(\mathrm{t})}_{q-\frac{1}{2}} 	+ (-1)^q \, 2 |\delta_q| \sin \phi_q	\right)		\overline{\chi}^q_2  \chi^q_2	\right.	\\	\nonumber
		&	\left.	+  (-1)^q \, 2 	|\delta_q| \cos \phi_q	\left(	\overline{\chi}^q_1  \chi^q_2	+	\overline{\chi}^q_2  \chi^q_1	\right)	\vphantom{\Delta^{(t)}_{q-\frac{1}{2}}}	\right]
\end{align}

Using the interaction Lagrangian above, one can readily obtain the Feynman rules for the coupling of the complex EWDM with gauge fields and scalars of the SM.

Feynman rule for the coupling of the charged particles to photon can be written as:
\begin{align}
	C^\mu ( \chi^\frac{n}{2}, \chi^\frac{n}{2} , A )	=	- \frac{i}{2} n e \,\gamma^\mu		\\	\nonumber
	C^\mu ( \chi^q_i, \chi^q_j, A )	=	-i e q \,\gamma^\mu \delta_{ij}
\end{align}

For vertices involving neutral electroweak $Z$ boson, one has:
\begin{align}
	C^\mu ( \chi^\frac{n}{2}, \chi^\frac{n}{2}, Z )	&=	- \frac{i }{2} g_\mathrm{z}	\left(n c_w^2 -1 \right) \gamma^\mu	\\		\nonumber
	C^\mu ( \chi^q_1, \chi^q_1, Z )	&=				- \frac{i }{2} g_\mathrm{z}		\left( 2 c_w^2 q +\cos \phi_q \right) \gamma^\mu	\\		\nonumber
	C^\mu ( \chi^q_2, \chi^q_2, Z )	&=				- \frac{i }{2} g_\mathrm{z}		\left( 2 c_w^2 q -\cos \phi_q \right) \gamma^\mu	\\		\nonumber
	C^\mu ( \chi^q_1, \chi^q_2, Z )	&=				- \frac{i }{2} g_\mathrm{z}		\sin \phi_q \gamma^\mu	\\							\nonumber
	C^\mu (\, \overline{\chi}^0, \widetilde{\chi}^0, Z )	&=		- C^\mu (\, \overline{\widetilde{\chi}^0}, \chi^0, Z ) =			g_\mathrm{z} \gamma^\mu					
\end{align}

Note that $Z$ boson can change the flavour of the odd particles with the same electric charge.

Interactions of the fields with different charges are mediated by $W^-$ gauge boson through:
\begin{align}
	C^\mu ( \overline{\chi}^{n/2-1}_2, \chi^\frac{n}{2}, W^- )	&=		- \frac{i }{\sqrt{2}}	\sqrt{n-1}	\,e^{\frac{i}{2} \widehat{\lambda}_0}	c_{\frac{n}{2} -1}	\,g_w \gamma^\mu 	\\	\nonumber
	C^\mu ( \overline{\chi}^{n/2-1}_1, \chi^\frac{n}{2}, W^- )	&=		\frac{i }{\sqrt{2}}	\sqrt{n-1}	\,e^{\frac{i}{2} \widehat{\lambda}_0}	s_{\frac{n}{2} -1}	\,g_w \gamma^\mu 	\\	\nonumber
	C^\mu ( \overline{\chi}^{q-1}_1, \chi^q_1, W^- )	&=		\frac{i }{2\sqrt{2}}	\left(	\sqrt{ n^2 -4q^2} \, c_{q-1} c_q		-\sqrt{ n^2 -4(q-1)^2}	\, s_{q-1} s_q	\right)	\,g_w \gamma^\mu 	\\	\nonumber
	C^\mu ( \overline{\chi}^{q-1}_2, \chi^q_2, W^- )	&=		\frac{i }{2\sqrt{2}}	\left(	\sqrt{ n^2 -4q^2} \, s_{q-1} s_q		-\sqrt{ n^2 -4(q-1)^2}	\, c_{q-1} c_q	\right)	\,g_w \gamma^\mu 	\\	\nonumber
	C^\mu ( \overline{\chi}^{q-1}_1, \chi^q_2, W^- )	&=		\frac{i }{2\sqrt{2}}	\left(	\sqrt{ n^2 -4q^2} \, c_{q-1} s_q		+\sqrt{ n^2 -4(q-1)^2}	\, s_{q-1} c_q	\right)	\,g_w \gamma^\mu 	\\	\nonumber
	C^\mu ( \overline{\chi}^{q-1}_2, \chi^q_1, W^- )	&=		\frac{i }{2\sqrt{2}}	\left(	\sqrt{ n^2 -4q^2} \, s_{q-1} c_q		+\sqrt{ n^2 -4(q-1)^2}	\, c_{q-1} s_q	\right)	\,g_w \gamma^\mu 	\\	\nonumber
	C^\mu ( \overline{\chi^0}, \chi^+_1, W^- )			&=		\frac{1}{4}		\left(   n \ e^{-\frac{i}{2} \widehat{\lambda}_0} s_+   	
						- \sqrt{ n^2 -4} \ e^{\frac{i}{2} \widehat{\lambda}_0} c_+ \right)		\,g_w \gamma^\mu 	\\	\nonumber
	C^\mu ( \overline{\widetilde{\chi}^0}, \chi^+_2, W^- )		&=		- \frac{i}{4}				\left(   n \ e^{-\frac{i}{2} \widehat{\lambda}_0} c_+   
						- \sqrt{ n^2 -4} \ e^{\frac{i}{2} \widehat{\lambda}_0} s_+ \right)		\,g_w \gamma^\mu 	\\	\nonumber
	C^\mu ( \overline{\chi^0}, \chi^+_2, W^- )			&=		- \frac{1}{4}		\left(   n \ e^{-\frac{i}{2} \widehat{\lambda}_0} c_+   	
						+ \sqrt{ n^2 -4} \ e^{\frac{i}{2} \widehat{\lambda}_0} s_+ \right)		\,g_w \gamma^\mu 	\\	\nonumber
	C^\mu( \overline{\widetilde{\chi}^0}, \chi^+_1, W^- )		&=		\frac{i}{4}				\left(   n \ e^{-\frac{i}{2} \widehat{\lambda}_0} s_+   
						+ \sqrt{ n^2 -4} \ e^{\frac{i}{2} \widehat{\lambda}_0} c_+ \right)		\,g_w \gamma^\mu 
\end{align}

For the comple doublet we have:
\begin{align}
	C^\mu ( \overline{\chi^0}, \chi^+, W^- )		&=		- \frac{1}{2}	\,g_w \gamma^\mu 	\\	\nonumber
	C^\mu ( \overline{\widetilde{\chi}^0}, \chi^+, W^- )		&=		- \frac{i}{2}	\,g_w \gamma^\mu 
\end{align}

Factorising out the root of unity 
$C^\mu  \equiv -i C' \gamma^\mu$, 
Feynman rules for the conjugate charged field $W^+$ can be easily obtained from:
\begin{equation}
	C' (\, \overline{\chi}^q_i, \chi^{q-1}_j, W^+ )	= 	C'^* (\, \overline{\chi}^{q-1}_j, \chi^q_i, W^- )
\end{equation}

Feynman rules for cubic non-renormalisable couplings to the Higgs boson has the form:
\begin{align}
\label{eq:C_h3}
	C ( \overline{\chi}^\frac{n}{2}, \chi^\frac{n}{2}, h )		&=		-2 \frac{i}{\nu}	\,\Delta^{(\mathrm{t})}_{\frac{1}{2} (n-1)}
		=	\frac{i}{4}	\left( n-1 \right)		\nu	\frac{\lambda_c}{\Lambda}			\\	\nonumber
	C ( \overline{\chi}^q_1, \chi^q_1, h )		&=		-2 \frac{i}{\nu}	\left(	\Delta^{(\mathrm{t})}_{q-\frac{1}{2}} 	+ (-1)^{q+1} \, 2 |\delta_q| \sin \phi_q	\right)
		=	\frac{i}{2}	\nu	\left[	\left( q -\frac{1}{2} \right)	\frac{\lambda_c}{\Lambda}		+ (-1)^q \sqrt{n^2 -4q^2}	\sin \phi_q	\left| \frac{\lambda_0}{\Lambda} \right|	\right]	\\	\nonumber
	C ( \overline{\chi}^q_2, \chi^q_2, h )		&=		-2 \frac{i}{\nu}	\left(	\Delta^{(\mathrm{t})}_{q-\frac{1}{2}} 	+ (-1)^q \, 2 |\delta_q| \sin \phi_q	\right)
		=	\frac{i}{2}	\nu	\left[	\left( q -\frac{1}{2} \right)	\frac{\lambda_c}{\Lambda}		+  (-1)^{q+1}   \sqrt{n^2 -4q^2}		\sin \phi_q	\left| \frac{\lambda_0}{\Lambda} \right|	\right]	\\	\nonumber
	C ( \overline{\chi}^q_2, \chi^q_1, h )		&=			C ( \overline{\chi}^q_1, \chi^q_2, h )	=	 (-1)^{q+1} \, 4 \frac{i}{\nu}	|\delta_q|	\cos \phi_q
		=	 (-1)^{q+1} \,  \frac{i}{2}	\nu		\sqrt{n^2 -4q^2}	\cos \phi_q	\left| \frac{\lambda_0}{\Lambda} \right|		\\	\nonumber
	C ( \overline{\chi^0}, \chi^0, h )		&=		- \frac{i}{\nu}		\left(	\Delta^{(\mathrm{t})}_{-\frac{1}{2}}		- 2 \Re \,\delta_0		\right)
		=	- \frac{i}{4} \nu	\left(	\frac{\lambda_c}{2 \Lambda}	- n \Re \frac{\lambda_0}{\Lambda}	\right)		\\	\nonumber
	C ( \overline{\widetilde{\chi}^0}, \widetilde{\chi}^0, h )		&=		- \frac{i}{\nu}		\left(	\Delta^{(\mathrm{t})}_{-\frac{1}{2}}		+ 2 \Re \,\delta_0		\right)
		=	- \frac{i}{4} \nu	\left(	\frac{\lambda_c}{2 \Lambda}	+ n \Re \frac{\lambda_0}{\Lambda}	\right)				\\	\nonumber
	C ( \overline{\chi^0}, \widetilde{\chi}^0, h )		&=		C ( \overline{\widetilde{\chi}^0}, \chi^0, h )		=		i \frac{4}{\nu}	\Im \,\delta_0
		=	i \frac{n}{2}	\,\nu		\,\Im \frac{\lambda_0}{\Lambda}
\end{align}

Quartic non-renormalisable interactions of two Higgs bosons and two DM particles can be cast as:
\begin{align}
\label{eq:C_h4}
	C ( \overline{\chi}^\frac{n}{2}, \chi^\frac{n}{2}, h, h )		&=		- \frac{i}{\nu^2}	\,\Delta^{(\mathrm{t})}_{\frac{1}{2} (n-1)}
			=		\frac{i}{8}	\left( n-1 \right)		\frac{\lambda_c}{\Lambda}			\\	\nonumber
	C ( \overline{\chi}^q_1, \chi^q_1, h, h )		&=		- \frac{i}{\nu^2}	\left(	\Delta^{(\mathrm{t})}_{q-\frac{1}{2}} 	+ (-1)^{q+1} \, 2 |\delta_q| \sin \phi_q	\right)
			=		\frac{i}{4}	\left[	\left( q -\frac{1}{2} \right)	\frac{\lambda_c}{\Lambda}		+  (-1)^q \sqrt{n^2 -4q^2}	\sin \phi_q	\left| \frac{\lambda_0}{\Lambda} \right|	\right]	\\	\nonumber
	C ( \overline{\chi}^q_2, \chi^q_2, h, h )		&=		- \frac{i}{\nu^2}	\left(	\Delta^{(\mathrm{t})}_{q-\frac{1}{2}} 	+ (-1)^q \, 2 |\delta_q| \sin \phi_q	\right)
			=		\frac{i}{4}	\left[	\left( q -\frac{1}{2} \right)	\frac{\lambda_c}{\Lambda}		+  (-1)^{q+1} \sqrt{n^2 -4q^2}		\sin \phi_q	\left| \frac{\lambda_0}{\Lambda} \right|	\right]	\\	\nonumber
	C ( \overline{\chi}^q_2, \chi^q_1, h, h )		&=			C ( \overline{\chi}^q_1, \chi^q_2, h, h )	=	 (-1)^{q+1} \, 2 \frac{i}{\nu^2}	|\delta_q|	\cos \phi_q
			=	(-1)^{q+1} \, \frac{i}{4}	\sqrt{n^2 -4q^2}	\cos \phi_q	\left| \frac{\lambda_0}{\Lambda} \right|		\\	\nonumber
	C ( \overline{\chi^0}, \chi^0, h, h )		&=		- \frac{i}{\nu^2}		\left(	\frac{1}{2}	\Delta^{(\mathrm{t})}_{-\frac{1}{2}}		- \Re \,\delta_0		\right)
			=	- \frac{i}{8} 		\left(	\frac{\lambda_c}{2 \Lambda}	- n \Re \frac{\lambda_0}{\Lambda}	\right)		\\	\nonumber
	C ( \overline{\widetilde{\chi}^0}, \widetilde{\chi}^0, h, h )		&=		- \frac{i}{\nu^2}		\left(	\frac{1}{2}	\Delta^{(\mathrm{t})}_{-\frac{1}{2}}		+ \Re \,\delta_0		\right)
			=	- \frac{i}{8} 		\left(	\frac{\lambda_c}{2 \Lambda}	+ n \Re \frac{\lambda_0}{\Lambda}	\right)				\\	\nonumber
	C ( \overline{\chi^0}, \widetilde{\chi}^0, h, h )		&=		C ( \overline{\widetilde{\chi}^0}, \chi^0, h, h )		=		i \frac{2}{\nu^2}	\Im \,\delta_0
			=	i \frac{n}{4}		\,\Im \frac{\lambda_0}{\Lambda}
\end{align}


\section{Loop integrals}
\label{app:Loop}

In order to compute the Wilson coefficients in the full theory of EWDM, we need to introduce and evaluate new loop integrals.


\subsection{One-point function}

The scalar one-point function is given by \cite{Scalar_1loop}:
\begin{subequations}
\begin{align}
	\label{eq:A0}
	A_0^{(n)} (m)	&\equiv	\mu^{4-d}	\int 	\frac{1}{\left(k^2 -m^2 \right)^n}	\frac{\mathrm{d}^d k}{(2\pi)^d}			\\	\nonumber
					&=		\frac{(-1)^n i}{(4\pi)^{d/2}}		\, \mu^{4-d}	m^{d-2n}	\frac{\Gamma(n- d/2)}{\Gamma(n)}
	\end{align}

In $d=4$ dimensions, for $n=1,2$ the integral diverges. In \emph{dimensional regularisation} scheme, the dimension is taken to be $d-4 \ll 1$. The regularisation scale $\mu$ fixes the dimension of measure at 4. 
Therefore, to $\mathcal{O}(d-4)$ we get:
\begin{align}
	A_0^{(1)} (m)	&=	\frac{i}{(4\pi)^2}	\, m^2	\left( \Delta	-2 \ln \frac{m}{\mu}		+1	\right)		\\
	A_0^{(2)} (m)	&=	\frac{i}{(4\pi)^2}	\left( \Delta	-2 \ln \frac{m}{\mu}		\right)	
\end{align}
\end{subequations}

where the divergence is contained in 
$\Delta \equiv	\ln 4\pi	-\gamma_E 		+2/(4-d)$
with $\gamma_E$ being Euler-Mascheroni constant.

The scalar one-point correlator can be generalised to two mass scales:
\begin{subequations}
\begin{align}
	A_0^{(n_1,n_2)} (m_1,m_2)		&\equiv	\mu^{4-d}	\int 	\frac{1}{ \left(k^2 -m_1^2 \right)^{n_1}	\left(k^2 -m_2^2 \right)^{n_2} }	\frac{\mathrm{d}^d k}{(2\pi)^d}			\\	\nonumber
			&=		\frac{(-1)^n i}{(4\pi)^2}		\frac{\Gamma(n-2)}{\Gamma(n_1) \Gamma(n_2)}		\frac{1}{ \left( m_1^2 -m_2^2 \right)^{n-2}}
			\left[	\vphantom{\sum_{k=1}^{n-3}}		(-1)^{n_2 +1}	+ (-1)^{n_2} 2 	\ \frac{ (n_2 -1) m_1^2  +(n_1 -1) m_2^2}{ m_1^2 -m_2^2 }	\ \ln \frac{m_1}{m_2}		\right.		\\
			&+	\left.	\sum_{k=1}^{n-3}	\frac{(-1)^{n_2 +k}}{k}		\frac{ m_1^{2k} -m_2^{2k} }{ m_1^2 -m_2^2 }		\sum_{j= k-n_2}^{ \min(k, n_1) -1 }																\nonumber		\begin{pmatrix}		n_1 -1\\		j	\end{pmatrix}		\begin{pmatrix}	n_2 -1\\		k-j-1	\end{pmatrix}		\frac{1}{ m_1^{2(j-1)} -m_2^{2(	k-j)} }	\right] \,,
\end{align}

with $n \equiv n_1 + n_2 >2$. 
Obviously $A_0^{(n_2,n_1)} (m_2,m_1) = A_0^{(n_1,n_2)} (m_1,m_2)$.

In this paper, we are interested in the special case:
\begin{equation}
	A_0^{(1,2)} (m_1,m_2)		=		\frac{-i}{(4\pi)^2}		\frac{1}{m_1^2 -m_2^2}	\left( \frac{2 \ m_1^2}{m_1^2 -m_2^2}		\ \ln \frac{m_1}{m_2}	-1	\right)	\,.
\end{equation}
\end{subequations}


The one-point function can be further generalised to include another mass scale:
\begin{equation}
	A_0^{(n_1, n_2, n_3)} (m_1,m_2, m_3)		\equiv 	\mu^{4-d}	\int 	
	\frac{1}{ \left(k^2 -m_1^2 \right)^{n_1}	\left(k^2 -m_2^2 \right)^{n_2} 	\left(k^2 -m_3^2 \right)^{n_3}}	\frac{\mathrm{d}^d k}{(2\pi)^d}
\end{equation}

As a useful example, we explicitly calculate:
\begin{equation}
	A_0^{(1,1,2)} (0, m_1 ,m_2)		=		\frac{i}{(4\pi)^2}	\frac{1}{m_2^4}		\ln	\left[	\left(	\frac{m_2}{m_1}	\right)^2	+1	\right]		\,.
\end{equation}


\subsection{Two-point function}
\label{app:sec:2pt}

The scalar two-point function can be written as:
\begin{align}
	B_0^{(n_1, n_2)} (p^2 | m_1, m_2)		&\equiv	\mu^{4-d}	\int 	\frac{1}{ 	\left[(k+p)^2 -m_1^2 \right]^{n_1}		\left( k^2 -m_2^2 \right)^{n_2}	 }	
											\frac{\mathrm{d}^d k}{(2\pi)^d}		\,.
\end{align}


Note $B_0^{(n_2, n_1)} (p^2 | m_2, m_1)	=	B_0^{(n_1, n_2)} (p^2 | m_1, m_2)$, 
also $B_0^{(n,0)} (p^2 | m_1, m_2)	=	A_0^{(n)} (m_1)$.
For $n_1 = n_2 =1$, the loop integral is divergent. To order $\mathcal{O}(d-4)$ we are left with \cite{Denner_loop}:
\begin{subequations}
\begin{align}
	B_0 ( p^2 | m_1, m_2 )	&=	\frac{i}{(4\pi)^2}	\left[	\Delta	+2		- \ln \frac{m_1 m_2}{\mu^2}		+ \frac{1}{2}	\left( x_1 - x_2 \right)	\ln \frac{x_2}{x_1}
								- \kappa^2		K (x_1, x_2)		\right]		\,,		\\
	B_0 ( p^2 | m, m )	&=		\frac{i}{(4\pi)^2}	\left(	\Delta	+2		- 2 \ln \frac{m}{\mu}		- \kappa^2	K (x)		\right)		\,.
\end{align}
\end{subequations}

where $x_1 \equiv (m_1 /p)^2$, $x_2 \equiv (m_2 /p)^2$, 
$\kappa$ is the Kallen function $\kappa (x_1, x_2)	\equiv	\sqrt{( 1 -x_1 -x_2 )^2 	-4 x_1 x_2 }$, and we define $K$-function in \eqref{eq:K}. 

In the event of mass degeneracy in the propagators, we use the parameter $x \equiv (m/p)^2$. The Kallen function reduces to $\kappa (x) = \sqrt{1-4x}$, and the simplified $K$-function is defined in \eqref{eq:K_}.

The following special two-point functions facilitate computation of the loop diagrams encountered in the theory of EWDM:
\begin{subequations}
\begin{align}
	\label{eq:B12}										
	B_0^{(1,2)} (p^2 | m_1, m_2)	&=	\frac{i}{(4\pi)^2 p^2}	\left[	\frac{1}{2}	\ln \frac{x_1}{x_2}		
											+ (1 +x_1 -x_2)	K (x_1, x_2)	\right]	\,,		\\
	\label{eq:B12_m}										
	B_0^{(1,2)} (p^2 | m, m)		&=	\frac{i}{ (4\pi)^2 \, p^2}	\ 	K(x)	\,.
\end{align}
\end{subequations}

\begin{subequations}
\begin{align}
	B_0^{(2,2)} (p^2 | m_1, m_2)	&=	\frac{i}{ 2 (2\pi)^2}	\frac{1}{ p^2 \kappa^2}	\left[	(1 -x_1 -x_2)	K (x_1, x_2)		-1	\right]	\,,		\\
	\label{eq:B22}
	B_0^{(2,2)} (p^2 | m, m)		&=	\frac{i}{ 2 (2\pi)^2}	\frac{1}{ p^2 \kappa^2}	\ 	\left[	\left( 1 -2x \right)	K(x)	-1	\right]		\,.
\end{align}
\end{subequations}

\begin{subequations}
\begin{align}
	B_0^{(1,3)} (p^2 | m_1, m_2)	&=	\frac{i}{ 2 (4\pi)^2}	\frac{1}{ p^2 \kappa^2}		\left[	4 \, x_1 K (x_1, x_2)		- \frac{1}{x_2}(1 -x_1 -x_2)		\right]	\,,		\\
	\label{eq:B13}
	B_0^{(1,3)} (p^2 | m, m)		&=	\frac{i}{ 2 (4\pi)^2}	\frac{1}{ p^2 \kappa^2}		\left(	4 \, x K(x)	- \frac{1- 2x}{x}		\right)		\,.
\end{align}
\end{subequations}

\begin{subequations}
\begin{align}
	B_0^{(1,4)} (p^2 | m_1, m_2)	&=	\frac{i}{ 6 (4\pi)^2}	\frac{1}{ \kappa^4 m_1^4 p^2}		\left\{	12	x_1 x_2^2 (1 +x_1 -x_2 ) K (x_1, x_2)	
									+ \left( 1 -x_1 -x_2 \right)^3		- x_2 \left[	\left( 1 -x_2	\right)^2	+8 x_1	-9 x_1^2	\right]	\right\}	\,,		\\
	\label{eq:B14}
	B_0^{(1,4)} (p^2 | m, m)		&=	\frac{i}{ 6 (4\pi)^2}	\frac{1}{ \kappa^4 m^4 p^2}		\left[	12 \, x^3 K(x)	+ \left( 1-x \right)	\left( 1-6x \right)	\right]		\,.
\end{align}
\end{subequations}

One can use the following identity:
\begin{equation}
	i \ln \frac{ 1 -x_1 -x_2 +\kappa }{2 \sqrt{x_1 x_2}}		=	\atg \frac{|\kappa|}{x_1 +x_2 -1}	\,,
\end{equation}

in order to redefine the $K$-function when $\kappa^2 <0$, 
where $|\kappa| (x_1, x_2)		\equiv	 -i \kappa (x_1, x_2)	=	\sqrt{ 4 x_1 x_2	- ( 1 -x_1 -x_2 )^2 }$ 
is a positive real function:
\begin{equation}
	\label{eq:K}
	K (x_1, x_2) =
	\frac{1}{\kappa}		\ln \, \frac{ 1 -x_1 -x_2 +\kappa}{2 \sqrt{x_1 x_2} }	=		- \frac{1}{|\kappa|}		\atg \frac{|\kappa|}{x_1 +x_2 -1}	\,.
\end{equation}

In this way, the root of negative numbers and complex logarithms in the expressions for 2-point functions can be avoided.

In the case of equal masses $m_1 = m_2 \equiv m$, the equality below, serves the same purpose:
\begin{equation}
	i \ln \frac{1 +\kappa}{2 \sqrt{x}}		=		\acot |\kappa|		\,.
\end{equation}

here $|\kappa (x)| = -i \kappa (x) = \sqrt{4x-1}$. 
Note that this gives the useful identity:
\begin{equation}
	\label{eq:K_}
	\frac{1}{2}	K(x)	=	\frac{1}{\kappa}	\ln	\frac{1 +\kappa}{2 \sqrt{x}}		=		- \frac{1}{|\kappa|}	\acot |\kappa|	\,,
\end{equation}

The vector loop-integral 
\begin{equation}
		B_\mu^{(n_1, n_2)} (p^2 | m_1, m_2)		\equiv	\mu^{4-d}	\int 	\frac{p_\mu}{ 	\left[(k+p)^2 -m_1^2 \right]^{n_1}		\left( k^2 -m_2^2 \right)^{n_2}	 }	
											\frac{\mathrm{d}^d k}{(2\pi)^d}		\,,
\end{equation}

can be decomposed into external momentum $p_\mu$ as a Lorentz covariant quantity:
\begin{equation}
	B_\mu ^{(n_1, n_2)} 	=	p_\mu B_1 ^{(n_1, n_2)}	\,,
\end{equation}

where the scalar coefficient function is given by:
\begin{equation}
	B_1 ^{(n_1, n_2)}		=		\frac{1}{2 p^2}		\left(	B_0^{(n_1-1, n_2)}		- B_0^{(n_1, n_2-1)}	- f B_0^{(n_1, n_2)}	\right)		\,,
\end{equation}

where $f \equiv p^2 +m_2^2 -m_1^2$. In case of single dominator factor $n_i =1$, one should use the related one-point function in the equation above. For example:
\begin{equation}
	\label{eq:B1_1n}
	B_1 ^{(1, n)} (p^2 | m_1, m_2)		=		\frac{1}{2 p^2}		\left(	A_0^{(n)} (m_2)		- B_0^{(1, n-1)} (p^2 | m_1, m_2)	- f B_0^{(1, n)} (p^2 | m_1, m_2)	\right)		\,,
\end{equation}

Similarly, the tensor loop-integral 
\begin{equation}
	B_{\mu\nu}^{(n_1, n_2)} (p^2 | m_1, m_2)		\equiv	\mu^{4-d}	\int 	\frac{p_\mu p_\nu}{ 	\left[(k+p)^2 -m_1^2 \right]^{n_1}		\left( k^2 -m_2^2 \right)^{n_2}	 }	
											\frac{\mathrm{d}^d k}{(2\pi)^d}		\,,
\end{equation}

can be decomposed into tensors constructed from Lorentz covariant quantities that are external momentum $p_\mu$ and metric $\eta_{\mu\nu}$:
\begin{equation}
	B_{\mu\nu} ^{(n_1, n_2)} 	=	\eta_{\mu\nu} B_{00} ^{(n_1, n_2)}	 + p_\mu p_\nu B_{11} ^{(n_1, n_2)}		\,,
\end{equation}

and the scalar coefficients read:
\begin{subequations}
\begin{align}
	B_{00} ^{(n_1, n_2)} 	&=		\frac{1}{3}		\left[	B_0^{(n_1, n_2-1)}		+m_2^2 \, B_0^{(n_1, n_2)}	
								+\frac{1}{2}	\left(	B_1^{(n_1, n_2-1)}		+f B_1^{(n_1, n_2)}		\right)	\right]		\\
	B_{11} ^{(n_1, n_2)} 	&=		\frac{1}{3 p^2}		\left[	B_0^{(n_1, n_2-1)}		+m_2^2 \, B_0^{(n_1, n_2)}	
								+2	\left(	B_1^{(n_1, n_2-1)}		+f B_1^{(n_1, n_2)}		\right)	\right]
\end{align}
\end{subequations}



\section{Renormalisable Couplings}
\label{app:L_Ren}

\begin{table}[t]
	\centering
	{\setlength{\tabulinesep}{3pt}
	\begin{tabu}{|[1pt]c|c|[1pt]c|c|[1pt]}
		\tabucline[1pt]{-}
		$d_2^{(p)}$		&	0.118(3)				&		$\overline{d}_2^{(p)}$			&	0.037(3)		\\
		\hline
		$u_2^{(p)}$		&	0.223(3)				&		$\overline{u}_2^{(p)}$			&	0.036(2)		\\
		\hline
		$s_2^{(p)}$			&	0.0258(4)				&		$\overline{s}_2^{(p)}$			&	0.0258(4)		\\
		\hline
		$c_2^{(p)}$			&	0.0187(2)				&		$\overline{c}_2^{(p)}$			&	0.0187(2)		\\
		\hline
		$b_2^{(p)}$		&	0.0117(1)				&		$\overline{b}_2^{(p)}$			&	0.0117(1)		\\
		\tabucline[1pt]{-}							
	\end{tabu}}
	\caption{Second moments of the PDF's for proton computed at $\mu_\mathrm{uv} = m_\mathrm{z}$ \cite{q_2}. The digits in parentheses represent the statistical uncertainty.}
	\label{tab:q_2}
\end{table}

In this section, we revisit the computation of the electroweak dark matter - nucleon scattering cross-section through the usual renormalisable couplings. This allows us to calibrate our results with previous publication before studying the higher-dimensional operator extension of the electroweak theory of dark matter.

The SI effective interactions of EWDM with nucleon arising from a UV Lagrangian which only contains renormalisable operators, in leading order, can be written as:
\begin{align}
	\label{eq:L_Ren}
	\mathcal{L}_\mathrm{R}		&=	
		\sum_{\mathfrak{q}=d}^b			\left(	 m_\mathfrak{q}	 C _\mathfrak{q}^\mathrm{s}		\, \overline{\chi^0} \chi^0		\,\overline{\mathfrak{q}} \mathfrak{q}
			+		\frac{C_\mathfrak{q}^\mathrm{t_1}}{M_\chi}	\overline{\chi^0} i\partial_\mu \gamma_\nu \chi^0		\mathcal{O}^{\mu\nu}_\mathfrak{q}
			+		\frac{C_\mathfrak{q}^\mathrm{t_2}}{M_\chi^2}		\,\overline{\chi^0} i\partial_\mu i\partial_\nu \chi^0		\mathcal{O}^{\mu\nu}_\mathfrak{q}		\right)	\nonumber	\\		
					&+	\frac{\alpha_\mathrm{s}}{\pi}		C_\mathfrak{g}^\mathrm{s} 		\,\overline{\chi^0} \chi^0		G_{\mu\nu}^a G^{\mu\nu a}		\,,
\end{align}

where the Lagrangian is decomposed into a term representing interactions with quarks in the first line,%
\footnote{It should be noted that suppression by factors of $M_\chi^{-1}$ in the twist-two coupling in \eqref{eq:L_Ren}, will be cancelled out by the derivative of the DM field when computing the dark matter matrix element.
So, the effective amplitude arising from twist-2 operators has the same order of magnitude as those for scalar operators.} 
and another term in the second line for coupling to gluon \cite{Neutralino_nucleon}.%
\footnote{The twist-2 operators of gluon does not contribute in leading order of $\mathcal{O}(\alpha_\mathrm{s}^0)$ \cite{QCD_corrn}.}

The order of loop momenta in the scattering diagrams of the renormalisable theory is again around the weak scale \cite{Wino_Higgsino_DD}. This agrees with the similar observation in case of the non-renormalisable extension where we took the factorisation scale at $\mu_\mathrm{uv} \approx m_\mathrm{z}$.

In addition to the scalar interactions, the renormalisable Lagrangian \eqref{eq:L_Ren} apparently includes terms that couple to the quark twist-2 operators.

The matrix element of twist operators are defined through the \emph{parton-distribution functions} (PDF's) of quarks and anti-quarks. 
PDF $\mathfrak{q}^{(N)} (x)$ expresses the probability for a given parton species $\mathfrak{q}$ to carry a portion $x$, called \emph{longitudinal fraction}, of the total momentum of hadron $N$. 
Due to the symmetries linking proton and neutron, the distribution functions of neutron can be obtained from those of proton by interchanging the up and down quarks, e.g. $u^{(n)} = d^{(p)}$ etc.

The $\textsf{n}$th \emph{moment} of the PDF can be defined as:
\begin{equation}
	\mathfrak{q}^{(N)}_\textsf{n}	\equiv		\int_0^1 	x^{\textsf{n}-1}		\mathfrak{q}^{(N)} (x)		\ \mathrm{d}x 	\
\end{equation}

The probability distributions are normalised so that the first moments $\mathfrak{q}^{(N)}_1$ give the net number of the partons in the hadron:
\begin{equation}
	\int_0^1 	\left(	\mathfrak{q}^{(N)} (x)	 -\bar{\mathfrak{q}}^{(N)} (x)	\right)		\mathrm{d} x 		=	\ \#_\mathfrak{q}	\,.
\end{equation}

The second moments $\mathfrak{q}^{(N)}_2$ return the averaged longitudinal fraction:
\begin{equation}
	\int_0^1 	x 	\left(	\mathfrak{q}^{(N)} (x)	 +\bar{\mathfrak{q}}^{(N)} (x)	\right)		\mathrm{d} x 		=	\ \langle x \rangle_\mathfrak{q}	\,.
\end{equation}

One can expand the operator product of two quark currents in the deep inelastic scattering process $e + N \to e +X$, where $X$ is a generic hadronic final state. The leading terms in OPE is given by twist-2 operators, and higher twist operators get suppressed by factors of inverse momentum transfer. By equating the dispersion integrals with contours lying in physical and unphysical regions of $1/x$ complex plane, we drive \cite{Belyaev_Ioffe_1,*Belyaev_Ioffe_2}:
\begin{equation}
	\langle \mathcal{O}^\mathfrak{q}_{{\mu_1} \dots {\mu_\textsf{n}}}		\rangle	=		
			\frac{1}{m_N}		\left(	p_{\mu_1} \ldots p_{\mu_\textsf{n}}		-\text{trace}	\right)
			\left(	\mathfrak{q}^{(N)}_\textsf{n}		+ (-1)^\textsf{n}	\ \overline{\mathfrak{q}}^{(N)}_\textsf{n}		\right)
\end{equation}

These equations are termed as \emph{moment sum rules}, and relate the moments of the distribution functions to the matrix element of twist-2 operators. 
By including the radiative correction through operator rescaling, it can be shown that moments of PDF, and thus matrix elements of twist-2 operator are logarithmically scale dependent.%
\footnote{This result obtained from operator renormalisation analysis agrees with Dokshitzer–Gribov–Lipatov–Altarelli–Parisi (DGLAP) equations for evolution of the parton splitting functions \cite{DGLAP_1,DGLAP_2,DGLAP_3}.}

In the theory of EWDM, we are especially interested in the matrix element of spin-2 twist-2 operator which is obtained from the second moments of quark and anti-quark distributions:
\begin{equation}
	\langle \mathcal{O}^\mathfrak{q}_{\mu\nu}	\rangle	=		
		\frac{1}{m_N}		\left(	p_\mu p_\nu	-\frac{1}{4}	m_N^2	\eta_{\mu\nu}	\right)
		\left(	\mathfrak{q}^{(N)}_2		+ 	\ \overline{\mathfrak{q}}^{(N)}_2		\right)
\end{equation}

The PDF's used in calculation of the matrix element are available at different scales, so we only need to decide about the appropriate energy level to evaluate them. It turns out that due to asymptotic freedom, in lower energy regions the uncertainty arising from the perturbative expansion in $\alpha_\mathrm{s}$ increases \cite{EFT_DM}. Therefore, we evaluate the matrix element at the factorisation scale $\mu_\mathrm{uv} = m_\mathrm{z}$.

We use the second moments of PDF's for proton provided by \cite{q_2} evaluated at $\mu_\mathrm{uv} = m_\mathrm{z}$ which are presented in table \ref{tab:q_2}. 
It can be seen that 2\textsuperscript{nd} moments of valence quarks are an order of magnitude larger than those of the sea quarks.


\subsection{Wilson Coefficients}

The Wilson Coefficients and mass functions have been actually computed by several research groups, but for some reason there are some discrepancies in the literature. We review the calculations to assess the accuracy of the published works, and present our results in this section for comparison.%
\footnote{Unless explicitly specified otherwise, our calculations agrees with the \cite{DD_EWDM} results, including the factor of 2 correction for Z-boson contribution in \cite{QCD_corrn}. 
More explicitly, we reached to a faintly different Higgs mass function $g_h$, as explained in the footnote \ref{ft:g_h}.

Regarding \cite{EWDM_Essig}, In the large DM mass limit $m_\mathrm{v}/M_\chi \to 0$, both scalar and twist-2 quark coefficients give the same results as our calculation,  although the mass functions $f_\mathrm{II}$ and $f_\mathrm{III}$ are formulated in a slightly different form. 
About gluon contribution, the 2-loop scattering diagrams are presented, but not evaluated explicitly. 
All the Wilson coefficients in their work has the same sign, therefore the accidental cancellation between different terms (c.f. section \ref{app:L_Ren_Xn} for details) does not occur.
The higher-dimensional operator is explained, however it is not taken into account in computation of the scattering amplitude.

About reference \cite{Hill_I}, the quark scalar $c^{(0)}_{U/D}$ and twist-2 $c^{(2)}_{U/D}$ Wilson coefficients lead to the same outcome as our results, in the limit of heavy dark matter particle $m_\mathrm{v}/M_\chi \to 0$; whilst the evaluated integrals take a different form. 
This is also true of gluon coefficients $c^{(0)}_g$ except the top flavour contribution to the Z-bosons mediated interactions. 
It is the very last term in expression for $c^{(0)}_g$ and does not match our calculations of $C^\mathrm{S}_\mathfrak{g}$, in the same limit.  

As to \cite{MDM}, about the scalar operator, although the coefficient induced by the Box diagram is different, our results matches for the Higgs-mediated penguin diagram, up to an overall constant. 
The expression for the twist-2 operator is also different and has an opposite sign. The contribution from the gluon operator was not taken into account in this reference.}%

In contrast to the scalar bilinears, the twist operators are scale dependant even at leading order of $\alpha_\mathrm{s}$ \cite{CT2_RGE}. However, there is no need to re-evaluate the twist-2 coefficients at low energy region. Since, as discussed, we adopted PDF's at UV scale 
$\mu_\mathrm{uv} = m_\mathrm{z}$ where the effective Lagrangian is matched with the full theory.  This choice would also avoid the errors such as quark mass threshold, resulting from evolving down the Wilson coefficients.


\subsubsection{EWDM -- Quark Scattering}

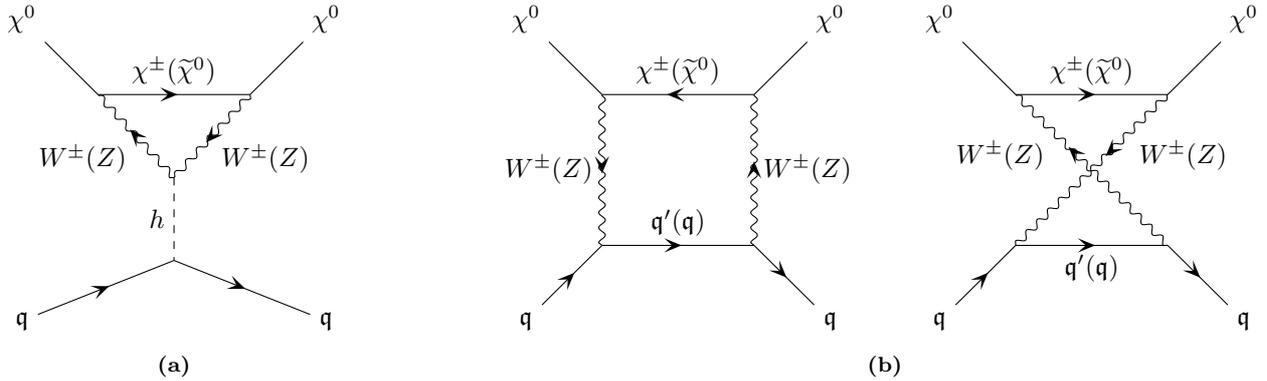
\begin{figure} [t] 				
	\begin{subfigure}{.3\linewidth}
		\centering

		\begin{tikzpicture}
		\begin{feynman}
			\vertex (a) at (0,0-.1);
			\vertex (b) at (-1,1);				
			\vertex (c) at (1,1);			
			\vertex (d) at (0,-1-.2);
			\vertex (i1) at (-2,2) {$\chi^0$};
			\vertex (f1) at (2,2) {$\chi^0$};
			\vertex (i2) at (-2,-2) {$\mathfrak{q}$};
			\vertex (f2) at (2,-2)  {$\mathfrak{q}$};
		\diagram* {
			(i1) -- [plain] (b)  -- [fermion, edge label=$\chi^\pm (\widetilde{\chi}^0)$] (c) -- [plain] (f1),
			(b) -- [anti charged boson, edge label'=$W^\pm (Z)$] (a) -- [anti charged boson, edge label'=$W^\pm (Z)$] (c),
			(a) -- [scalar, edge label'=$h$] (d),
			(i2) -- [fermion] (d) -- [fermion] (f2),
		};		
		\end{feynman}
		\end{tikzpicture}

		\subcaption{}
		\label{fig:Xq_R_triangle}
	\end{subfigure}%
	\begin{subfigure}{.8\linewidth}
		\centering
		\begin{subfigure}{.4\linewidth}
			\centering
		
			\begin{tikzpicture}
			\begin{feynman}
				\vertex (i1) at (-2,2) {$\chi^0$};
				\vertex (f1) at (2,2)  {$\chi^0$};
				\vertex (a) at (-1,1);
				\vertex (b) at (1,1);
				\vertex (c) at (-1,-1);			
				\vertex (d) at (1,-1);			
				\vertex (i2) at (-2,-2) {$\mathfrak{q}$};
				\vertex (f2) at (2,-2) {$\mathfrak{q}$};
			\diagram* {
				(i1) -- [plain] (a)  -- [anti fermion, edge label=$\chi^\pm (\widetilde{\chi}^0)$] (b) -- [plain] (f1),
				(a) -- [charged boson, edge label'=$W^\pm (Z)$] (c),
				(b) -- [anti charged boson, edge label=$W^\pm (Z)$] (d),			
				(i2) -- [fermion] (c)  -- [fermion, edge label=$\mathfrak{q}'(\mathfrak{q})$] (d) -- [fermion] (f2),
			};				
			\end{feynman}
			\end{tikzpicture}
			
		\end{subfigure}%
		\begin{subfigure}{.4\linewidth}
			\centering
		
			\begin{tikzpicture}
			\begin{feynman}
				\vertex (i1) at (-2,2) {$\chi^0$};
				\vertex (f1) at (2,2)  {$\chi^0$};
				\vertex (a) at (-1,1);
				\vertex (b) at (1,1);
				\vertex (c) at (-1,-1);			
				\vertex (d) at (1,-1);			
				\vertex (i2) at (-2,-2) {$\mathfrak{q}$};
				\vertex (f2) at (2,-2) {$\mathfrak{q}$};
			\diagram* {
				(i1) -- [plain] (a)  -- [fermion, edge label=$\chi^\pm (\widetilde{\chi}^0)$] (b) -- [plain] (f1),
				(a) -- [anti charged boson4, edge label'=$W^\pm (Z)$, near start] (d),
				(b) -- [charged boson4, edge label=$W^\pm (Z)$, near start] (c),			
				(i2) -- [fermion] (c)  -- [fermion, edge label'=$\mathfrak{q}'(\mathfrak{q})$] (d) -- [fermion] (f2),
			};				
			\end{feynman}
			\end{tikzpicture}
			
		\end{subfigure}%
		
		\subcaption{}
		\label{fig:Xq_R_Box}
	\end{subfigure}
	\caption{One-loop diagrams that contribute to the effective interactions of EWDM with quarks at leading order through renormalisable couplings. 
	 For real representations, only $Z$ vector can mediate the interaction; whereas in pseudo-real models, both neutral $Z$ and charged $W^\pm$ gauge bosons are allowed.}
	\label{fig:Xq_R}
\end{figure}

Since no tree-level interactions are allowed via renormalisable couplings to gauge bosons, dark matter scatters off quarks at one loop level. As shown in figure \ref{fig:Xq_R}, this happens through Higgs-exchange Penguin  \eqref{fig:Xq_R_triangle} and Box \eqref{fig:Xq_R_Box} diagrams \cite{EWDM_Essig}.%
\footnote{There also other penguin diagrams with $Z$ and $\gamma$ mediators. It turns out that the amplitude for these diagrams vanish, and therefore they are not shown in the figure.}

The effective couplings are model dependent, and for a general n-tuplet with hyper-charge $Y$ are given by:%
\begin{subequations}
	\label{eq:Cq_R}
\begin{gather}
	\label{eq:Cq_S}
	C_\mathfrak{q}^\mathrm{s} =	\frac{\alpha_\mathrm{w}^2}{4 m_h^2}	\left[ \frac{1}{8m_\mathrm{w}} (n^2 - 4y^2 -1) \, g_\mathrm{h}(w)	
		+\frac{1}{2 \mathrm{c}_\mathrm{w}^4 m_\mathrm{z}} y^2 g_\mathrm{h}(z) \right]	+
		\frac{ 3 \alpha_\mathrm{w}^2 }{ \mathrm{c}_\mathrm{w}^4 m_\mathrm{z}^3 } \, y^2 ( {c_\mathfrak{q}^\mathrm{v}}^2 - {c_\mathfrak{q}^\mathrm{a}}^2 ) \, g_\mathrm{box}(z)	\,,	\\
	C_\mathfrak{q}^{\mathrm{t}_i} =	\frac{\alpha_\mathrm{w}^2}{8 m_\mathrm{w}^3}	 (n^2 - 4y^2 -1) \, g_{\mathrm{t}_i}(w)	+	
		\frac{ \alpha_\mathrm{w}^2 }{ \mathrm{c}_\mathrm{w}^4 m_\mathrm{z}^3 } \, y^2 ( {c_\mathfrak{q}^\mathrm{v}}^2 +  {c_\mathfrak{q}^\mathrm{a}}^2 ) \, g_{\mathrm{t}_i}(z)	\,,
				\qquad		i=1,2 \,.
\end{gather}
\end{subequations}

where $w \equiv (m_\mathrm{w} / M_\chi)^2$ and $z \equiv (m_\mathrm{z} / M_\chi)^2$. Quark-Z boson vector and axial-vector couplings are 
$c_\mathfrak{q}^\mathrm{v} 	\equiv T^{(3)}_\mathfrak{q} 		-2 Q_\mathfrak{q} \mathrm{s}^2_\mathrm{w}$ and 
$c_\mathfrak{q}^\mathrm{a} =  		T^{(3)}_\mathfrak{q}$ 
respectively. 

The terms proportional to $g_\mathrm{H}$ in \eqref{eq:Cq_S} are generated by the penguin diagram \ref{fig:Xq_R_triangle} mediated by Higgs boson , and the rest of the terms in equations \eqref{eq:Cq_R} are induced by the box diagrams \ref{fig:Xq_R_Box}.

Scalar Higgs $g_\mathrm{h}$, box $g_\mathrm{box}$, twist-1 $g_{\mathrm{t}_1}$ and twist-2 $g_{\mathrm{t}_2}$ mass functions are defined as:%
\footnote{\label{ft:g_h}
The Higgs mass function $g_\mathrm{h}$ is slightly different from $g_H$ in \cite{DD_EWDM}. It originates from a sign difference in the expression for reduction of the vector integral to scalar correlation functions\\
$ 2 M_\chi^2 \ B_1 ^{(1,2)} (M_\chi^2 | m_\mathfrak{q}, m_\mathfrak{q} )	=	A_0^{(2)} (m_\mathfrak{q})		-B_0	\pm \left( 2 M_\chi^2 -m_\mathfrak{q}^2 \right) B_0 ^{(1,2)} \,,$ 
where we used minus sign for the last term. In the large DM mass limit $M_\chi \gg m_\mathrm{w}$, there is a discrepancy of a factor of a few, so the final results are not significantly affected.}
\begin{subequations}
\begin{align}
	g_\mathrm{h} (x)		&=	2 \sqrt{x} \left( 1 -\ln x \right)		- \frac{4}{\sqrt{4-x}} ( 3-x )	\atg \sqrt{ 4/x -1}	\,,		\\
	g_\mathrm{box} (x)		&=	\frac{\sqrt{x}}{24}	\left(  2 -x \ln x  \right)		+ \frac{1}{12 \sqrt{4-x}}		\left( 4 -2x +x^2 \right)		\atg \sqrt{ 4/x -1}	\,,	\\
	g_\mathrm{t_1} (x)		&=	\frac{\sqrt{x}}{12}	\left[  1 -2x 	-x(2-x)\ln x  \right]		+ \frac{1}{6} \sqrt{4-x}		\left( 2 +x^2 \right)		\atg \sqrt{ 4/x -1}	\,,	\\	
	g_\mathrm{t_2} (x)		&=	- \frac{\sqrt{x}}{4}	\left[  1 -2x 	-x(2-x)\ln x  \right]		+ \frac{x}{2 \sqrt{4-x}}		\left( 2 -4x +x^2 \right)		\atg \sqrt{ 4/x -1}	\,.	
\end{align}
\end{subequations}

In the limit of large dark matter mass $x \to 0$, the mass functions take the following values:
\begin{subequations}
\begin{align}
	g_\mathrm{h} (x)		&\to 	- 3 \pi	\,,			\\
	g_\mathrm{box} (x)		&\to 	\frac{\pi}{12}	\,,		\\
	g_\mathrm{t_1} (x)		&\to 	\frac{\pi}{3}	\,,		\\
	g_\mathrm{t_2} (x)		&\to 	0	\,.			
\end{align}
\end{subequations}

Consequently, even if dark matter is considerably heavier than electroweak gauge bosons $M_\chi \ll m_\mathrm{w}$, the quark Wilson coefficients except $C_\mathfrak{q}^{\mathrm{t}_2}$ are not suppressed to leading order $\mathcal{O}(\alpha^\mathrm{s})$.


\subsubsection{EWDM -- Gluon Scattering}

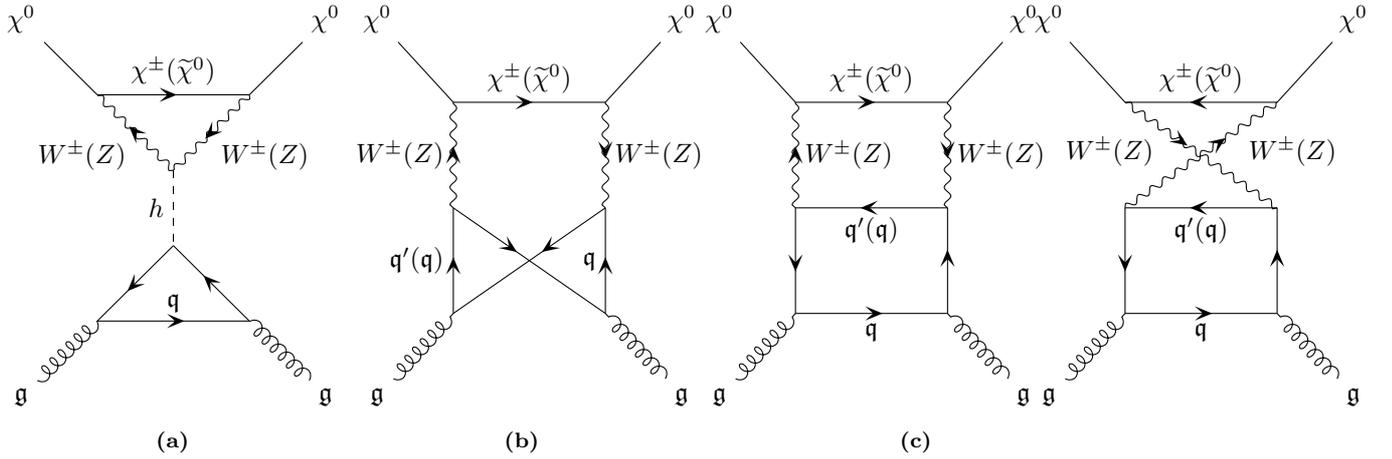
\begin{figure} [t] 				
	\begin{subfigure}{.27\linewidth}	
		\centering

		\begin{tikzpicture}
		\begin{feynman}
			\vertex (a) at (0,0);
			\vertex (i1) at (-2,3) {$\chi^0$};
			\vertex (f1) at (2,3)  {$\chi^0$};
			\vertex (c) at (-1,2);
			\vertex (d) at (1,2);
			\vertex (b) at (0,1);
			\vertex (a) at (0,0);			
			\vertex (e) at (-1,-1);			
			\vertex (f) at (1,-1);															
			\vertex (i2) at (-2,-2) {$\mathfrak{g}$};
			\vertex (f2) at (2,-2) {$\mathfrak{g}$};
		\diagram* {
			(i1) -- [plain] (c)  -- [fermion, edge label=$\chi^\pm (\widetilde{\chi}^0)$] (d) -- [plain] (f1),
			(c) -- [anti charged boson, edge label'=$W^\pm (Z)$] (b) -- [anti charged boson, edge label'=$W^\pm (Z)$] (d),
			(b) -- [scalar, edge label'=$h$] (a),
			(a) -- [anti fermion] (f) -- [anti fermion, edge label'=$\mathfrak{q}$] (e)-- [anti fermion] (a),	
			(i2) -- [gluon] (e),
			(f2) -- [gluon] (f),
		};		
		\end{feynman}
		\end{tikzpicture}

		\subcaption{}
		\label{fig:Xg_R_triangle}
	\end{subfigure}%
	\begin{subfigure}{.27\linewidth}		
		\centering

		\begin{tikzpicture}
		\begin{feynman}
			\vertex (i1) at (-2,4) {$\chi^0$};
			\vertex (f1) at (2,4)  {$\chi^0$};
			\vertex (a) at (-1,3-.1);
			\vertex (b) at (1,3-.1);
			\vertex (c) at (-1,1.5);			
			\vertex (d) at (1,1.5);
			\vertex (e) at (-1,0+.1);			
			\vertex (f) at (1,0+.1);			
			\vertex (i2) at (-2,-1) {$\mathfrak{g}$};
			\vertex (f2) at (2,-1) {$\mathfrak{g}$};
		\diagram* {
			(i1) -- [plain] (a)  -- [fermion, edge label=$\chi^\pm (\widetilde{\chi}^0)$] (b) -- [plain] (f1),
			(a) -- [anti charged boson, edge label'=$W^\pm (Z)$] (c),
			(b) -- [charged boson, edge label=$W^\pm (Z)$] (d),
			(c) -- [fermion4] (f) -- [fermion, edge label=$\mathfrak{q}$] (d)--  [fermion4] (e)--[fermion, edge label=$\mathfrak{q}'(\mathfrak{q})$] (c),	
			(i2) -- [gluon] (e),
			(f2) -- [gluon] (f),
		};				
		\end{feynman}
		\end{tikzpicture}

		\subcaption{}
		\label{fig:Xg_R_Box_1g}
	\end{subfigure}%
	\begin{subfigure}{.34\linewidth}	
		\centering

		\begin{subfigure}{.75\linewidth}
		\centering
		
		\begin{tikzpicture}
		\begin{feynman}
			\vertex (i1) at (-2,4) {$\chi^0$};
			\vertex (f1) at (2,4)  {$\chi^0$};
			\vertex (a) at (-1,3-.1);
			\vertex (b) at (1,3-.1);
			\vertex (c) at (-1,1.5);			
			\vertex (d) at (1,1.5);
			\vertex (e) at (-1,0+.1);			
			\vertex (f) at (1,0+.1);			
			\vertex (i2) at (-2,-1) {$\mathfrak{g}$};
			\vertex (f2) at (2,-1) {$\mathfrak{g}$};
		\diagram* {
			(i1) -- [plain] (a)  -- [fermion, edge label=$\chi^\pm (\widetilde{\chi}^0)$] (b) -- [plain] (f1),
			(a) -- [anti charged boson, edge label=$W^\pm (Z)$] (c),
			(b) -- [charged boson, edge label=$W^\pm (Z)$] (d),
			(c) -- [anti fermion, edge label'=$\mathfrak{q}'(\mathfrak{q})$] (d) -- [anti fermion] (f)-- [anti fermion, edge label=$\mathfrak{q}$] (e)-- [anti fermion] (c),	
			(i2) -- [gluon] (e),
			(f2) -- [gluon] (f),
		};					
		\end{feynman}
		\end{tikzpicture}

		\end{subfigure}%
		\begin{subfigure}{.3\linewidth}		
		\centering
		
		\begin{tikzpicture}
		\begin{feynman}
			\vertex (i1) at (-2,4) {$\chi^0$};
			\vertex (f1) at (2,4)  {$\chi^0$};
			\vertex (a) at (-1,3-.1);
			\vertex (b) at (1,3-.1);
			\vertex (c) at (-1,1.5);			
			\vertex (d) at (1,1.5);
			\vertex (e) at (-1,0+.1);			
			\vertex (f) at (1,0+.1);			
			\vertex (i2) at (-2,-1) {$\mathfrak{g}$};
			\vertex (f2) at (2,-1) {$\mathfrak{g}$};
		\diagram* {
			(i1) -- [plain] (a)  -- [anti fermion, edge label=$\chi^\pm (\widetilde{\chi}^0)$] (b) -- [plain] (f1),
			(a) -- [charged boson4, edge label'=$W^\pm (Z)$, near start] (d),
			(b) -- [anti charged boson4, edge label=$W^\pm (Z)$, near start] (c),			
			(c) -- [anti fermion, edge label'=$\mathfrak{q}'(\mathfrak{q})$] (d) -- [anti fermion] (f)-- [anti fermion, edge label=$\mathfrak{q}$] (e)-- [anti fermion] (c),	
			(i2) -- [gluon] (e),
			(f2) -- [gluon] (f),
		};				
		\end{feynman}
		\end{tikzpicture}
		
		\end{subfigure}%

		\subcaption{}
		\label{fig:Xg_R_Box_2g}
	\end{subfigure}
	\caption{Two-loop diagrams that contribute to the interactions of dark matter with gluon via renormalisable couplings.
	In real models, only $Z$-boson can mediate the reactions; whereas in pseudo-real representations, both neutral $Z$ and charged $W^\pm$ gauge vectors are involved.}
	\label{fig:Xg_R}
\end{figure}

Vector fields cannot mediate the Penguin diagram of figure \ref{fig:Xg_h3}, so at the level of renormalisable couplings, EWDM will only interact with gluons through two-loop processes.

Figure \ref{fig:Xg_R}, presents the Feynman diagrams that yield the effective interactions of EWDM with gluons. 
The double-triangle diagram \ref{fig:Xg_R_triangle} obviously only gives rise to long-distance contributions as there is no other scale than mass of quarks circulating in the quark triangle integral. 
The same holds for double-box diagrams \ref{fig:Xg_R_Box_2g} where only one quark propagator emits the two gluon fields. That is because, the external momentum which is predominantly of order of the weak gauge boson mass cannot contribute to the quark box loop more than what virtual quark masses do. 
In contrast, the reaction \ref{fig:Xg_R_Box_1g} where each quark propagator is attached with one gluon field, yields short distance contribution. In this case, the mass of gauge bosons as the external momentum dominates the quark loop integral.

At first step, we need to evaluate the loop integral attached with gluon fields, in order to derive the polarisation functions of the neutral and charged week bosons in the gluon background. 
Due to current conservation, the longitudinal part of the two-point functions do not contribute \cite{Extl_field_SVZ}. Finally, the transversal polarisation function is used to compute the second loop, and thus to find the Wilson coefficient for the gluon scalar operator which is given by:
\begin{align}
	\label{eq:Cg_R}
	C^\mathrm{s}_\mathfrak{g}		&=		- \frac{\alpha^2_\mathrm{w}}{48 \, m_h^2}	\left[ \frac{1}{8m_\mathrm{w}} (n^2 - 4y^2 -1) \, g_\mathrm{h}(w)	
			+\frac{1}{2 \mathrm{c}_\mathrm{w}^4 m_\mathrm{z}} \, y^2 g_\mathrm{h}(z) \right]		\\		\nonumber
		&+	\frac{\alpha_\mathrm{w}^2}{4}	 \left\{		\vphantom{\sum_{\mathfrak{q}=u}^b}	
			\frac{1}{8 m_\mathrm{w}^3}		(n^2 - 4y^2 -1)	\left[ 2 g_\mathrm{box} (w)	+g_t (w,t)	\right]	\right.	\\	\nonumber
		&+	\left.	\frac{1}{2 \mathrm{c}^4_\mathrm{w} m_z^3}	\, y^2	\left[	\sum_{\mathfrak{q}=u}^b		
			( {c_\mathfrak{q}^\mathrm{v}}^2 + {c_\mathfrak{q}^\mathrm{a}}^2 )	\, g_\mathrm{box} (z)
			+{c_t^\mathrm{v}}^2	\left(	g_\mathrm{v} +I_\mathrm{v}	\right)	(z,t)	+{c_t^\mathrm{a}}^2	\left(	g_\mathrm{a} +I_\mathrm{a}	\right)	(z,t)	\right]	\right\}				
\end{align}

where $t \equiv (m_t / M_\chi)^2$, and mass of all flavours except top quark has been ignored.

The first line of the expression \eqref{eq:Cg_R} for the gluonic coefficient is induced by the two-triangle diagram \ref{fig:Xg_R_triangle}. It can be verified that this long distance contribution arising from the top quark loop is related to the effective coupling with the top flavour in diagram \ref{fig:Xq_R_triangle} through $C_\mathfrak{g}^\mathrm{s}	=	- C_t^\mathrm{s} /12$. 
The last two lines come from the two double-box diagrams \ref{fig:Xg_R_Box_1g} and \ref{fig:Xg_R_Box_2g}. 
In each case, the first term (the 2\textsuperscript{nd} line) is mediated by $W$ bosons, but the second one (3\textsuperscript{th} line) is generated by the $Z$-boson exchanges. For $W$-boson mediated interactions (2\textsuperscript{nd} line) the first term comes from the 1\textsuperscript{st} and 2\textsuperscript{nd} generations, whereas the mass function $g_t$ in the second term formulates the contribution from the 3\textsuperscript{th} generation of quarks:
\begin{align}
	g_t (x,y) 	&=	\frac{- x^{3/2}}{12 (y-x)}		- \frac{x^{3/2} y^2}{24 (y-x)^2}  \ln y		+ \frac{x^{5/2} (2y -x)}{24 (y-x)^2}  \ln x
			- \frac{x^{3/2} \sqrt{y} (2+y) \sqrt{4-y}}{12 (y-x)^2}  	\atg \sqrt{ \frac{4-y}{y}}			\nonumber	\\	
			&+ \frac{x  [ 2y (2 +2x -x^2)  +x (4 -2x +x^2) ] }{ 12 (y-x)^2 \sqrt{4-x}}	\atg \sqrt{ \frac{4-x}{x}}	
\end{align}

Concerning the $Z$-boson exchange reactions (3\textsuperscript{nd} line), the first term is generated by all quarks lighter than top flavour. The contribution from top quark is further decomposed into vector mass functions with analytical and non-analytical form of:
\begin{subequations}
\begin{align}
	g_\mathrm{v} (x,y)	&=		-  \frac{ \sqrt{x} \left( 12y^2 -xy +x^2 \right) }{ 12 \left( 4y -x \right)^2 }
		+ \frac{ x^{3/2} \left( 48 y^3  -20 xy^2  +12 x^2 y  -x^3 \right) }{ 24 \left( 4y -x \right)^3 }	\ln x	
		+ \frac{ x^{3/2} y^2 \left( 4y -7x \right) }{ 6 \left( 4y -x \right)^3 }	\ln 4y		\nonumber	\\
	&- \frac{ x^{3/2} \sqrt{y} \left( 16 y^3 -4 \left( 2 +7x \right) y^2  +14 \left( 2+x \right) y  +5x  \right]}{ 12 \left( 4y -x \right)^3 \sqrt{1-y}}	\atg \sqrt{ \frac{1-y}{y} }	\\	\nonumber
	&- \frac{ 48 \left( x^2 -2x +4 \right) y^3	-4x \left( 5x^2 -10 x +44 \right) y^2	+12 x^3 \left( x-2 \right) y	-x^3 \left( x^2 -2x +4 \right) }{ 12 \left( 4y -x \right)^3 \sqrt{4-x}}	\atg \sqrt{ \frac{4-x}{x}}	\,,	\\
	I_\mathrm{v} (x,y)	&=		- x^{3/2} y^2	\int_0^\infty		\frac{ \left(l +2y \right)	\left[ \left(2 -l \right)   \sqrt{l+4}	+l \sqrt{l}	\right]}{ l \left( l+x \right)^2	\left( l +4y	\right)^{5/2}}
								\ln 	\frac{ \sqrt{ l +4x } + \sqrt{l}}{2 \, \sqrt{x}}		\ \mathrm{d} l 		\,,
\end{align}

and axial vector analytical and non-analytical mass functions:
\begin{align}
	g_\mathrm{a} (x,y)	&=	\frac{ \sqrt{x} \left( 2y -x \right) }{ 4 \left( 4y -x \right) } 		+ \frac{ x^{3/2} \left( 8y^2 -8xy +x^2 \right) }{ 8 \left( 4y -x \right)^2}	\ln x
		- \frac{ x^{3/2} y^2 }{ \left( 4y -x \right)^2} 	\ln 4y			- \frac{ x^{3/2} \sqrt{y} \left( 2y^2 -y -1 \right)}{ \left( 4y -x \right)^2 \sqrt{1-y}}	\atg \sqrt{ \frac{1-y}{y} }	\nonumber	\\	
			&- \frac{ 8 x \left( x^2 -2x +1 \right) y	- \left( x^2 -2x +4 \right) \left( 8y^2 +x^2 \right)  }{ 4 \left( 4y -x \right)^2 \sqrt{4-x}}	\atg \sqrt{ \frac{4-x}{x}}	\,,	\\
	I_\mathrm{a} (x,y)	&=		x^{3/2} y^2	\int_0^\infty		\frac{ \left(l +4y \right)	\left[ \left(2 -l \right)   \sqrt{l+4}	+l \sqrt{l}	\right]}{ l \left( l+x \right)^2	\left( l +4y	\right)^{5/2}}
								\ln 	\frac{ \sqrt{ l +4x } + \sqrt{l}}{2 \, \sqrt{x}}		\ \mathrm{d} l 		\,. 	
\end{align}
\end{subequations}	

In the high DM mass limit $x,y \to 0$, the mass functions reduce to:
\begin{subequations}
\begin{align}
	g_t (x,y)				&\to 	\frac{\pi}{12}	\frac{1}{\left( 1 +r^2 \right)^2} 	\,,			\\
	g_\mathrm{v} (x,y)		&\to 	\frac{\pi}{24}	\ \frac{ -2 +5r +28r^3 -88 r^4 +96 r^6 }{ \left( 1 -4r^2 \right)^3 }	\,,		\\
	I_\mathrm{v} (z,t)		&\to 	- 0.19	\,,		\\
	g_\mathrm{a} (x,y)		&\to 	\frac{\pi}{4}	\ \frac{ 1 -2r -2r^2 +8r^4 }{ \left( 1 -4r^2 \right)^2 }	\,,		\\
	I_\mathrm{a} (z,t)		&\to 	0.36	\,,
\end{align}
\end{subequations}

with $ r \equiv \sqrt{x/y}$. 
Therefore, the Wilson coefficient for gluon operator will not be suppressed, in the limit of large dark matter mass.



\subsection{Cross-section}
\label{app:L_Ren_Xn}

The effective amplitude for EWDM - nucleon scattering is obtained form the S-matrix element of the scalar and twist interactions induced by $\mathcal{L}_\mathrm{R}$ in the non-relativistic limit:
\begin{equation}
	f_N^\mathrm{R}	=	\langle \mathcal{L}_\mathrm{R}	\rangle		=
			m_N \left[	\sum_{\mathfrak{q}	=d,u,s}		C_\mathfrak{q}^\mathrm{s}	 \, f_{\mathrm{T}\mathfrak{q}}
			+ \frac{3}{4} 	\sum_{\mathfrak{q} = d}^b 		\left(	\mathfrak{q}^{(N)}_2		+ 	\ \overline{\mathfrak{q}}^{(N)}_2		\right)
						\left(	C_\mathfrak{q}^{\mathrm{t}_1}	+	C_\mathfrak{q}^{\mathrm{t}_2}	\right)
			- \frac{8}{9}	C_\mathfrak{g}^\mathrm{s}		\, f_{\mathrm{T}\mathfrak{g}}		\right]
\end{equation}

Notice the difference in the number of active flavours for scalar and twist-2 quark contributions as the associated Wilson coeffifcents $C_\mathfrak{q}^\mathrm{s}$ and $C_\mathfrak{q}^{\mathrm{t}_i}$ are evaluated at different scales of $\mu_\mathrm{had}$ and $\mu_\mathrm{uv}$ respectively.


The spin-independent scattering cross-section with nucleon can be obtained from the effective amplitude:
\begin{equation}
	\sigma_N 	=	\frac{4}{\pi}	m_{\chi N}^2	\left| f_N^\mathrm{R} \right|^2	\,.
\end{equation}

As discussed in the previous sections, the Wilson coefficients generated by renormalisable couplings depend on EWDM mass through the mass functions. However, when dark matter is much heavier than the electroweak gauge mediators $M_\chi \ll m_\mathrm{w}$, these effective coefficients become independent of $M_\chi$. As a result, the spin-independent scattering cross-section will not be sensitive to EWDM mass.

It is observed that the effective coefficients generated by twist-2 operators are positive whereas other coefficients are negative. Additionally, their amplitudes are comparable with each other. This leads to an accidental cancellation among different contributions which reduces the scattering cross-section by an order of magnitude. As a consequence, the SI cross-section remains below the current direct detection bounds \cite{DD_Wino}.

The complex doublet model has the lowest dimension of representation and thus receives the smallest $W$-boson contribution due to the factor of $[n^2 - (4y^2 +1)]/8$. In addition, $\mathbb{C}2$ multiplet lacks those diagrams that include the light charged dark propagator $\chi^+_1$ in the loops. As a result, the effective coupling significantly reduces, and the spectrum becomes more dependent on DM mass, especially below TeV scale. The scattering cross-section for the pseudo-real model, therefore stays well below the neutrino floor.


\nocite{apsrev42Control}
\bibliography{DD_EWDM_ref, DD_EWDMnotes}		

\end{document}